\documentclass[sigplan,10pt,nonacm]{acmart}
\renewcommand\footnotetextcopyrightpermission[1]{}
\AtBeginDocument{%
  }

\setcopyright{acmlicensed}
\copyrightyear{2026}
\acmYear{2026}
\acmDOI{xxxxxxx.xxxxxxx}
\acmISBN{978-1-4503-xxxx-x/xxxx/xx}

\usepackage[dvipsnames,table]{xcolor}

\usepackage{xspace}
\usepackage{amsmath}
\usepackage{bold-extra}
\usepackage{graphbox}

\usepackage[export]{adjustbox}
\usepackage{setspace}

\usepackage{etoolbox}
\makeatletter
\preto{\@verbatim}{\topsep=1pt \partopsep=1pt}
\makeatother

\usepackage[compact]{titlesec}
\titlespacing*{\section}{0pt}{5pt}{2pt}
\titlespacing*{\subsection}{0pt}{4pt}{1pt}
\titlespacing*{\subsubsection}{0pt}{4pt}{1pt}

\usepackage[ruled,linesnumbered]{algorithm2e}
\SetArgSty{textup}

\usepackage{float}
\usepackage{enumitem}
\SetLabelAlign{center}{\strut\smash{\parbox[t]\labelwidth{\centering#1}}}

\usepackage{caption}
\setlist{leftmargin=9pt,itemsep=2pt,topsep=3pt}

\usepackage[labelformat=simple]{subcaption}

\usepackage{xurl}
\usepackage{hyperref}
\hypersetup{
    colorlinks,
    linkcolor={green!70!black},
    citecolor={red!70!black},
    urlcolor={blue!70!black}
}

\usepackage{tikz}
\usetikzlibrary{shapes.geometric}
\usetikzlibrary{arrows.meta}

\usepackage{bm}
\usepackage{wasysym}
\usepackage{amsthm}

\usepackage{array}
\usepackage{makecell}
\usepackage{multirow}

\usepackage{lipsum}
\usepackage{duckuments}

\usepackage{listings}
\usepackage{pdfpages}

\usepackage[symbol*]{footmisc}

\newcommand{\darkred}[1]{\textcolor{BrickRed}{#1}\xspace}

\newcommand{\revise}[1]{\textcolor{black}{#1}\xspace}

\newcommand{\crossword}{\textsc{Crossword}\xspace}
\newcommand{\summerset}{\revise{Gazette\xspace}}
\newcommand{\tlaplus}{$\textrm{TLA}^{+}$\xspace}

\newcommand{\paratitle}[1]{\vspace{0.8pt}\noindent\textbf{#1}.}
\newcommand{\paratitlenodot}[1]{\vspace{0.8pt}\noindent\textbf{#1}}

\newcommand*\circleb[1]{\tikz[baseline=(char.base)]{
    \node[shape=circle,fill,inner sep=.5pt](char){\textcolor{white}{\small #1}};}}
\newcommand*\circlew[1]{\tikz[baseline=(char.base)]{
    \node[shape=circle,draw,inner sep=.5pt](char){\textcolor{black}{\small #1}};}}
    
\newcommand{\floatcap}[2]{\caption[#1]{\textbf{#1.} \textit{#2}}}
\newcommand{\floatcapnodot}[2]{\caption[#1]{\textbf{#1} \textit{#2}}}
\newcommand{\floatcapoffig}[2]{\captionof{figure}{\textbf{#1.} \textit{#2}}}

\begin{document}

\widowpenalty=0
\displaywidowpenalty=0
\clubpenalty=0

\title{\vspace*{-22pt}\LARGE \crossword: \revise{Adaptive Consensus for Dynamic Data-Heavy Workloads}}


\author{\vspace*{-11pt}Guanzhou Hu}
\affiliation{%
  \institution{Univeristy of Wisconsin--Madison}
  \city{}
  \country{}}

\author{\vspace*{-11pt}Yiwei Chen}
\affiliation{%
  \institution{Carnegie Mellon University\footnotemark}
  \city{}
  \country{}}

\author{\vspace*{-7pt}Andrea C. Arpaci-Dusseau}
\affiliation{%
  \institution{Univeristy of Wisconsin--Madison}
  \city{}
  \country{}}

\author{\vspace*{-7pt}Remzi H. Arpaci-Dusseau}
\affiliation{%
  \institution{Univeristy of Wisconsin--Madison}
  \city{}
  \country{}}

\renewcommand{\shortauthors}{Paper \#XXX}

\begin{abstract}
  We present \crossword, a flexible consensus protocol for \revise{dynamic data-heavy workloads}, a rising challenge in the cloud \revise{where replication payload sizes span a wide spectrum and introduce sporadic bandwidth stress}. \crossword applies per-instance erasure coding and distributes coded shards intelligently to reduce critical-path data transfer significantly \revise{when desirable}. Unlike previous approaches that statically assign shards to servers, \crossword enables an adaptive tradeoff between the assignment of shards and quorum size in reaction to dynamic workloads and network conditions, while always retaining the availability guarantee of classic protocols. \crossword handles leader failover gracefully by employing a lazy follower gossiping mechanism that incurs minimal impact on critical-path performance.


We implement \crossword (along with relevant protocols) in \summerset, a distributed, replicated, and protocol-generic key-value store written in async Rust. We evaluate \crossword comprehensively to show that it matches the best performance among previous protocols (MultiPaxos, Raft, RSPaxos, and CRaft) in static scenarios, and outperforms them by up to 2.3x under dynamic workloads and network conditions. Our integration of \crossword with CockroachDB brings 1.32x higher aggregate throughput to TPC-C under 5-way replication. We will open-source \summerset upon publication.




\end{abstract}

\maketitle
\pagestyle{plain}

\footnotetext{$^*$Work done when at UW--Madison.}

\section{Introduction}
\label{sec:introduction}

\textit{Consensus protocols} are at the heart of most fault-tolerant distributed systems. They enable the fundamental operation of consistent replication, often with a log of state machine commands~\cite{smr-approach, raft}, across machines over an asynchronous network~\cite{paxos-parliament}. Protocols such as MultiPaxos~\cite{paxos-made-simple} and Raft~\cite{raft} prove indispensable in guarding the critical data of widely-deployed distributed systems, including but not limited to distributed file systems~\cite{gfs, colossus, polarfs}, object stores~\cite{s3-strong-consistency, hadoop-s3guard}, coordination services~\cite{chubby, zookeeper, kafka, redpanda}, cluster managers~\cite{etcd, firescroll, kubernetes, blackwater-raft}, and cloud databases~\cite{spanner, f1, cockroachdb, tidb, scylladb}.


Due to the common assumption that consensus workloads carry small values (e.g., \textasciitilde100 byte state machine commands~\cite{etcd}), network transport delay has been considered the major bottleneck in consensus protocols~\cite{paxos-made-simple}. Previous protocols were thus mainly designed to reduce message round-trips~\cite{paxos-made-simple, viewstamped-replication, raft, fast-paxos} and minimize communication delay~\cite{generalized-paxos, mencius, epaxos, epaxos-revisited, non-blocking-raft, copilots, consistency-aware-durability, skyros-nil-externality}, largely neglecting bandwidth constraints on the critical path.

Unfortunately, bandwidth constraints can no longer be neglected as modern distributed systems place increasingly heavier workloads onto consensus protocols. For example, cloud databases~\cite{spanner, f1, cockroachdb, tidb, scylladb} use Paxos or Raft to replicate redo/undo actions, each carrying KBs or even MBs of data per operation. When consensus instances carry large payloads, broadcasting them over the network and persisting them durably stresses the bandwidth aspect of the system.

\begin{figure}[t]
    \centering
    \includegraphics[align=c,width=0.74\linewidth]{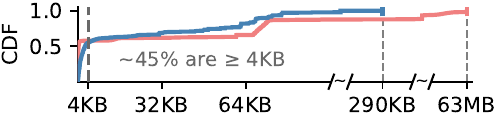}
    \hfill
    \includegraphics[align=c,width=0.24\linewidth]{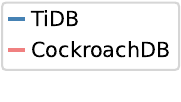}
    \vspace{-7pt}
    \floatcap{Raft replication payload size CDF in modern cloud databases}{Profiled by running 200-warehouses TPC-C.}
    \label{fig:db-motivation}
    \vspace{-10pt}
\end{figure}

Figure~\ref{fig:db-motivation} shows a payload size profile of the Raft module in two modern databases, TiDB~\cite{tidb} and CockroachDB~\cite{cockroachdb}, running TPC-C~\cite{tpc-c}. A considerable portion of the payloads are large ($\geqslant$ 4KB) and span a wide range of up to 290KB and 63MB, respectively. The dynamicity and diversity in hardware environment conditions and workload sizes further complicate the problem: small payloads are likely delay-bounded, while large payloads in a bandwidth-limited environment necessitate optimizations that reduce data size.

Previous research has integrated consensus protocols with \textit{erasure coding}~\cite{rs-coding, erasure-coding-azure}, allowing the leader node to transform the payload into shards, each having a fractional size of the original. These protocols, including RSPaxos~\cite{rs-paxos}, CRaft~\cite{craft} and variants~\cite{crs-raft, ecraft, hraft, flex-raft}, reduce bandwidth pressure by sending exactly one shard to each follower. However, they provide a degraded availability guarantee, exhibit ungraceful leader failover, and provide no flexibility in reaction to dynamic workloads and hardware conditions.

Instead of fixing the number of coded shards assigned per server to one, we treat it as a new dimension in the design space. Specifically, we establish an availability-preserving tradeoff between the number of shards per server ($c$) and the minimum accept quorum size ($q$). Using this intuitive but powerful result, we propose \crossword, a \textit{bandwidth-adaptive} consensus protocol that operates dynamically on the set of valid $[c,q]$ configurations, reducing the data transfer volume in bandwidth-constrained cases and minimizing the quorum size in delay-dominant scenarios. Moreover, \crossword employs a \textit{follower gossiping} mechanism to keep followers updated without interfering with critical-path operations, permitting graceful handling of leader failover.


\paratitle{Summary of Contributions}
\circleb{1} We recognize the increasing significance of dynamic data-heavy consensus workloads, and demonstrate the insufficiency of previous protocols under such workloads. \circleb{2} We propose \crossword, the first consensus protocol to our knowledge that establishes a runtime-dynamic tradeoff between data volume and quorum size using erasure coding; \crossword retains the availability guarantee and graceful failover behavior of classic protocols. \circleb{3} We build \summerset, a protocol-generic replicated key-value store, and implement six consensus protocols in a sum of 27.3k lines of async Rust. \circleb{4} We evaluate \crossword comprehensively to show that \crossword matches the best performance among previous protocols in static scenarios, and outperforms MultiPaxos/Raft by up to 2.3x and RSPaxos/CRaft by up to 1.9x under dynamic mixed workloads. \crossword recovers promptly after leader failover and sustains a consistent performance gain under macro-benchmarks. Integration with CockroachDB~\cite{cockroachdb} delivers 1.32x higher throughput to 200-warehouses TPC-C under 5-way replication. \circleb{5} We append a \tlaplus specification.


The rest of this paper is organized as follows. \S\ref{sec:background} details background and motivation.
\S\ref{sec:design} and \S\ref{sec:implementation} presents design and implementation, respectively.
\S\ref{sec:evaluation} presents evaluation results. \S\ref{sec:related-work} discusses related work and \S\ref{sec:conclusion} concludes.

\section{Background and Motivation}
\label{sec:background}

We provide background context and motivation.


\subsection{The Consensus Problem}
\label{sec:bg-consensus-problem}

Consensus is a fundamental problem that confronts any distributed system seeking consistency. We consider the common fail-stop model with an asynchronous network~\cite{paxos-made-simple}.


\paratitle{State Machine Replication (SMR)}
Given the ubiquity of the state machine approach to designing fault-tolerant distributed systems~\cite{smr-approach}, practical protocols often agree upon an ordered \textit{log} of state machine commands. This yields a particularly interesting variant of multi-decree consensus called \textit{state machine replication} (SMR). The replicated log is divided, either explicitly~\cite{paxos-made-simple} or implicitly~\cite{raft}, into single-decree \textit{instances}. Each instance carries a batch of state machine commands (e.g., \texttt{Get}s/\texttt{Put}s on a hash map) to be replicated and executed on all nodes. We assume \textit{linearizability}~\cite{linearizability} across instances; weaker consistency levels exist~\cite{practical-consistency-summary, sequential-vs-linearizability, regular-sequential-consistency, session-guarantees, eventual-consistency, tact-continuous, stalestores} but do not guarantee real-time serial ordering, a semantic that many systems demand.

\paratitle{Availability Requirements}
Beyond correctness, a practical consensus protocol must also offer \textit{availability}. This is usually measured in terms of the maximum number of node failures the protocol can tolerate (assuming ill-networked nodes fail) while allowing progress~\cite{practical-consistency-summary}. Most consensus protocols tolerate $f = \lfloor \frac{n}{2} \rfloor$ faults, where $n$ is an odd total number of nodes; this result is rooted in the use of majority quorums and is a property that should be preserved.




\subsection{Classic Consensus Protocols}
\label{sec:bg-typical-protocols}


For simplicity of terminology, we assume a collection of servers with one being the \textit{leader} responsible for serving client requests and proposing commands to \textit{followers}.

\paratitlenodot{MultiPaxos}~\cite{paxos-made-simple} is a composition of single-decree Paxos~\cite{paxos-parliament} instances with a ``covering-all'' prepare phase, which allows a leader to use a single round of \texttt{Prepare} messages to settle for a ballot number that stays valid in the absence of competing leaders. On the critical path, only accept phases happen, where the leader broadcasts \texttt{Accept} messages carrying the prepared ballot and a complete copy of the next (batch of) client request(s). Upon receiving a majority of replies, the leader commits this instance, executes contained commands, and replies to clients. To connect with classic terminology, one may think of the leader as the distinguished proposer.




\paratitlenodot{Raft}~\cite{raft} shares the same underlying mechanisms with MultiPaxos but introduces two differences. First, Raft has strong leadership; a follower node must convert to a candidate role and go through a \texttt{RequestVote} round to be elected as the leader with a higher term. Second, Raft operates on finer-grained log entries and groups entries into consensus instances implicitly through \texttt{AppendEntries} messages.


In this paper, we primarily use Paxos-style narration; all the design decisions are applicable to Raft variants due to their inherent duality~\cite{parallel-paxos-raft}. We use \textit{data} or \textit{payload} to refer to the commands contained in a consensus instance.


\subsection{Dynamic Data-Heavy Workloads}
\label{sec:bg-data-heavy-workloads}

Most previous consensus protocols are designed to minimize the impact of network delay. Precisely speaking, their performance metrics are the number of message rounds; empirical evaluations mostly assume byte-scale payloads~\cite{paxos-made-simple, raft, fast-paxos, generalized-paxos, epaxos, pbft, hotstuff}. These metrics make sense when data sizes are small and delay is the dominant bottleneck. However, this assumption is no longer accurate as modern distributed systems generate \textit{dynamic data-heavy} workloads \revise{where payload sizes span a wide range, causing bandwidth stress for some (but not all) instances}. We give examples below.


\paratitle{(a) Cloud Databases}
Cloud HTAP databases implement consensus protocols to provide a strongly-consistent storage layer abstraction to the SQL layer. F1/Spanner~\cite{f1, spanner} uses Paxos to maintain a consistent mapping of tablet data. CockroachDB~\cite{cockroachdb}, TiDB~\cite{tidb}, ScyllaDB~\cite{scylladb}, and rqlite~\cite{rqlite} use Raft to replicate user transactions, key-value updates, or the database redo/undo log itself. These workloads drive consensus protocols with up to MBs of data per instance.


\paratitle{(b) Object Storage}
Consensus protocols have seen extensive usage in object/key-value storage systems, including research proposals~\cite{gaios, craq} and industrial standards~\cite{s3-replication, daos, bigtable, dynamo}. Recent studies have reported large value sizes ranging from 10KB to 128MB in these systems~\cite{rocksdb-study, s3-hpc-workloads}.

\paratitle{(c) Metadata of Large-Scale Systems}
Many systems rely on consensus to manage critical metadata, either internally~\cite{gfs, colossus, etcd, firescroll} or through a coordination service~\cite{chubby, zookeeper, kafka, redpanda}. As the scales of modern systems increase, metadata workloads themselves become considerably heavier. For example, an Apache Storm study reported that ZooKeeper becomes a significant bottleneck as scheduling decisions exceed 1MB as the cluster grows beyond 100 nodes~\cite{zookeeper-storm}.


\paratitle{(d) Request Batching}
Batching is a ubiquitous technique used in systems expecting high levels of concurrency~\cite{paxos-made-live}. In the context of consensus, batching collects multiple client requests into a single instance, typically at millisecond-scale intervals, to prevent overloading the system with excessive coordination overhead. The presence of batching amplifies payload sizes, as now a consensus instance carries all client requests that arrive during one batching interval.

\paratitle{Current Workarounds}
Some systems separate data off to a weakly-consistent multi-version data store~\cite{niobe, gnothi, s3-replication, gfs, colossus} or opt in for chain replication~\cite{chain-replication, craq, chain-paxos}, both sacrificing latency by a multiplicative factor for improved throughput. A bandwidth-aware consensus protocol can retain one-round latency and potentially remove the need for these workarounds. Further discussions can be found in \S\ref{sec:related-bandwidth-techniques}.

\paratitle{Implications of Dynamic Data-Heavy Workloads}
Under data-heavy workloads, a third performance metric comes into play, which is the size of data to be transferred to and persisted by each node on the critical path. Assume a network link and a storage device both offer 400Mbps bandwidth for a consensus job. The lower-bound time to pass a single 1MB payload would be $\frac{1\text{MB}}{400\text{Mbps}} \times 2 \approx 43\text{ms}$ not including any overheads, which is comparable to wide-area RTTs.

Real workloads impose more complexity as they are a dynamic mix of light/heavy workloads (as shown in Figure~\ref{fig:db-motivation}) and can change over time. Furthermore, network and storage hardware conditions may span a wide spectrum and be delay-bounded and/or bandwidth-constrained. The significance of bandwidth pressure varies greatly across different situations.

\subsection{Erasure-Coded Consensus}
\label{sec:bg-previous-work}


\textit{Erasure coding} has been widely applied to networks~\cite{ecc-survey, network-coding}, storage~\cite{raid, erasure-coding-azure, geometric-partitioning, stair-codes, openec, oceanstore}, and memory hardware~\cite{memory-errors-eval}.


\paratitlenodot{Reed-Solomon (RS) code} is a \revise{standard} type of erasure code~\cite{rs-coding}. It splits data into $d$ shards and uses Galois fields to compute an adjustable number of $p$ parity shards of the same size, forming a \textit{codeword} of $n = d + p$ shards; by convention, this is an $(n, d)$-coding scheme. The original data can be reconstructed with $\leqslant p$ shards erased or with $\leqslant t = \frac{p}{2}$ shards corrupted. We are interested only in its erasure aspect. Other algorithms such as LRC~\cite{lrc-codes, erasure-coding-azure, wide-lrcs} exist \revise{but sacrifice recoverability (for faster reconstruction under single failures)}.


Erasure coding introduces less redundancy than full-copy replication, but it alone does not grant any ability to maintain consistency. Existing systems rely on a coordinator responsible for managing the distribution of shards~\cite{raid, erasure-coding-azure, wide-lrcs}. However, opportunities to combine it with replication exist. Several proposals have been made to integrate it with consensus; we describe two representative protocols.


\begin{figure}[t]
    \centering
    \includegraphics[width=0.9\linewidth]{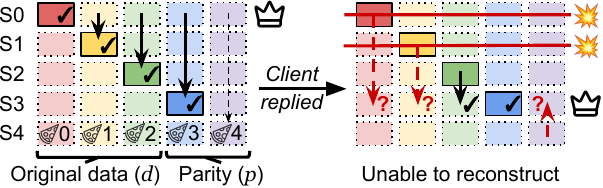}
    \vspace{-10pt}
    \floatcap{RSPaxos or CRaft under failures}{Both vulnerable to temporally close failures leaving the number of reachable shards $< d$, even with fallback mechanisms. See \S\ref{sec:bg-previous-work}.}
    \label{fig:concurrent-failures}
    \vspace{-12pt}
\end{figure}

\paratitlenodot{RSPaxos}~\cite{rs-paxos} is the first protocol equipping consensus with erasure coding. Suppose a cluster of $n$ servers where $n$ is odd. To start a new instance, an RSPaxos leader divides the payload into $d = m = \lceil \frac{n}{2} \rceil$ shards where $m$ represents the size of a simple majority, and appends $p = n - m$ parity shards to form a codeword of $n$ shards. It then transfers shard $i$ to follower $i$ and lets it persist that shard. Doing so reduces bandwidth consumption and storage cost to nearly $\frac{1}{m}$. Note that the leader still has the complete command in memory.

When collecting replies, however, a simple majority quorum is no longer enough to assert an instance as safe and alive. If the leader fails after acknowledging the client, a new leader may not be able to reach $d$ shards to reconstruct the data and execute the instance, hanging the system forever. \revise{Specifically,} RSPaxos provides a degraded fault-tolerance level of $f = \lfloor \frac{p}{2} \rfloor$ while waiting for a larger quorum size of $m + \lceil \frac{p}{2} \rceil$. This is a significant weakening to availability. For example, with 5 nodes, RSPaxos offers tolerance of only 1 node failure with a fixed quorum size of 4.


\paratitlenodot{CRaft}~\cite{craft} and variants~\cite{flex-raft, hraft} apply the same idea to Raft and behave identically on the critical path. To alleviate the degraded availability guarantee, CRaft introduces a fallback mechanism to switch to full-copy replication when a failure is detected. However, this does not fully solve the availability issue, as failures that happen before the completion of the fallback still lead to unavailability. Figure~\ref{fig:concurrent-failures} demonstrates this using the RS codeword space notation we introduce in \S\ref{sec:design-codeword-space}. Leader S0 commits an instance according to an S0{\textasciitilde}S3 quorum and the message to S4 is lost. If S0 and S1 both fail, a new leader cannot reach $d$ shards to reconstruct the acknowledged instance, effectively still offering $f = \lfloor \frac{p}{2} \rfloor$.

\begin{figure*}[t]
    \centering
    \begin{subfigure}[t]{0.185\linewidth}
        \centering
        \includegraphics[width=0.88\columnwidth]{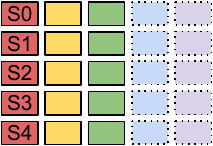}
        \captionsetup{justification=centering}
        \caption{MultiPaxos \& Raft}
        \label{fig:policy-multipaxos}
    \end{subfigure}\hspace{10pt}%
    \begin{subfigure}[t]{0.185\linewidth}
        \centering
        \includegraphics[width=0.88\columnwidth]{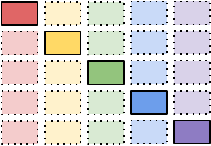}
        \captionsetup{justification=centering}
        \caption{RSPaxos \& CRaft,\\also Balanced RR, $c=1$}
        \label{fig:policy-rspaxos}
    \end{subfigure}\hspace{10pt}%
    \begin{subfigure}[t]{0.185\linewidth}
        \centering
        \includegraphics[width=0.88\columnwidth]{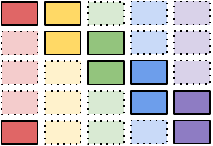}
        \captionsetup{justification=centering}
        \caption{\crossword,\\Balanced RR, $c=2$}
        \label{fig:policy-balanced_rr_2}
    \end{subfigure}\hspace{10pt}%
    \begin{subfigure}[t]{0.185\linewidth}
        \centering
        \includegraphics[width=0.88\columnwidth]{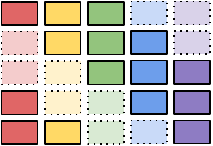}
        \captionsetup{justification=centering}
        \caption{\crossword,\\Balanced RR, $c=3$}
        \label{fig:policy-balanced_rr_3}
    \end{subfigure}\hspace{10pt}%
    \begin{subfigure}[t]{0.178\linewidth}
        \centering
        \includegraphics[width=0.88\columnwidth]{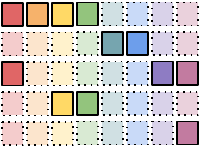}
        \captionsetup{justification=centering}
        \caption{\crossword,\\Unbalanced case}
        \label{fig:policy-unbalanced}
    \end{subfigure}
    
    \addtocounter{figure}{-1}  
    \captionsetup{justification=raggedright}
    \vspace{-4pt}
    \floatcapoffig{Assignment policy examples}{\textbf{(a)} assigns the original data to all servers. \textbf{(b)} assigns shard $i$ to server $i$ diagonally, which is itself also a balanced Round-Robin (RR) assignment with shard count $c=1$. \textbf{(c)} shows a balanced RR assignment with $c=2$. \textbf{(d)} shows another balanced RR with $c=3$, which is equivalent to \textbf{(a)}. \textbf{(e)} is an example of an unbalanced assignment.}
    \label{fig:assignment-policies}
    \vspace*{-3pt}
\end{figure*}

\paratitle{Drawbacks of Previous Coded Consensus Protocols}
The aforementioned protocols open an interesting design space but have three drawbacks, rendering them incomplete for practical use. \circleb{1} They sacrifice the availability guarantee and offer a reduced fault-tolerance level. \circleb{2} They cannot handle leader failover gracefully, since followers do not see the complete payload and hence cannot execute commands in committed instances. A leader failover triggers significant reconstruction traffic to reassemble those instances on any new leader. \circleb{3} They always use a disjoint shard assignment scheme and cannot adapt with delay-optimized configurations when desirable. This missing flexibility is crucial when payload sizes and network environments are dynamic.

\section{Design}
\label{sec:design}

We present the design of \crossword, a protocol that extends previous solutions with flexible erasure code shard assignment policies, preserving availability and enabling tradeoffs between delay- and bandwidth-friendly configurations. \crossword employs gossiping to keep followers up-to-date with minimal impact on critical-path performance.



\subsection{Reed-Solomon (RS) Codeword Space}
\label{sec:design-codeword-space}

We start by recognizing that the mappings from erasure code shards to server nodes need not be disjoint and symmetric, i.e., shard $i$ to node $i$. Instead, the mappings can be chosen from a 2-dimensional space. To visualize this, we introduce a new per-instance notation called an RS \textit{codeword space}. Refer to the left half of Figure~\ref{fig:concurrent-failures} for an example RS codeword space with a $(5, 3)$-coding scheme across 5 servers.

An RS codeword space is a 2-dimensional grid. Each row represents a server and lays out the shards of a conceptual RS codeword with data shards on the left. Each column corresponds to one particular shard of the codeword that could be replicated on any of the servers. We label shards starting with index 0 from left to right and name the servers similarly from top to bottom. For a given instance, this notation identifies the shards distributed across server nodes after erasure coding. The coding scheme does not require the total number of shards to equal the number of servers.


With the codeword space notation, we can mark which shards must be replicated onto which servers for a given instance. Specifically, we say that shard $i$ is \textit{assigned} to server $s$ if we require server $s$ to receive shard $i$ and durably remember its content. A shard could be assigned to multiple servers, meaning that all those servers should receive and persist the same bytes. A shard could also be assigned to no servers. A parity shard has the same power as a data shard.


\subsection{Shard Assignment Policies}
\label{sec:design-shard-assignment-policies}

Having an RS codeword space creates new possibilities in how shards can be assigned to servers. When a \crossword leader initiates the accept phase of a new instance (whose payload is a batch of client requests received in the last batching interval), it computes the RS codeword for that payload and decides which subset of shards to assign to each follower. We call such a decision a \textit{shard assignment policy}.

An assignment policy depicts which specific chunks of bytes in the erasure-coded payload need to be carried in the critical-path \texttt{Accept} messages to each follower. Accordingly, those are the bytes that follower must persist before replying with a positive vote. We use the word ``assign'' to capture both meanings. An assignment policy also includes what the leader itself must persist, though this involves no network traffic. Note that the leader is assumed to have the complete payload codeword in its main memory. Also, note that an assignment policy is solely restricted to one individual instance and is independent of other slots of the log.




We visualize an assignment policy by marking assigned cells with a darker color in the RS codeword space. Figure~\ref{fig:assignment-policies} demonstrates interesting examples of assignment polices in a 5-node cluster. A valid shard assignment policy has to satisfy certain constraints, which we derive in \S\ref{sec:design-constraint-boundary}; some policies (e.g., assigning zero shards to all replicas) are useless.


\paratitle{Previous Protocols}
Classic protocols and previous coded consensus protocols described in \S\ref{sec:background} can be represented as special cases of shard assignment policies. MultiPaxos~\cite{paxos-made-simple} and Raft~\cite{raft} map to Figure~\ref{fig:policy-multipaxos}, where all data shards are assigned to all the servers. In other words, a full copy of the original payload is replicated onto all servers, and parity shards are not used. RSPaxos~\cite{rs-paxos} and CRaft~\cite{craft} map to Figure~\ref{fig:policy-rspaxos}. They are on the other extreme, where only one disjoint shard is assigned to each server in a diagonal pattern.

\begin{figure*}[t]
    \centering
    \includegraphics[width=\textwidth]{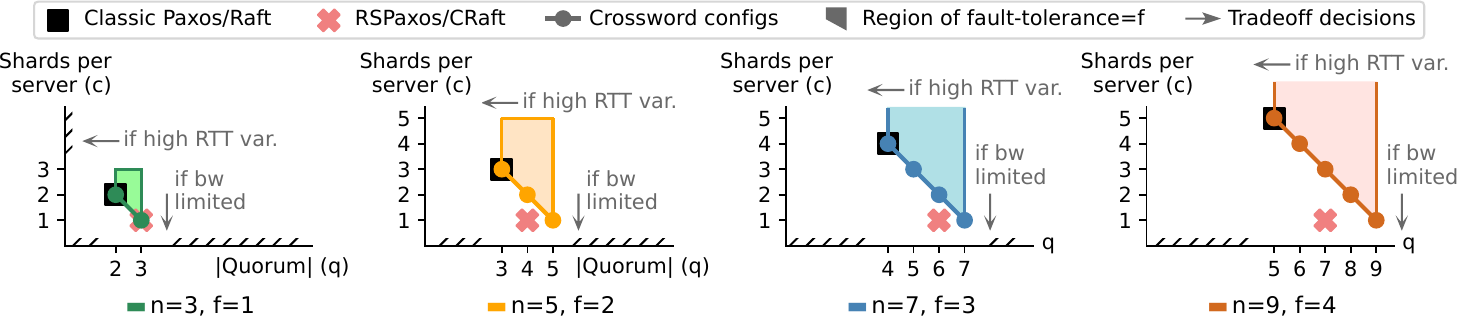}
    \vspace{-21pt}
    \floatcap{Availability constraint boundary and tradeoff line assuming balanced Round-Robin assignments}{See \S\ref{sec:design-constraint-boundary}.}
    \label{fig:constraint-boundary}
    \vspace*{-4pt}
\end{figure*}

\vspace{-0.5pt}
\paratitle{Balanced Round-Robin (RR) Assignments}
To generalize over \ref{fig:policy-multipaxos}--\ref{fig:policy-rspaxos} and bridge the gap between the two extremes, one needs a category of assignment policies that follow a consistent pattern. The key intuition is to spread the assigned shards to cover as many columns as possible in the codeword space, so that the number of reachable shards is maximized in the presence of failures. Specifically, consider assigning to server $s$ the shards $s \sim s+c$ rounding back to 0 if necessary, where $c \in [1, m]$. We call this category \textit{balanced Round-Robin} (RR) assignment policies. Figure~\ref{fig:policy-balanced_rr_2} shows such an assignment policy with $c = 2$. Notice that \ref{fig:policy-rspaxos} is also a balanced RR assignment with $c = 1$. Figure \ref{fig:policy-balanced_rr_3} gives a $c=3$ assignment equivalent to \ref{fig:policy-multipaxos}, except that a shifting subset of shards instead of the same set of data shards is assigned to each server. \S\ref{sec:design-constraint-boundary} shows why this unified family of assignment policies is useful for our adaptability and availability goals.

\paratitle{Unbalanced Assignments}
The assignment policies mentioned above are balanced, meaning servers are assigned the same number of shards. It is also possible and potentially useful to make \textit{unbalanced} assignments, where servers receive different numbers of shards. Figure~\ref{fig:policy-unbalanced} gives an example of this based on a $(8, 5)$-coding scheme. Unbalanced assignments share similarities with weighted voting~\cite{weighted-voting} \revise{(but with linearizability constraints)} and cannot be described by a single numeric parameter. \revise{Latest work has recognized asymmetric failure probabilities in replication~\cite{uncertain-consensus}, where \crossword could offer an effective solution.} We found that balanced RR policies meet our main goals and, hence, leave deeper exploration of unbalanced ones as future work.

\subsection{Availability Constraint Boundary}
\label{sec:design-constraint-boundary}

We derive the necessary constraints on shard assignment policies that categorize which of them maintain a desired availability guarantee. We start with a general definition and derive a constraint formula that can bound any assignment policy. We then narrow the scope to balanced RR assignments and present a more concise constraint.

To discuss the usefulness of an assignment policy, we denote the set of \texttt{Accept} replies received from followers as an \textit{acceptance pattern}. A reply from follower $s$ in an acceptance pattern essentially conveys the following statement: ``$s$ votes yes to the ballot number of this \texttt{Accept} message and has durably remembered the shards it carried.'' We implicitly include the leader itself in an acceptance pattern.



\paratitle{Constraints in the General Form}
Suppose an acceptance pattern $ap$ is observed by the leader during its wait on \texttt{Accept} replies. How can we determine when it is safe (both in terms of correctness and availability) to commit? To answer this, we define the following metrics on $ap$. Let:
\begin{itemize}
    \item $Nodes(ap)$ denote the number of replies in $ap$, i.e., how many server nodes have replied (including the leader).
    \item $Cover(ap)$ denote the \textit{shard coverage} of $ap$, which is the number of distinct shards that the replies cover. For example, suppose a balanced RR assignment with $c = 2$ as in Figure~\ref{fig:policy-balanced_rr_2}, and suppose $ap$ contains replies from servers S0, S1, and S4; $Cover(ap)$ is $4$ because shards 0, 1, 2, and 4 are covered by at least one reply, while shard 3 is not.
\end{itemize}
$Nodes(ap)$ is what classic consensus protocols use when making commit decisions. In particular, it is safe for them to commit an instance if at least a majority have replied, i.e.,
\begin{equation} \label{eqn:constraint-nodes} \tag{C1}
    Nodes(ap) \ge m.
\end{equation}
This constraint remains necessary in the presence of sharding, as the majority quorum intersection property is still required to establish consistency.



One may attempt to assert that it is safe to commit as long as $Cover(ap)$ reaches the number of data shards $d$. However, this may violate the availability guarantee as shown in \S\ref{sec:bg-previous-work} and Figure~\ref{fig:concurrent-failures}. When each follower holds fewer than $d$ shards, losing the leader may decrease the number of reachable shards below $d$, preventing the new leader from reconstructing the payload. To capture potential failures, a more sophisticated metric on $ap$ is needed. Let:
\begin{itemize}
    \item $SubCover(ap, f)$ denote the \textit{subset coverage} of $ap$, which is the minimum coverage among subsets of $ap$ with $f$ replies removed, where $f$ is the number of tolerable failures.
\end{itemize}
It is straightforward to see that to preserve the desired level of availability, i.e., allow progress when at most $f$ servers fail, the following constraint must hold besides \ref{eqn:constraint-nodes}:
\begin{equation} \label{eqn:constraint-subcover} \tag{C2}
    SubCover(ap, f) \ge d.
\end{equation}
For classic consensus protocols, this trivially holds, because any single server is assigned $d$ shards, meaning any set of $f+1$ replies satisfies this condition. For RSPaxos and CRaft, i.e., Figure~\ref{fig:policy-rspaxos}, one can also validate that they offer a fault-tolerance level $f = 1$ when waiting for a quorum of $Nodes(ap) = 4$ replies in a 5-node cluster by plugging in $d = m = 3$. For more general assignment policies, this constraint can be programmatically checked by the leader when an \texttt{Accept} reply is received.

\paratitle{Specific Form for Balanced RR Assignments}
Since we focus on balanced RR assignment policies, we give a more concise form of constraints for them. For an acceptance pattern $ap$, denote $q = Nodes(ap)$, the quorum size. $q$ obviously cannot be larger than the total number of servers. Recall that each server is assigned $c$ shards in an overlapping round-robin fashion using an $(n, m)$-coding scheme. One can see that $SubCover(ap, f) \ge q - f + c - 1$, with the equal sign taken when all replies are from adjacent servers. This gives a combined constraint of:
\begin{equation} \label{eqn:constraint-specific} \tag{C3}
    n \ge q \ge m  \quad \wedge \quad  q - f + c - 1 \ge m.
\end{equation}
The protocol must retain the same fault-tolerance level as classic protocols, i.e., $f = n - m$, giving:
\begin{equation} \label{eqn:constraint-specific-simplified} \tag{C4}
    n \ge q \ge m  \quad \wedge \quad  q + c \ge n + 1.
\end{equation}


Figure~\ref{fig:constraint-boundary} visualizes the derived availability constraint \ref{eqn:constraint-specific-simplified} for balanced RR assignments with four different cluster sizes. In each subfigure, every point $(q, c)$ in the grid maps to a potential \textit{configuration} for a given consensus instance, where the protocol uses a balanced RR assignment policy with $c$ shards per server and commits upon receiving $q$ replies. The set of valid configurations satisfying the desired availability guarantee form the colored region. Notice that MultiPaxos and Raft (black squares) are at the bottom-left corner of the region because they always assign a full copy of original data to all servers and expect a simple majority quorum. RSPaxos and CRaft (red crosses) are outside of the region (except for when $n=3$) due to always assigning exactly one shard per server and waiting for an insufficient quorum size that results in a degraded availability guarantee.

\subsection{Performance Tradeoff}
\label{sec:design-perf-tradeoff}

Among the set of valid configurations in Figure~\ref{fig:constraint-boundary}, those on the bottom boundary line (satisfying $q + c = n + 1$) are particularly interesting. Configurations above this line in the availability constraint region deliver strictly worse performance; for any quorum size $q$, one should pick the smallest number of shards $c$ per server. We call these \textit{candidate configurations} and highlight them with circular dots.

Across the candidate configurations, there exists a tradeoff between the quorum size and the number of shards assigned to each server. Choosing a smaller $c$ reduces the size of data to be transferred to and persisted on each follower at the cost of requiring more \texttt{Accept} replies, and vice versa. The tradeoff decisions will be affected by both the runtime hardware environment and the payload size of the instance. On the one hand, a small payload on a high-RTT, jittery network favors smaller $q$, because slower replies take substantially longer to wait for. On the other hand, a large payload on a bandwidth-constrained network favors smaller $c$, since the time saved by streaming less data overshadows the fluctuation in arrival times of replies.


\crossword is a consensus protocol that operates along the line of candidate configurations. For each instance, it picks the best configuration among the candidates according to the instance's payload size and the real-time hardware conditions. We describe a regression-based method as the default heuristic for choosing configurations in \S\ref{sec:impl-perf-monitoring}; more sophisticated solutions are possible~\cite{perf-eval-monitor}, as well as simpler heuristics such as payload size or user-given hints.

\subsection{Follower Gossiping}
\label{sec:design-follower-gossiping}

Besides adaptability, another goal of \crossword is to retain the graceful leader failover of classic consensus. After a leader's failure, a newly elected leader should be able to quickly recover all committed state and start serving incoming requests. This is not the case in RSPaxos and CRaft because, during normal-case operation, followers always receive a partial piece of the codeword and cannot assemble committed client commands. To overcome this, \crossword employs lazy \textit{follower gossiping} to let followers exchange their knowledge of shards without interrupting the leader.

\begin{figure}[t]
    \centering
    \includegraphics[width=\linewidth]{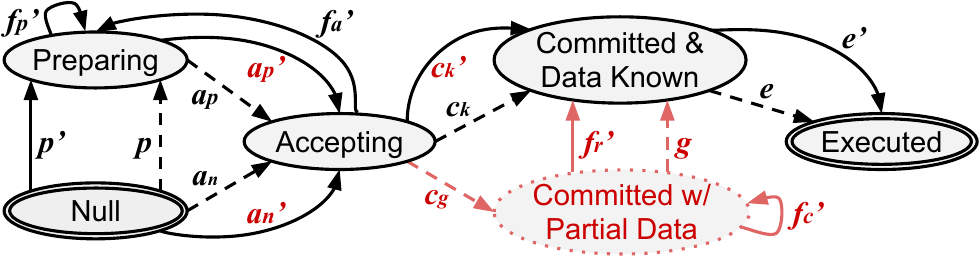}
    \vspace{-8pt}
    \newline
    \fontsize{9.5}{11}\selectfont
    \setlength{\tabcolsep}{5pt}
    \setlength{\doublerulesep}{4pt}
    \begin{tabular}{c|l|l}
        \hline
        \tikz[baseline={([yshift=-.5ex]current bounding box.center)}]{
            \draw [-latex,thick](0,0) -- (0.5,0);
        }
            & \multicolumn{1}{c|}{\textbf{Trigger on leader}}
            & \multicolumn{1}{c}{\textbf{Action by leader}}
            \\ \hline
        \textit{\textbf{p'}}
            & client req, unprepared
            & broadcast \texttt{Prepare}
            \\ \hline
        \textit{\textbf{\darkred{a\textsubscript{p}'}}}
            & decide \darkred{prepared value}
            & \makecell[l]{broadcast \texttt{Accept}\darkred{, each}\\\darkred{w/ (a subset of) shards}}
            \\ \hline
        \textit{\textbf{\darkred{a\textsubscript{n}'}}}
            & \makecell[l]{client req, ballot\\already prepared}
            & \makecell[l]{broadcast \texttt{Accept}\darkred{, each}\\\darkred{w/ (a subset of) shards}}
            \\ \hline
        \textit{\textbf{\darkred{c\textsubscript{k}'}}}
            & reach \darkred{commit condition}
            & commit instance
            \\ \hline
        \textit{\textbf{e'}}
            & instance committed
            & execute, reply to client
            \\ \hline
        \textit{\textbf{f\textsubscript{p}'}}
            & new leader after failover
            & redo with higher ballot
            \\ \hline
        \textit{\textbf{f\textsubscript{a}'}}
            & new leader after failover
            & redo with higher ballot
            \\ \hline
        \hline
        \tikz[baseline={([yshift=-.5ex]current bounding box.center)}]{
            \draw [-latex,thick,dashed](0,0) -- (0.5,0);
        }
            & \multicolumn{1}{c|}{\textbf{Trigger on followers}}
            & \multicolumn{1}{c}{\textbf{Action by followers}}
            \\ \hline
        \textit{\textbf{p}}
            & recv \texttt{Prepare}
            & send \texttt{Prepare} reply
            \\ \hline
        \textit{\textbf{a\textsubscript{p}}}
            & recv \texttt{Accept}
            & send \texttt{Accept} reply
            \\ \hline
        \textit{\textbf{a\textsubscript{n}}}
            & recv \texttt{Accept}
            & send \texttt{Accept} reply
            \\ \hline
        \textit{\textbf{c\textsubscript{k}}}
            & \makecell[l]{leader committed,\\payload fully known}
            & commit instance
            \\ \hline
        \textit{\textbf{e}}
            & committed, full payload
            & execute commands
            \\ \hline
        \hline
        \tikz[baseline={([yshift=-.5ex]current bounding box.center)}]{
            \draw [-latex,thick,BrickRed](0,0.08) -- (0.5,0.08);
            \draw [-latex,thick,dashed,BrickRed](0,-0.08) -- (0.5,-0.08);
        }
            & \multicolumn{2}{c}{\textbf{New gossiping-related transitions}}
            \\ \hline
        \darkred{\textit{\textbf{f\textsubscript{c}'}}}
            & \darkred{new leader after failover}
            & \darkred{do reconstruction reads}
            \\ \hline
        \darkred{\textit{\textbf{f\textsubscript{r}'}}}
            & \darkred{enough shards received}
            & \darkred{re-assemble the payload}
            \\ \hline
        \darkred{\textit{\textbf{c\textsubscript{g}}}}
            & \makecell[l]{\darkred{leader committed,}\\\darkred{payload partially known}}
            & \makecell[l]{\darkred{commit instance,}\\\darkred{schedule gossiping}}
            \\ \hline
        \darkred{\textit{\textbf{g}}}
            & \darkred{enough shards gossiped}
            & \darkred{re-assemble the payload}
            \\ \hline
    \end{tabular}
    \vspace{-8pt}
    \floatcap{Instance status transition diagram}{Solid edges represent transitions on the leader and dashed edges represent those on followers. Differences and additions of \crossword w.r.t. classic Paxos highlighted in \textcolor{Maroon}{red} color. See \S\ref{sec:design-follower-gossiping}-\ref{sec:design-complete-protocol}.}
    \label{fig:status-diagram}
    \vspace{-13pt}
\end{figure}

\paratitle{Status Transition Diagram}
We summarize servers' actions in a consensus instance as a transition diagram in Figure~\ref{fig:status-diagram}. For now, ignore the ``Committed w/ Partial Data'' status and related transitions. A MultiPaxos instance may undergo the following status transitions: Null (i.e., empty instance), Preparing (only after leader changes), Accepting, Committed (i.e., chosen and ready to be learned), and Executed (i.e., commands applied and clients replied). We use edges to represent actions that a server will do to the instance, triggered by certain conditions; actions labeled with prime are those carried out by the leader, while others are by followers. On the critical path, the following actions happen to an instance: [\textbf{\textit{a\textsubscript{n}'}}, \textbf{\textit{a\textsubscript{n}}}, \textbf{\textit{c\textsubscript{k}'}}, \textbf{\textit{e'}}]. The Preparing status and failure-related actions only appear after a leader failover.

\paratitle{Committed with Partial Data \& Gossiping}
\crossword introduces a new status, ``Committed w/ Partial Data'', that could appear on followers and newly elected leaders. Correspondingly, we rename "Committed" to ``Committed \& Data Known''. On the critical path on a follower, unless it receives $m$ shards (e.g., when using a balanced RR assignment policy of $c = m$), they take the \textbf{\textit{c\textsubscript{g}}} transition and schedule gossiping with other followers. During gossiping, followers share their assigned shards to fill others' missing ones. When a follower receives enough shards, it takes the \textbf{\textit{g}} transition, which allows commands to be executed on the follower.



Follower gossiping happens entirely among followers in the background, and the gossiped shards stay purely in followers' memory. The gossiping of an instance can happen arbitrarily late after its commitment; in practice, some delay is desired, as discussed in \S\ref{sec:impl-gossiping}. Note that a newly elected leader may see instances in the ``Committed w/ Partial Data'' status at the end of its log; in this case, special actions \textbf{\textit{f\textsubscript{c}'}} and \textbf{\textit{f\textsubscript{r}'}} reconstruct those instances synchronously.


\paratitle{Benefits of Follower Gossiping}
Follower gossiping essentially moves the replication of a significant portion of payloads off the leader-to-follower critical path and turns it into a flexible, asynchronous, follower-to-follower background task. This brings three benefits. \circleb{1} We gain critical-path improvements while retaining graceful leader failover behavior (\S\ref{sec:eval-failover}). \circleb{2} \texttt{Reconstruct} messages make use of follower-to-follower bandwidth whenever idle and carry batched payloads of each gossiping cycle, improving data transfer efficiency (\S\ref{sec:eval-gossip-params}). \circleb{3} Followers prioritize critical-path messages over gossiping messages, allowing improved performance even in cases when follower-to-follower bandwidth is sporadically saturated in a keyspace-partitioned system (\S\ref{sec:eval-macrobench}).


\subsection{\crossword: The Complete Protocol}
\label{sec:design-complete-protocol}

We wrap up the design of \crossword and condense it into well-defined extensions to classic MultiPaxos, highlighted in red in Figure~\ref{fig:status-diagram}. Below is a summary of the differences.

\begin{itemize}
    \item \textbf{\textit{a\textsubscript{n}'}}: To initiate an instance, leader computes the RS codeword of the payload (or glues together pieces pre-computed by clients) and adaptively assigns to each follower a subset of shards through \texttt{Accept} messages. (\S\ref{sec:design-shard-assignment-policies}, \S\ref{sec:design-perf-tradeoff})
    \item \textbf{\textit{a\textsubscript{p}'}}: In the prepare phase, a corner-case occurs if the leader finds $< d$ shards with the highest ballot number among $\ge m$ \texttt{Prepare} replies; in this case, the leader safely uses the next client command batch as the prepared value since that payload could not have been chosen. If $\ge d$ shards are found, that payload is used as in classic Paxos.
    \item \textbf{\textit{c\textsubscript{k}'}}: Upon receiving an \texttt{Accept} reply, leader checks constraints \ref{eqn:constraint-nodes} and \ref{eqn:constraint-subcover}, or the simplified formula \ref{eqn:constraint-specific-simplified} for balanced RR assignments, to decide whether to commit the instance given the received replies. (\S\ref{sec:design-constraint-boundary})
    \item \textbf{\textit{c\textsubscript{g}}} and \textbf{\textit{g}}: Followers gossip each other's missing shards of committed instances in the background. (\S\ref{sec:design-follower-gossiping})
    \item \textbf{\textit{f\textsubscript{c}'}} and \textbf{\textit{f\textsubscript{r}'}}: If a newly elected leader sees partially committed instances at the end of its log, it broadcasts reconstruction reads to grab enough shards for re-assembly. (\S\ref{sec:design-follower-gossiping})
\end{itemize}
Based on this per-instance diagram, \crossword assembles a multi-decree SMR protocol, as MultiPaxos does over Paxos. The same design can also be applied to Raft-style protocols.

\section{Implementation}
\label{sec:implementation}

We provide details of our \crossword implementations.

\paratitle{The \summerset Replicated KV-store}
We create \summerset, a distributed, replicated, and protocol-generic key-value store, as a fair codebase for implementing and evaluating consensus protocols. \summerset is built with async Rust/\texttt{tokio}, and adopts a lock-less channel-based architecture. We do not stack our implementation directly atop codebases from previous work~\cite{epaxos, chain-paxos}, as we found that \circleb{1} they were not extensible enough and \circleb{2} their language runtime overheads were noticeable, leading to unfair disadvantages.


The codebase contains 12.7k lines of code as infrastructure. We have implemented six protocols (with individual lines of code reported): Chain Replication (1.1k), MultiPaxos (2.3k), Raft (2.3k), RSPaxos (2.5k), CRaft (2.6k), and \crossword (3.4k). All protocol implementations have passed extensive unit tests and fuzz tests. We will open-source our complete codebase to welcome future research.


\paratitle{CockroachDB Raft Integration}
We have also implemented a Go prototype of \crossword in CockroachDB v24.3.0a, a production OLTP database~\cite{cockroachdb}, by patching its sophisticated Raft package with \textasciitilde1.6k lines of changes. This version reuses CockroachDB's production-quality infrastructure and contains all the core \crossword mechanisms, but does not include the regression-based config chooser mentioned below; instead, we use payload size thresholds as simple guides.

\subsection{Choosing the Best Configuration}
\label{sec:impl-perf-monitoring}

Per-instance configurations can be chosen based on any appropriate heuristics, e.g., payload sizes or user-specified hints. \crossword adopts a simple linear-regression-based performance monitoring approach as a good default to adaptively select among balanced RR assignment policies. The leader bookkeeps a sliding window (over 2 seconds) of two statistics---data size and response time---of internal message rounds with each follower. The messages include \texttt{Accept} messages and periodic heartbeats; heartbeats are messages carrying zero-size payload that track server health. 

\paratitle{Linear Regression of Size-Time Mappings}
With the response time statistics, the leader maintains an ordinary least squares model~\cite{ordinary-least-squares} for each follower, updated at 200 ms intervals. Each model uses datapoints in the current sliding window, with message data sizes ($v$) treated as x-axis values and response times ($t$) as y-axis values, with the highest 5\% discarded as outliers. Doing so produces a per-follower linear estimate of recent performance: $t_s(v) = d_s + \frac{1}{b_s} \cdot v$, where $s$ is the follower, $d_s$ is an overall delay estimate (the learned intercept), $b_s$ is an overall bandwidth estimate (the reciprocal of the learned slope), $v$ is the payload size, and $t_s(v)$ is the overall response time given payload size $v$. Computing such ordinary least squares incurs negligible overhead.

This linear model captures an end-to-end latest estimation of the time to send a message with data size $v$ to the follower, let it persist $v$ amount of data, and receive its reply. When choosing a configuration for a new instance with payload size $v_p$, the leader enumerates choices of $c \in [1, m]$ allowed by balanced RR assignments, and for each $c$, computes the set $T_c = \{t_1(\frac{v_p}{c}), \dots, t_{n-1}(\frac{v_p}{c})\}$. The $(q-1)$-th smallest value in $T_c$ represents the time to wait for the last reply with a quorum size of $q$, and hence determines the estimated completion time of the \texttt{Accept} phase. The leader chooses the $c$ that yields the smallest estimated completion time.


\paratitle{Limitations}
Performance monitoring and estimation is a complex topic~\cite{perf-eval-monitor}. The simple linear-regression-based approach worked well for us, but has limitations. \circleb{1} It produces the best choice among balanced RR policies and currently does not automate unbalanced assignment policies, where the space of candidate configurations is considerably larger. \circleb{2} It does not react to sharp fluctuations within small time frames (e.g., less than 1 second). More sophisticated methods can be applied~\cite{perf-eval-monitor} but are outside of the scope of this paper.

\subsection{Follower Gossiping Implementation}
\label{sec:impl-gossiping}

\begin{figure}[t]
    \centering
    \includegraphics[width=\linewidth]{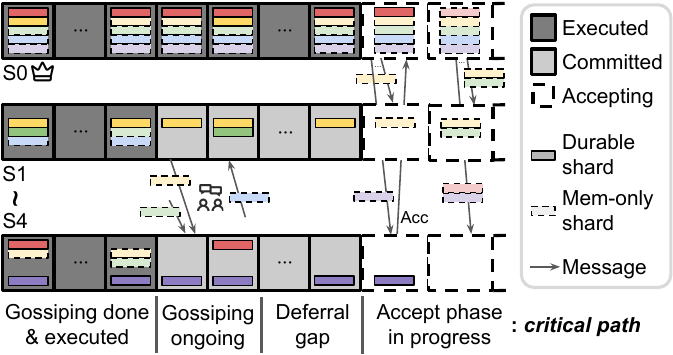}
    \vspace{-22pt}
    \floatcap{Demonstration of the replicated log in action across \crossword servers}{Shows an example view on a cluster of 5 servers with S0 being the leader. Each slot is a consensus instance using (5,3)-coding on its data. See \S\ref{sec:impl-gossiping}.}
    \label{fig:log-in-action}
    \vspace{-8pt}
\end{figure}

Figure~\ref{fig:log-in-action} visualizes an example runtime state of the replicated log across 5 \crossword servers. On the right-most end are instances on their critical path, whose operations have been covered in previous sections. This section covers how \crossword enables followers to push their execution forward for committed instances through follower gossiping.

A \crossword leader embeds the shard assignment of an instance in its \texttt{Accept} messages. The embedding is a compact array of bitmaps representing the RS codeword space: assigned cells are marked as 1 and others as 0.


Using this information, each follower $s$ checks its peers starting with $s+1$ rounding back to 0 (skipping the leader), and maintains a monotonically growing set of expected shard indices, which initially contains only the shards that $s$ itself holds. For each peer checked, the follower enqueues a \texttt{Reconstruct} message containing a set of shard indices to request from that peer and adds those indices into the expected set. This loop ends when the size of the set reaches $d$. \texttt{Reconstruct} messages are batched across multiple instances for better bandwidth utilization. Upon receiving a \texttt{Reconstruct}, a follower sends back the shards it knows within the requested set if it has committed an instance.

\paratitle{Introducing a Deferral Gap}
Followers trigger gossiping for partially-committed instances periodically at \textasciitilde20 ms intervals. However, attempting to gossip immediately for a just-committed instance is not ideal, because it is likely that followers not in the committing quorum have not yet fully received their assigned shards from leader. \crossword introduces a configurable \textit{deferral gap} that specifies the \revise{accumulated size of payloads} to skip at the end of a follower's log when attempting gossiping. The deferral gap defaults to \revise{a 400KB threshold}; it helps restrict gossiping to instances that everyone has likely committed. In the case of stragglers, a follower \revise{skips requesting shards from a peer entirely} if it has not heard of its \texttt{Reconstruct} reply for 10 gossiping cycles.


\subsection{Other Practicality Features}
\label{sec:impl-practicality-details}

We also implemented other common protocol features.

\paratitle{Heartbeats}
A \crossword leader broadcasts heartbeats at $\ge$20 ms intervals. Heartbeats carry the leader's latest committed slot index to notify followers of this information asynchronously. A follower attempts to step up as a new leader if it has not received a valid heartbeat from the current leader for a randomly chosen timeout between 300-600 ms. Followers reply to the leader's heartbeats to help the leader track their health status as well; in the case of follower failures, the number of healthy followers bounds the largest $q$ we should choose from possible configurations.

\paratitle{Snapshots}
\crossword servers autonomously take periodic snapshots of executed instances of their log to avoid unbounded growth of log length. Thanks to follower gossiping, followers can take snapshots without requesting \textit{state-sending} snapshot messages from the leader~\cite{raft}, which RSPaxos and CRaft required (but did not implement~\cite{rs-paxos, craft}, and neither did we for them).


\paratitle{Leases for Read-only Commands}
As is common practice~\cite{chubby, paxos-made-live}, we implement simple time-based read leases for all five protocols by assuming an upper bound of clock drift, e.g., a few seconds, across servers. When holding the lease, the leader serves read-only \texttt{Get} commands locally without placing them into the next instance.

\section{Evaluation}
\label{sec:evaluation}

\begin{figure}[t]
    \centering
    \captionsetup{}
    \begin{subfigure}[t]{\linewidth}
        \centering
        \includegraphics[height=12pt]{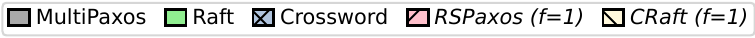}
        \vspace{-8pt}
    \end{subfigure}
    \begin{subfigure}[t]{0.06\linewidth}
        \centering
        \includegraphics[width=\linewidth]{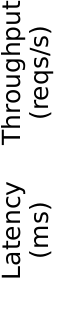}
    \end{subfigure}
    \hspace{1pt}
    \begin{subfigure}[t]{0.29\linewidth}
        \hfill
        \includegraphics[align=r,width=0.97\columnwidth]{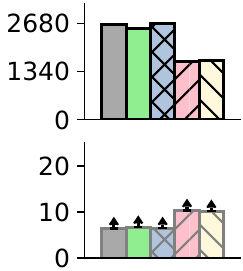}
        \captionsetup{justification=centering}
        \caption{Region, 8B}
        \label{fig:critical-small-1dc}
    \end{subfigure}\hspace{4pt}%
    \begin{subfigure}[t]{0.29\linewidth}
        \hfill
        \includegraphics[align=r,width=0.91\columnwidth]{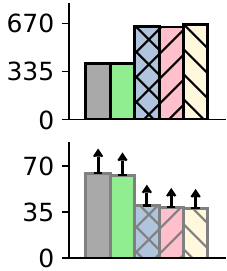}
        \captionsetup{justification=centering}
        \caption{Region, 128KB}
        \label{fig:critical-large-1dc}
    \end{subfigure}\hspace{4pt}%
    \begin{subfigure}[t]{0.29\linewidth}
        \hfill
        \includegraphics[align=r,width=0.97\columnwidth]{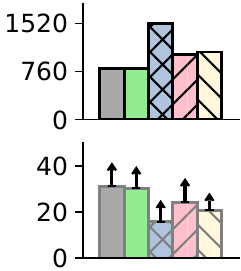}
        \captionsetup{justification=centering}
        \caption{Region, mixed}
        \label{fig:critical-mixed-1dc}
    \end{subfigure}
    \hfill
    \vspace{10pt}
    \begin{subfigure}[t]{0.06\linewidth}
        \centering
        \includegraphics[width=\linewidth]{figures/ylabels-critical.pdf}
    \end{subfigure}
    \hspace{1pt}
    \begin{subfigure}[t]{0.29\linewidth}
        \hfill
        \includegraphics[align=r,width=0.91\columnwidth]{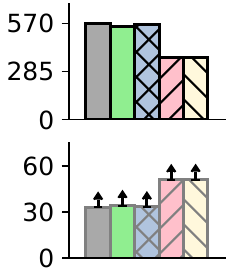}
        \captionsetup{justification=centering}
        \caption{WAN, 8B}
        \label{fig:critical-small-wan}
    \end{subfigure}\hspace{4pt}%
    \begin{subfigure}[t]{0.29\linewidth}
        \hfill
        \includegraphics[align=r,width=0.91\columnwidth]{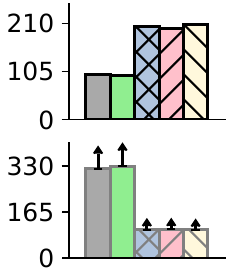}
        \captionsetup{justification=centering}
        \caption{WAN, 128KB}
        \label{fig:critical-large-wan}
    \end{subfigure}\hspace{4pt}%
    \begin{subfigure}[t]{0.29\linewidth}
        \hfill
        \includegraphics[align=r,width=0.91\columnwidth]{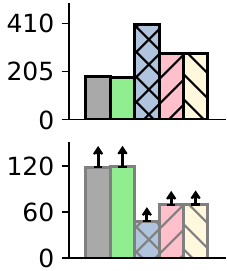}
        \captionsetup{justification=centering}
        \caption{WAN, mixed}
        \label{fig:critical-mixed-wan}
    \end{subfigure}%
    \vspace{-4pt}
    \floatcapnodot{Critical path performance}{under different environments and workload sizes. See \S\ref{sec:eval-critical-path} for parameters used.}
    \label{fig:critical-perf}
    \vspace{-5pt}
\end{figure}

We evaluate \crossword and previous consensus protocols on \summerset to answer the following questions:
\begin{itemize}[itemsep=1pt]
    \item How well does \crossword perform under various network environments and workload sizes? (\S\ref{sec:eval-critical-path})
    \item Can \crossword adapt promptly to workload size shifts and hardware condition changes? (\S\ref{sec:eval-adaptability})
    \item Can \crossword handle leader failover gracefully? (\S\ref{sec:eval-failover})
    \item Can \crossword take advantage of unbalanced assignments to handle asymmetric performance scenarios? (\S\ref{sec:eval-unbalanced})
    \item What effects do gossiping-related parameters have? (\S\ref{sec:eval-gossip-params})
    \item Does \crossword work under realistic macro-benchmarks with partitioned keys and in CockroachDB? (\S\ref{sec:eval-macrobench} \& \S\ref{sec:eval-cockroach})
    \item How much overhead does computing RS code incur? (\S\ref{sec:eval-rs-coding})
\end{itemize}


\paratitle{Experimental Setup}
We use CloudLab~\cite{cloudlab} machines running Linux kernel v6.1.64 as our testbed. We use a cluster of \texttt{c220g2}-type machines with 40 CPU cores and 160GB memory each. The network connection between each pair of nodes is 1Gbps bandwidth and 4ms average delay, which is representative of a regional replication system~\cite{wan-latency-estimator}. For some experiments, we also include results on a more wide-area setting spanning multiple CloudLab datacenters, with an average of 200Mbps bandwidth and 30ms delay between nodes. All the server and client processes are pinned to disjoint CPU cores. Clients are launched on the same set of machines and are distributed evenly across machines. Servers apply 1ms-interval request batching.


\subsection{Critical Path Performance}
\label{sec:eval-critical-path}

We compare \crossword against previous protocols using 5 servers and 15 closed-loop clients running microbenchmarks. We examine both regional and wide-area network environments as described in the setup. We generate 50\% \texttt{Get}s which carry only 8B keys and are served by the lease-holding leader locally, plus 50\% \texttt{Put}s whose payload sizes are varied: 8B, 128KB, and a half-half mix of the two. To add realistic variations to the workloads, for every \texttt{Put} request, a client will sample a value size from a normal distribution with the given size as mean and 10\% of it as standard deviation. Throughput is aggregated over all clients, and latency is averaged per request; the tips of arrows mark P95 latency.

\begin{figure}[t]
    \centering
    \captionsetup{}
    \begin{subfigure}[t]{\linewidth}
        \centering
        \includegraphics[height=12pt]{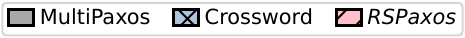}
        \vspace{-5pt}
    \end{subfigure}
    \begin{subfigure}[t]{\linewidth}
        \centering
        \includegraphics[width=0.99\linewidth]{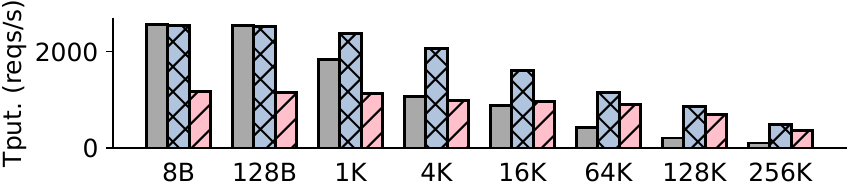}
        \vspace{-2pt}
        \captionsetup{justification=centering}
        \caption{Mean value size}
        \label{fig:critical-payload-size}
        \vspace{9pt}
    \end{subfigure}
    \begin{subfigure}[t]{0.56\linewidth}
        \centering
        \includegraphics[width=\linewidth]{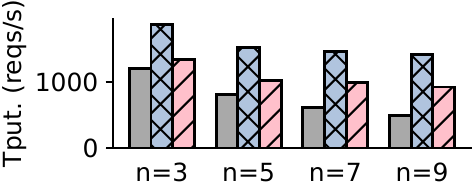}
        \captionsetup{justification=centering}
        \caption{Cluster size}
        \label{fig:critical-cluster-size}
    \end{subfigure}
    \begin{subfigure}[t]{0.41\linewidth}
        \centering
        \includegraphics[width=\linewidth]{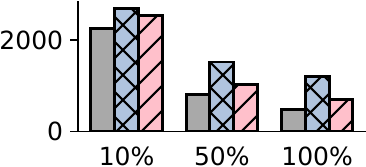}
        \captionsetup{justification=centering}
        \caption{Put request ratio}
        \label{fig:critical-write-ratio}
    \end{subfigure}
    \addtocounter{figure}{-1}  
    \vspace{-4pt}
    \floatcapoffig{Throughput with varying value size, cluster size, and put ratio}{See \S\ref{sec:eval-critical-path} payload size and sensitivity.}
    \label{fig:critical-cluster-n-ratio}
\end{figure}

\begin{figure}[t]
    \centering
    \includegraphics[align=c,width=0.53\linewidth]{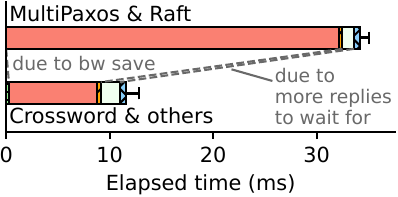}
    \hfill
    \includegraphics[align=c,width=0.46\linewidth]{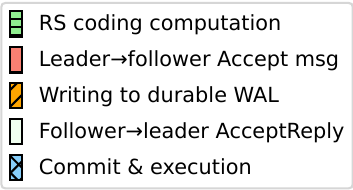}
    \vspace{-10pt}
    \floatcapnodot{Latency breakdown}{of time spent in different steps of a bandwidth-bounded instance. See \S\ref{sec:eval-critical-path} last paragraph.}
    \label{fig:perf-breakdown}
    \vspace{-5pt}
\end{figure}

\begin{figure*}[t]
    \centering
    \captionsetup{}
    \begin{minipage}{0.39\textwidth}
        \includegraphics[width=\linewidth]{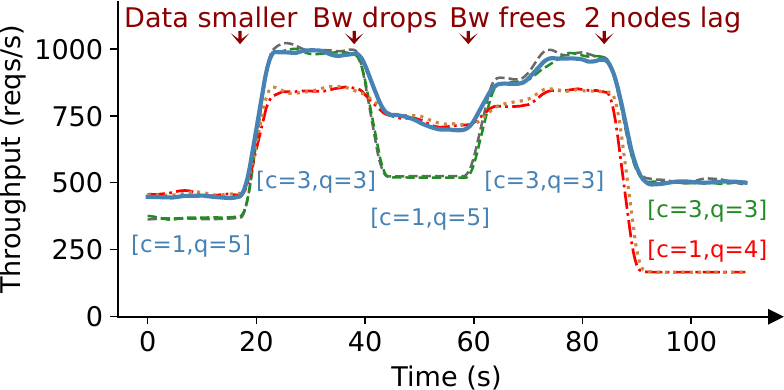}
        \vspace{-23pt}
        \floatcap{Real-time adaptability}{See \S\ref{sec:eval-adaptability}.}
        \label{fig:adaptability}
    \end{minipage}\hspace{5pt}%
    \begin{minipage}{0.595\textwidth}
        \includegraphics[align=c,width=0.705\linewidth]{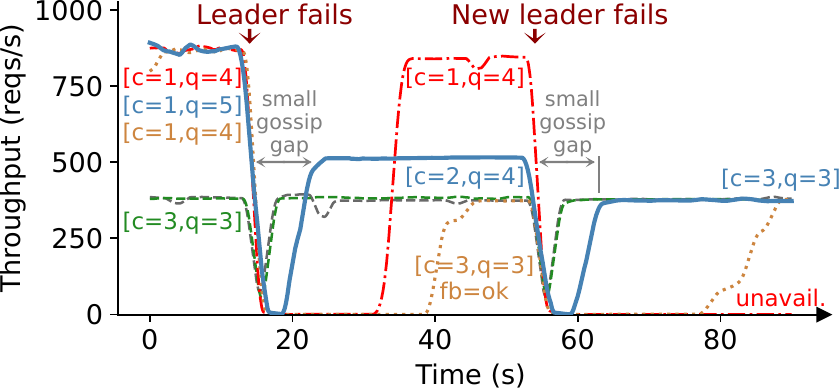}
        \hfill
        \includegraphics[align=c,width=0.275\linewidth]{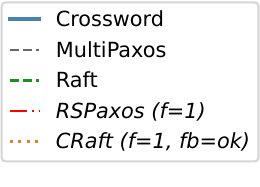}
        \vspace{-11pt}
        \floatcap{Real-time comparison of leader failover behavior}{See \S\ref{sec:eval-failover}.}
        \label{fig:failover}
    \end{minipage}
\end{figure*}

The results presented in Figure~\ref{fig:critical-perf} yield several observations. \circleb{1} In \ref{fig:critical-small-1dc} and \ref{fig:critical-small-wan}, payloads are small and bandwidth is relatively abundant, favoring configurations with smaller quorum sizes, i.e., MultiPaxos/Raft. \crossword performs as well as them and delivers better throughput than RSPaxos/CRaft by 1.9x. \circleb{2} In contrast, \ref{fig:critical-large-1dc} and \ref{fig:critical-large-wan} favor fewer shards per server, and thus \crossword performs as well as RSPaxos/CRaft and outperforms MultiPaxos/Raft by 2.0x. \circleb{3} In \ref{fig:critical-mixed-1dc} and \ref{fig:critical-mixed-wan}, \crossword outperforms all four others by up to 2.1x thanks to its adaptability in choosing the best per-instance configuration according to payload size. \circleb{4} For all cases, latency numbers match the inverse of bandwidth numbers due to the closed-loop nature of clients.

\paratitle{Varying Payload Size in Finer Grains}
Figure~\ref{fig:critical-payload-size} varies the mean value size from 8B to 256KB, while keeping other settings the same as in the regional setting. \crossword matches MultiPaxos on the left end and outperforms both when payloads are around the bandwidth-constraining threshold, which appears to be \textasciitilde4KB in our setup. With $\ge$ 64KB, \crossword tends to prefer $c=1$ and approach RSPaxos, while both outperform MultiPaxos by larger ratios ($>$4x).

\paratitle{Sensitivity to Cluster Size \& Put Ratio}
Figure~\ref{fig:critical-cluster-size} and \ref{fig:critical-write-ratio} verify the effectiveness of \crossword across four different cluster sizes and three different \texttt{Put} request ratios under the regional setting. Results show that \crossword consistently delivers improved performance of up to 2.3x over MultiPaxos and 1.4x over RSPaxos for all cluster sizes. \crossword brings larger improvement to higher \texttt{Put} ratios, since all \texttt{Get}s are treated in the same way across all protocols.

\paratitle{Performance Breakdown}
To further dissect the differences between configurations, we present in Figure~\ref{fig:perf-breakdown} a breakdown of the response time of an average instance carrying 64KB payloads. In this case, \crossword chooses $c=1$ and brings 71\% reduction to the time spent in leader-to-follower \texttt{Accept} messages and durability. As a tradeoff, a larger reply quorum is required, introducing a slight overhead to that segment. Computing the RS code incurs negligible overhead.

We also observe that in this experiment, \crossword and RSPaxos/CRaft consume 913MB of space for the durable log, while MultiPaxos/Raft consume 1467MB. This shows a side benefit of log storage space reduction by 38\%.


\subsection{Dynamic Adaptability}
\label{sec:eval-adaptability}

To demonstrate \crossword's runtime adaptability to both workload size shifts and network environment changes, we trace the real-time aggregated throughput of 15 closed-loop clients on a 5-node cluster while changing relevant parameters along the way. For runtime network performance changes, we use \texttt{tc-netem}, a Linux kernel built-in traffic control queuing discipline for network emulations~\cite{tc-netem}. Clients initially run 100\% \texttt{Put} workloads with 64KB payloads. Throughput values are profiled at 1-second intervals and presented in Figure~\ref{fig:adaptability}. The four changes in order are: \circlew{1} average payload size reduces from 64KB to 4KB, \circlew{2} network bandwidth drops from 1Gbps to 100Mbps per link, \circlew{3} network bandwidth returns to 1Gbps per link, and \circlew{4} two nodes in the cluster experience lag with 10x worse delay and bandwidth. The changes happen at 15 seconds apart from each other. MultiPaxos and Raft always use a $[c=3,q=3]$ configuration, while RSPaxos and CRaft always use $[c=1,q=4]$. \crossword adapts to the best configuration among valid ones and matches the best performance among the rest of the protocols at all times.

\begin{figure*}
    \centering
    \captionsetup{}
    \begin{minipage}{0.39\textwidth}
        \includegraphics[align=c,width=0.48\linewidth]{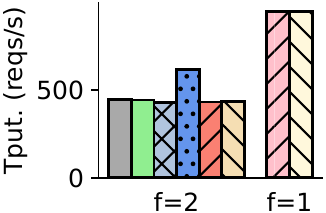}
        \hfill
        \includegraphics[align=c,width=0.50\linewidth]{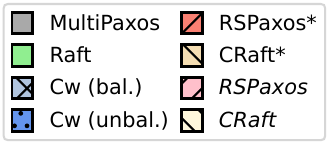}
        \vspace{-8pt}
        \floatcapnodot{Unbalanced assignment advantage}{in an asymmetric case. $^{*}$: $q = 5$ forced. See \S\ref{sec:eval-unbalanced}.}
        \label{fig:unbalanced}
    \end{minipage}\hspace{14pt}%
    \begin{minipage}{0.31\textwidth}
        \includegraphics[align=c,width=0.635\linewidth]{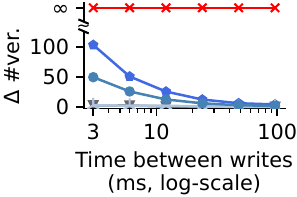}
        \hfill
        \includegraphics[align=c,width=0.345\linewidth]{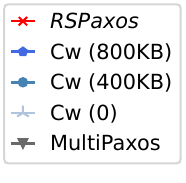}
        \vspace{-11pt}
        \floatcapnodot{Avg. staleness of follower reads}{with gossiping gaps. See \S\ref{sec:eval-gossip-params}.}
        \label{fig:staleness}
    \end{minipage}\hspace{13pt}%
    \begin{minipage}{0.23\textwidth}
        \includegraphics[align=c,width=0.445\linewidth]{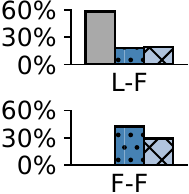}
        \hfill
        \includegraphics[align=c,width=0.535\linewidth]{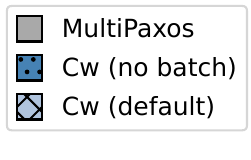}
        \vspace{-11pt}
        \floatcapnodot{Bandwidth usage}{w/ or w/o gossip batching, fixing throughput. See \S\ref{sec:eval-gossip-params}.}
        \label{fig:bw_utils}
    \end{minipage}\hspace{5pt}%
    \vspace{-3pt}
\end{figure*}

\subsection{Graceful Leader Failover}
\label{sec:eval-failover}

To show the performance impact of leader failover to all five protocols, we trace the real-time aggregated throughput of 15 closed-loop clients making 64KB requests to a 5-node cluster in Figure~\ref{fig:failover}, while crashing the leader node at 15 and 55 seconds. To make failover gaps easier to observe, for this experiment specifically, we increase the heartbeat timeout on followers to 1.5 seconds and turn off snapshotting for all protocols. \circleb{1} Classic protocols MultiPaxos and Raft recover from leader failover instantly; the only sources of delay are the heartbeat timeout and the leader election round. \circleb{2} \crossword exhibits similar graceful failover behavior with \textasciitilde2x longer gap; the additional delay is introduced by the reconstruction time of not-yet-gossiped instances at the end of the new leader's log. Also, note that after the first failover, \crossword operates with a $[c=2,q=4]$ configuration---the best choice then. \circleb{3} RSPaxos experiences significantly longer downtime after the first failover due to the inevitable reconstruction work to fill the new leader's log with complete data. In practice, this downtime is bounded by the interval between expensive state-sending snapshots, or could be unbounded otherwise. RSPaxos returns to the original throughput level due to keeping $c=1$; this leads to it being totally unavailable after a second failure. \circleb{4} CRaft shows an even longer downtime after the first failure due to the additional work of falling back to full-copy replication, after which it comes back to the same throughput level as Raft. We make the second failure late enough so that fallback is successful (fb=ok). The same pattern repeats after the second failure.




\begin{figure}[t]
    \centering
    \includegraphics[align=c,width=0.67\linewidth]{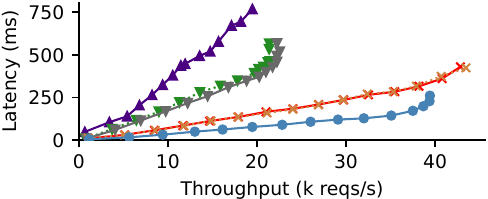}
    \hfill
    \includegraphics[align=c,width=0.28\linewidth]{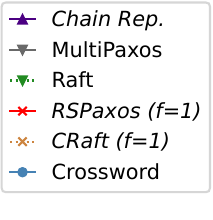}
    \vspace{-11pt}
    \floatcapnodot{Throughput-Latency curves}{with YCSB-A key trace, TiDB payload size profile, and partitioning. See \S\ref{sec:eval-macrobench}.}
    \label{fig:macrobench}
    \vspace{-5pt}
\end{figure}

\subsection{Unbalanced Assignment Policy}
\label{sec:eval-unbalanced}

\crossword also supports unbalanced assignment policies. To demonstrate potential benefit, we use \texttt{tc-netem} to set up a static yet asymmetric network environment across 5 nodes where followers S1 and S2 have 1Gbps bandwidth connected to the leader, S3 has 400Mbps, and S4 has 100Mbps. We run the same 64KB-value workload and configure \crossword to use an assignment policy that assigns 5 shards to each of S1 and S2, 3 shards to S3, and 1 shard to S4. Figure~\ref{fig:unbalanced} shows that \crossword (unbalanced) establishes a better match between the amount of assigned load and the link bandwidth, delivering a higher throughput of \textasciitilde1.4x over MultiPaxos/Raft and balanced policy. The default settings of RSPaxos/CRaft yield better throughput with lower fault-tolerance of $f=1$.


\subsection{Gossiping-Related Parameters}
\label{sec:eval-gossip-params}

We evaluate the impact of two gossiping-related factors: gossip gap length and gossip batching.

\paratitle{Follower Read Staleness with Gossiping Gaps}
Systems that apply multi-versioning to objects may allow local reads served directly by a follower with the newest value it knows. The replies are not linearizable but sequentially consistent, possibly stale versions in the history~\cite{sequential-consistency, sequential-vs-linearizability}.

Figure~\ref{fig:staleness} shows the average staleness of follower reads (measured as the version difference from the latest value at leader) on an object of 8B size (thus least favorable to \crossword), varying the frequency of writes. MultiPaxos consistently delivers close-to-zero staleness, while RSPaxos cannot support such follower reads. We force \crossword to use a $c=1$ configuration to expose its worst-case staleness, and try three different gossiping gap lengths. A larger gossiping gap leads to higher staleness if writes are frequent; a gap length of zero converges to MultiPaxos but would lead to premature gossiping attempts as discussed in \S\ref{sec:impl-gossiping}. We default to a reasonably small gap length of \revise{400KB}.

\paratitle{Effectiveness of Gossip Batching}
Figure~\ref{fig:bw_utils} presents the network bandwidth utilization percentage of leader-to-follower links (L-F) and follower-to-follower links (F-F) under a 64KB-value workload. Utilization is profiled by accumulating the total size of messages transferred divided by the link's bandwidth, while fixing end-to-end user throughput to the same as MultiPaxos. \crossword moves $\frac{2}{3}$ of the payloads into background F-F communication, and can further reduce F-F bandwidth usage by 13\% through batching all gossips of each \textasciitilde20ms cycle into a single round of \texttt{Reconstruct}s.

\subsection{YCSB with Keyspace Partitioning}
\label{sec:eval-macrobench}

We run a macro-benchmark using YCSB-A with Zipfian~\cite{ycsb} to generate a trace of key accesses out of 1k keys; we treat each request as a generic \texttt{Put} and sample a payload size from the TiDB workload profile presented in \S\ref{sec:introduction}. We partition the keys into 5 disjoint ranges and run one consensus group per partition on the same set of 5 regional cluster machines with rotating leaders (i.e., machine $i$ serves as the leader of partition $i$ and a follower in the other 4). This architecture matches how data systems deploy consensus~\cite{gaios, tidb, cockroachdb}. We include a Chain Replication implementation~\cite{chain-replication}.

Figure~\ref{fig:macrobench} presents the throughput-vs-latency curves measured by varying the number of clients. We make the following observations. \circleb{1} When total system bandwidth is not saturated (lower-left part of each curve), Chain Replication exhibits the highest latency due to chain propagation; \crossword outperforms other protocols, matching \S\ref{sec:eval-critical-path}. \circleb{2} All protocols (except Chain Replication) exhibit throughput limitations. \crossword delivers higher maximum throughput than MultiPaxos due to follower gossiping being transparently postponed in hot partitions in favor of critical-path messages. RSPaxos/CRaft have an even higher throughput limit because they simply have no gossiping. Chain Replication has not hit its throughput limit due to its chain structure, but latency would ramp up further as load increases.

\begin{figure}[t]
    \centering
    \vspace{6pt}
    \includegraphics[align=c,width=0.78\linewidth]{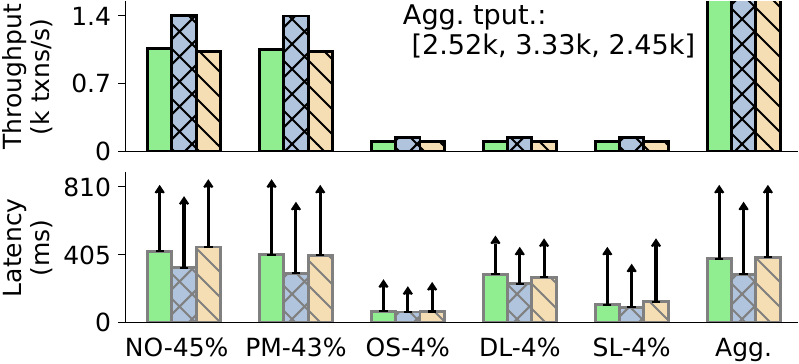}
    \hfill
    \includegraphics[align=c,width=0.205\linewidth]{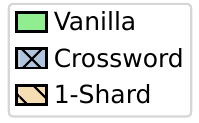}
    \vspace{-10pt}
    \floatcap{TPC-C over CockroachDB}{Txn types: NO-NewOrder, PM-Payment, OS-OrderStatus, DL-Delivery, SL-StockLevel, with ratio in the mix. Agg.: aggregated. See \S\ref{sec:eval-cockroach}.}
    \label{fig:cockroach}
    \vspace{-5pt}
\end{figure}

\subsection{TPC-C over CockroachDB}
\label{sec:eval-cockroach}

To demonstrate the end-to-end performance improvement that \crossword brings to a database system, we run TPC-C over CockroachDB~\cite{cockroachdb} v24.3.0a across 12 nodes in the regional setting, with 200 warehouses input, 5 replicas per table key-range, and 400 concurrent workers. We compare its vanilla Raft implementation with our \crossword patch and a 1-shard-forced configuration (that mimics CRaft). We keep default Cockroach settings except for turning off leader/leaseholder rebalancing and transactional write pipelining features, for both compatibility with our patch and more deterministic results. We pick 4KB/8KB as payload thresholds for choosing $c=2$/$c=1$ configurations, respectively.

Figure~\ref{fig:cockroach} shows the throughput (in txns/s), per-txn latency (in ms), and P95 latency (tips of arrows). \crossword brings up to 44\% speedup to read-write transactions (NO, PM, and DL) and less improvement to read-only transactions (OS and SL). Throughput follows similar relative improvement with per-txn latency, but the absolute numbers are adjusted by their percentage in the workload mix. \crossword brings 1.32x higher aggregate throughput. The 1-shard configuration leads to unchanged (and sometimes slightly worse) performance due to wasteful sharding for small payloads.

\subsection{RS Code Computation Overhead}
\label{sec:eval-rs-coding}


Table~\ref{tab:rs-coding-perf} presents the time taken to compute $(5,3)$-coding on inputs of different sizes using SIMD on 32 CPU cores (including memory allocation and serialization overheads) \revise{and the overall per-request CPU/memory usage overhead at the leader}. Even for 1MB, this takes no longer than 1ms. Compared to the latency values in Figure~\ref{fig:critical-perf}, computing RS code contributes < 1\% \revise{and introduces negligible CPU/memory overhead (besides the memory space for parity shards)}.



\section{Related Work and Discussion}
\label{sec:related-work}



We discuss related work and their insufficiency for dynamic data-heavy workloads or their orthogonality to \crossword.

\begin{table}[t]
    \centering
    \small
    \renewcommand{\arraystretch}{1.0}  
    \setlength\tabcolsep{2.6pt}
    \begin{tabular}{c|c|c|c|c|c|c}
        \hline
        \textbf{\revise{Payload} size} & 4KB & 16KB & 64KB & 256KB & 1MB & 4MB \\\hline
        \textbf{\revise{Time taken}} & 1$\mu$s & 4$\mu$s & 16$\mu$s & 77$\mu$s & 1ms & 6ms \\\hline
        \textbf{\revise{CPU usage}} & 1.25\% & 1.24\% & 1.24\% & 1.26\% & 1.25\% & 1.26\% \\\hline
        \textbf{\revise{Mem usage}} & 1.6KB & 6.4KB & 25.6KB & 102.4KB & 409.6KB & 1.6MB \\\hline
    \end{tabular}
    \renewcommand{\arraystretch}{1}
    \vspace{5pt}
    \floatcap{RS $(5,3)$-coding computation overhead}{See \S\ref{sec:eval-rs-coding}.}
    \label{tab:rs-coding-perf}
    \vspace{-30pt}
\end{table}

\subsection{Erasure-Coded Consensus}
\label{sec:related-coded-consensus}

RSPaxos~\cite{rs-paxos} is to our knowledge the first proposal on integrating erasure coding with consensus. CRSRaft~\cite{crs-raft} and adRaft~\cite{adraft} apply the same idea to Raft. CRaft~\cite{craft} is a recent proposal that falls back to full-copy replication upon failures. ECRaft~\cite{ecraft} and HRaft~\cite{hraft} propose smoother fallback mechanisms by gradually replenishing shards on healthy nodes. FlexRaft~\cite{flex-raft} achieves a similar goal by altering the RS-coding scheme. All of these variants do not fully address degraded availability, shard assignment rigidity, and ungraceful failover. \textsc{Pando}~\cite{tradeoffs-geo-distributed} is a higher-level, WAN-optimized, coded replication protocol that emphasizes a latency-to-storage-cost tradeoff, which is a different goal from ours (run-time dynamicity); it assumes a topology with frontends, uses a pre-deployment planner to assign quorum memberships, and does not yet support reconfigurations. \revise{Racos~\cite{racos} applies erasure coding to Rabia~\cite{rabia}, a leaderless randomized consensus protocol, to reduce leader load.}



\subsection{Bandwidth-Aware Techniques}
\label{sec:related-bandwidth-techniques}


\paratitle{Address Space Partitioning}
Gaios~\cite{gaios} is one of the first systems that deploy Paxos in a scalable manner to support data storage. It does so by partitioning the keys into disjoint regions and assigning them to \textit{Paxos groups}. Each group spans multiple servers and each server can host multiple agents of different Paxos groups; the placement of groups is managed by a separate fault-tolerant service. This design has been adopted by modern systems~\cite{tidb, cockroachdb, fine-grained-rsms, derecho, spanner, consul}, oftentimes as \textit{Raft groups}. In such architectures, \crossword can easily be applied to each of the consensus groups.


\paratitle{Master/Metadata Replication}
Distributed storage systems may split data and metadata into two separate layers: a weakly-consistent, possibly erasure-coded data store and a strongly-consistent, consensus-backed metadata service~\cite{niobe, gnothi, gfs, s3-replication, sdpaxos, springfs}. This design provides a workaround for uniformly data-heavy workloads but comes with three drawbacks. \circleb{1} It entails at least two rounds of operations for any request, one through the data store and the other through the metadata service (in strict order), bringing higher latency. \circleb{2} To provide overall linearizability, it requires a multi-versioned data store with carefully-timed garbage collection so as to ensure all the in-use data references held by metadata stay valid; this increases system complexity even when unnecessary. \circleb{3} It does not help when the metadata themselves are heavy~\cite{zookeeper-storm, cocytus}, in which case \crossword still applies. Overall, we believe erasure-coded consensus provides an attractive alternative of less error-prone designs.


\paratitle{Pipelining \& Chain Replication}
\textit{Pipelining} is an iconic technique for building throughput-optimized systems. Chain Replication~\cite{chain-replication, craq} is a classic protocol offering consistent high-throughput replication by organizing replicas as a one-directional chain. RingPaxos~\cite{ring-paxos}, ChainPaxos~\cite{chain-paxos}, and others~\cite{pigpaxos, pull-based-consensus-mongodb, chainreaction} apply similar ideas to derive higher throughput and to simplify membership management. These protocols are purely bandwidth-optimized but have two significant downsides. \circleb{1} They amplify latency by a multiplicative factor. \circleb{2} They are particularly vulnerable to stragglers and performance asymmetry along the chain.


\paratitle{\revise{Data Relaying or Dissemination}}
\revise{Several works such as PigPaxos~\cite{pigpaxos}, S-Paxos~\cite{s-paxos}, and Autobahn~\cite{autobahn} explored \textit{data relaying} or \textit{block dissemination} techniques that relieve contention at the leader by making payload transfer multi-hop or asynchronous, albeit still on the critical path. These techniques could be combined with \crossword's gossiping path to further improve scalability under constant high load.}


\subsection{Orthogonal Consensus Designs}
\label{sec:related-consensus-designs}

Various designs address orthogonal problems: read leases~\cite{quorum-leases, megastore, paxos-made-live}, geo-scale optimizations~\cite{epaxos, epaxos-revisited, mencius, generalized-paxos, atlas, fpaxos, swiftpaxos}, membership management~\cite{viewstamped-replication, raft, chain-paxos}, fail-slow tolerance~\cite{copilots, occult, gray-failures}, hardware acceleration~\cite{derecho, swarm-disaggregated-memory, aceso-disaggregated-memory, p4xos, nopaxos, netpaxos, off-path-smartnic, fisslock, apus-rdma-paxos, mu-consensus, hydra-network-ordering, speculative-paxos}, SMR scalability~\cite{scalable-smr, compartmentalization, insanely-scalable-smr, dynastar, pigpaxos}, exploiting API semantics~\cite{consistency-aware-durability, skyros-nil-externality, gryff, fuzzylog, lazylog, tango, corfu, delos-virtual-consensus, paxos-made-transparent, consistency-based-sla}, relaxed consistency~\cite{regular-sequential-consistency, occult, eventual-consistency, noctua, cops, stalestores}, formal methods~\cite{I4, sift-veri, ironfleet, verus, anvil}, and BFT~\cite{pbft, hotstuff, ubft, basil, autobahn, bidl-blockchain}.


\subsection{High-End Network Hardware}
\label{sec:high-bandwidth-hardware}

Hardware advancements have brought 40/100GbE or higher-bandwidth network devices. Unfortunately, they offer limited help to replication. \circleb{1} These network interfaces and interchange devices are usually deployed in intra-datacenter networks and have limited availability in wider area~\cite{ec2-network-guide}. \circleb{2} Replication is only one part of a system; it often shares the network with other data-heavy application logic~\cite{tidb, cockroachdb, gfs, kubernetes, electrode}. \circleb{3} Payload sizes keep growing into the MBs or even GBs. We believe protocol-level improvements prove useful.


As multi-NIC servers become more popular~\cite{ec2-network-guide}, \crossword will bring higher gains, because background follower gossiping can take advantage of separate NICs and impose zero interference with critical-path messages.


\section{Conclusion}
\label{sec:conclusion}

We present \crossword, \revise{an adaptive consensus protocol} for dynamic data-heavy workloads by integrating erasure coding with flexible shard assignment policies, retaining the availability and failover behavior of classic consensus. We envision that dynamic data-heavy workloads will be a rising challenge facing linearizable replication systems; we take \crossword as a first step towards optimal solutions that address the dynamicity and bandwidth constraints imposed.



\clearpage
\bibliographystyle{ACM-Reference-Format}
\bibliography{references}


\begin{thebibliography}{144}


\ifx \showCODEN    \undefined \def \showCODEN     #1{\unskip}     \fi
\ifx \showISBNx    \undefined \def \showISBNx     #1{\unskip}     \fi
\ifx \showISBNxiii \undefined \def \showISBNxiii  #1{\unskip}     \fi
\ifx \showISSN     \undefined \def \showISSN      #1{\unskip}     \fi
\ifx \showLCCN     \undefined \def \showLCCN      #1{\unskip}     \fi
\ifx \shownote     \undefined \def \shownote      #1{#1}          \fi
\ifx \showarticletitle \undefined \def \showarticletitle #1{#1}   \fi
\ifx \showURL      \undefined \def \showURL       {\relax}        \fi
\providecommand\bibfield[2]{#2}
\providecommand\bibinfo[2]{#2}
\providecommand\natexlab[1]{#1}
\providecommand\showeprint[2][]{arXiv:#2}

\bibitem[Aguilera et~al\mbox{.}(2020)]%
        {mu-consensus}
\bibfield{author}{\bibinfo{person}{Marcos~K. Aguilera}, \bibinfo{person}{Naama Ben-David}, \bibinfo{person}{Rachid Guerraoui}, \bibinfo{person}{Virendra~J. Marathe}, \bibinfo{person}{Athanasios Xygkis}, {and} \bibinfo{person}{Igor Zablotchi}.} \bibinfo{year}{2020}\natexlab{}.
\newblock \showarticletitle{Microsecond Consensus for Microsecond Applications}. In \bibinfo{booktitle}{\emph{14th USENIX Symposium on Operating Systems Design and Implementation (OSDI 20)}}. \bibinfo{publisher}{USENIX Association}, \bibinfo{pages}{599--616}.
\newblock
\showISBNx{978-1-939133-19-9}
\urldef\tempurl%
\url{https://www.usenix.org/conference/osdi20/presentation/aguilera}
\showURL{%
\tempurl}


\bibitem[Aguilera et~al\mbox{.}(2023)]%
        {ubft}
\bibfield{author}{\bibinfo{person}{Marcos~K. Aguilera}, \bibinfo{person}{Naama Ben-David}, \bibinfo{person}{Rachid Guerraoui}, \bibinfo{person}{Antoine Murat}, \bibinfo{person}{Athanasios Xygkis}, {and} \bibinfo{person}{Igor Zablotchi}.} \bibinfo{year}{2023}\natexlab{}.
\newblock \showarticletitle{UBFT: Microsecond-Scale BFT Using Disaggregated Memory}. In \bibinfo{booktitle}{\emph{Proceedings of the 28th ACM International Conference on Architectural Support for Programming Languages and Operating Systems, Volume 2}} (Vancouver, BC, Canada) \emph{(\bibinfo{series}{ASPLOS 2023})}. \bibinfo{publisher}{Association for Computing Machinery}, \bibinfo{address}{New York, NY, USA}, \bibinfo{pages}{862–877}.
\newblock
\showISBNx{9781450399166}
\href{https://doi.org/10.1145/3575693.3575732}{doi:\nolinkurl{10.1145/3575693.3575732}}


\bibitem[Almeida et~al\mbox{.}(2013)]%
        {chainreaction}
\bibfield{author}{\bibinfo{person}{S\'{e}rgio Almeida}, \bibinfo{person}{Jo\~{a}o Leit\~{a}o}, {and} \bibinfo{person}{Lu\'{\i}s Rodrigues}.} \bibinfo{year}{2013}\natexlab{}.
\newblock \showarticletitle{ChainReaction: a causal+ consistent datastore based on chain replication}. In \bibinfo{booktitle}{\emph{Proceedings of the 8th ACM European Conference on Computer Systems}} (Prague, Czech Republic) \emph{(\bibinfo{series}{EuroSys '13})}. \bibinfo{publisher}{Association for Computing Machinery}, \bibinfo{address}{New York, NY, USA}, \bibinfo{pages}{85–98}.
\newblock
\showISBNx{9781450319942}
\href{https://doi.org/10.1145/2465351.2465361}{doi:\nolinkurl{10.1145/2465351.2465361}}


\bibitem[Attiya and Welch(1994)]%
        {sequential-vs-linearizability}
\bibfield{author}{\bibinfo{person}{Hagit Attiya} {and} \bibinfo{person}{Jennifer~L. Welch}.} \bibinfo{year}{1994}\natexlab{}.
\newblock \showarticletitle{Sequential Consistency versus Linearizability}.
\newblock \bibinfo{journal}{\emph{ACM Trans. Comput. Syst.}} \bibinfo{volume}{12}, \bibinfo{number}{2} (\bibinfo{date}{may} \bibinfo{year}{1994}), \bibinfo{pages}{91–122}.
\newblock
\showISSN{0734-2071}
\href{https://doi.org/10.1145/176575.176576}{doi:\nolinkurl{10.1145/176575.176576}}


\bibitem[AWS(2023)]%
        {s3-replication}
\bibfield{author}{\bibinfo{person}{AWS}.} \bibinfo{year}{2023}\natexlab{}.
\newblock \bibinfo{title}{Amazon S3 Replication}.
\newblock
\newblock
\shownote{\url{https://aws.amazon.com/s3/features/replication/}, Last accessed on 2023-11-25}.


\bibitem[AWS(2024)]%
        {ec2-network-guide}
\bibfield{author}{\bibinfo{person}{AWS}.} \bibinfo{year}{2024}\natexlab{}.
\newblock \bibinfo{title}{Amazon EC2 instance network bandwidth}.
\newblock
\newblock
\shownote{\url{https://docs.aws.amazon.com/AWSEC2/latest/UserGuide/ec2-instance-network-bandwidth.html}, Last accessed on 2024-09-06}.


\bibitem[Baker et~al\mbox{.}(2011)]%
        {megastore}
\bibfield{author}{\bibinfo{person}{Jason Baker}, \bibinfo{person}{Chris Bond}, \bibinfo{person}{James~C. Corbett}, \bibinfo{person}{JJ Furman}, \bibinfo{person}{Andrey Khorlin}, \bibinfo{person}{James Larson}, \bibinfo{person}{Jean-Michel Leon}, \bibinfo{person}{Yawei Li}, \bibinfo{person}{Alexander Lloyd}, {and} \bibinfo{person}{Vadim Yushprakh}.} \bibinfo{year}{2011}\natexlab{}.
\newblock \showarticletitle{Megastore: Providing Scalable, Highly Available Storage for Interactive Services}. In \bibinfo{booktitle}{\emph{Proceedings of the Conference on Innovative Data system Research (CIDR)}}. \bibinfo{pages}{223--234}.
\newblock
\urldef\tempurl%
\url{http://www.cidrdb.org/cidr2011/Papers/CIDR11_Paper32.pdf}
\showURL{%
\tempurl}


\bibitem[Balakrishnan et~al\mbox{.}(2020)]%
        {delos-virtual-consensus}
\bibfield{author}{\bibinfo{person}{Mahesh Balakrishnan}, \bibinfo{person}{Jason Flinn}, \bibinfo{person}{Chen Shen}, \bibinfo{person}{Mihir Dharamshi}, \bibinfo{person}{Ahmed Jafri}, \bibinfo{person}{Xiao Shi}, \bibinfo{person}{Santosh Ghosh}, \bibinfo{person}{Hazem Hassan}, \bibinfo{person}{Aaryaman Sagar}, \bibinfo{person}{Rhed Shi}, \bibinfo{person}{Jingming Liu}, \bibinfo{person}{Filip Gruszczynski}, \bibinfo{person}{Xianan Zhang}, \bibinfo{person}{Huy Hoang}, \bibinfo{person}{Ahmed Yossef}, \bibinfo{person}{Francois Richard}, {and} \bibinfo{person}{Yee~Jiun Song}.} \bibinfo{year}{2020}\natexlab{}.
\newblock \showarticletitle{Virtual Consensus in Delos}. In \bibinfo{booktitle}{\emph{14th USENIX Symposium on Operating Systems Design and Implementation (OSDI 20)}}. \bibinfo{publisher}{USENIX Association}, \bibinfo{pages}{617--632}.
\newblock
\showISBNx{978-1-939133-19-9}
\urldef\tempurl%
\url{https://www.usenix.org/conference/osdi20/presentation/balakrishnan}
\showURL{%
\tempurl}


\bibitem[Balakrishnan et~al\mbox{.}(2013a)]%
        {corfu}
\bibfield{author}{\bibinfo{person}{Mahesh Balakrishnan}, \bibinfo{person}{Dahlia Malkhi}, \bibinfo{person}{John~D. Davis}, \bibinfo{person}{Vijayan Prabhakaran}, \bibinfo{person}{Michael Wei}, {and} \bibinfo{person}{Ted Wobber}.} \bibinfo{year}{2013}\natexlab{a}.
\newblock \showarticletitle{CORFU: A distributed shared log}.
\newblock \bibinfo{journal}{\emph{ACM Trans. Comput. Syst.}} \bibinfo{volume}{31}, \bibinfo{number}{4}, Article \bibinfo{articleno}{10} (\bibinfo{date}{Dec.} \bibinfo{year}{2013}), \bibinfo{numpages}{24}~pages.
\newblock
\showISSN{0734-2071}
\href{https://doi.org/10.1145/2535930}{doi:\nolinkurl{10.1145/2535930}}


\bibitem[Balakrishnan et~al\mbox{.}(2013b)]%
        {tango}
\bibfield{author}{\bibinfo{person}{Mahesh Balakrishnan}, \bibinfo{person}{Dahlia Malkhi}, \bibinfo{person}{Ted Wobber}, \bibinfo{person}{Ming Wu}, \bibinfo{person}{Vijayan Prabhakaran}, \bibinfo{person}{Michael Wei}, \bibinfo{person}{John~D. Davis}, \bibinfo{person}{Sriram Rao}, \bibinfo{person}{Tao Zou}, {and} \bibinfo{person}{Aviad Zuck}.} \bibinfo{year}{2013}\natexlab{b}.
\newblock \showarticletitle{Tango: Distributed Data Structures over a Shared Log}. In \bibinfo{booktitle}{\emph{Proceedings of the Twenty-Fourth ACM Symposium on Operating Systems Principles}} (Farminton, Pennsylvania) \emph{(\bibinfo{series}{SOSP '13})}. \bibinfo{publisher}{Association for Computing Machinery}, \bibinfo{address}{New York, NY, USA}, \bibinfo{pages}{325–340}.
\newblock
\showISBNx{9781450323888}
\href{https://doi.org/10.1145/2517349.2522732}{doi:\nolinkurl{10.1145/2517349.2522732}}


\bibitem[Barr(2023)]%
        {s3-strong-consistency}
\bibfield{author}{\bibinfo{person}{Jeff Barr}.} \bibinfo{year}{2023}\natexlab{}.
\newblock \bibinfo{title}{Amazon S3 Update – Strong Read-After-Write Consistency}.
\newblock
\newblock
\shownote{\url{https://aws.amazon.com/blogs/aws/amazon-s3-update-strong-read-after-write-consistency/}, Last accessed on 2023-11-19}.


\bibitem[Berlekamp(1980)]%
        {ecc-survey}
\bibfield{author}{\bibinfo{person}{E.R. Berlekamp}.} \bibinfo{year}{1980}\natexlab{}.
\newblock \showarticletitle{The technology of error-correcting codes}.
\newblock \bibinfo{journal}{\emph{Proc. IEEE}} \bibinfo{volume}{68}, \bibinfo{number}{5} (\bibinfo{year}{1980}), \bibinfo{pages}{564--593}.
\newblock
\href{https://doi.org/10.1109/PROC.1980.11696}{doi:\nolinkurl{10.1109/PROC.1980.11696}}


\bibitem[Bezerra et~al\mbox{.}(2014)]%
        {scalable-smr}
\bibfield{author}{\bibinfo{person}{Carlos~Eduardo Bezerra}, \bibinfo{person}{Fernando Pedone}, {and} \bibinfo{person}{Robbert Van~Renesse}.} \bibinfo{year}{2014}\natexlab{}.
\newblock \showarticletitle{Scalable State-Machine Replication}. In \bibinfo{booktitle}{\emph{2014 44th Annual IEEE/IFIP International Conference on Dependable Systems and Networks}}. \bibinfo{pages}{331--342}.
\newblock
\href{https://doi.org/10.1109/DSN.2014.41}{doi:\nolinkurl{10.1109/DSN.2014.41}}


\bibitem[Biely et~al\mbox{.}(2012)]%
        {s-paxos}
\bibfield{author}{\bibinfo{person}{Martin Biely}, \bibinfo{person}{Zarko Milosevic}, \bibinfo{person}{Nuno Santos}, {and} \bibinfo{person}{André Schiper}.} \bibinfo{year}{2012}\natexlab{}.
\newblock \showarticletitle{S-Paxos: Offloading the Leader for High Throughput State Machine Replication}. In \bibinfo{booktitle}{\emph{2012 IEEE 31st Symposium on Reliable Distributed Systems}}. \bibinfo{pages}{111--120}.
\newblock
\href{https://doi.org/10.1109/SRDS.2012.66}{doi:\nolinkurl{10.1109/SRDS.2012.66}}


\bibitem[Bolosky et~al\mbox{.}(2011)]%
        {gaios}
\bibfield{author}{\bibinfo{person}{William~J. Bolosky}, \bibinfo{person}{Dexter Bradshaw}, \bibinfo{person}{Randolph~B. Haagens}, \bibinfo{person}{Norbert~P. Kusters}, {and} \bibinfo{person}{Peng Li}.} \bibinfo{year}{2011}\natexlab{}.
\newblock \showarticletitle{Paxos Replicated State Machines as the Basis of a {High-Performance} Data Store}. In \bibinfo{booktitle}{\emph{8th USENIX Symposium on Networked Systems Design and Implementation (NSDI 11)}}. \bibinfo{publisher}{USENIX Association}, \bibinfo{address}{Boston, MA}.
\newblock
\urldef\tempurl%
\url{https://www.usenix.org/conference/nsdi11/paxos-replicated-state-machines-basis-high-performance-data-store}
\showURL{%
\tempurl}


\bibitem[Breitenfeld et~al\mbox{.}(2017)]%
        {daos}
\bibfield{author}{\bibinfo{person}{M.~Scot Breitenfeld}, \bibinfo{person}{Neil Fortner}, \bibinfo{person}{Jordan Henderson}, \bibinfo{person}{J{\'e}rome Soumagne}, \bibinfo{person}{Mohamad Chaarawi}, \bibinfo{person}{Johann Lombardi}, {and} \bibinfo{person}{Quincey Koziol}.} \bibinfo{year}{2017}\natexlab{}.
\newblock \showarticletitle{DAOS for Extreme-scale Systems in Scientific Applications}.
\newblock \bibinfo{journal}{\emph{ArXiv}}  \bibinfo{volume}{abs/1712.00423} (\bibinfo{year}{2017}).
\newblock
\urldef\tempurl%
\url{https://api.semanticscholar.org/CorpusID:28851378}
\showURL{%
\tempurl}


\bibitem[Burke et~al\mbox{.}(2020)]%
        {gryff}
\bibfield{author}{\bibinfo{person}{Matthew Burke}, \bibinfo{person}{Audrey Cheng}, {and} \bibinfo{person}{Wyatt Lloyd}.} \bibinfo{year}{2020}\natexlab{}.
\newblock \showarticletitle{Gryff: Unifying Consensus and Shared Registers}. In \bibinfo{booktitle}{\emph{17th USENIX Symposium on Networked Systems Design and Implementation (NSDI 20)}}. \bibinfo{publisher}{USENIX Association}, \bibinfo{address}{Santa Clara, CA}, \bibinfo{pages}{591--617}.
\newblock
\showISBNx{978-1-939133-13-7}
\urldef\tempurl%
\url{https://www.usenix.org/conference/nsdi20/presentation/burke}
\showURL{%
\tempurl}


\bibitem[Burrows(2006)]%
        {chubby}
\bibfield{author}{\bibinfo{person}{Mike Burrows}.} \bibinfo{year}{2006}\natexlab{}.
\newblock \showarticletitle{The Chubby Lock Service for Loosely-Coupled Distributed Systems}. In \bibinfo{booktitle}{\emph{Proceedings of the 7th Symposium on Operating Systems Design and Implementation}} (Seattle, Washington) \emph{(\bibinfo{series}{OSDI '06})}. \bibinfo{publisher}{USENIX Association}, \bibinfo{address}{USA}, \bibinfo{pages}{335–350}.
\newblock
\showISBNx{1931971471}


\bibitem[Cao et~al\mbox{.}(2018)]%
        {polarfs}
\bibfield{author}{\bibinfo{person}{Wei Cao}, \bibinfo{person}{Zhenjun Liu}, \bibinfo{person}{Peng Wang}, \bibinfo{person}{Sen Chen}, \bibinfo{person}{Caifeng Zhu}, \bibinfo{person}{Song Zheng}, \bibinfo{person}{Yuhui Wang}, {and} \bibinfo{person}{Guoqing Ma}.} \bibinfo{year}{2018}\natexlab{}.
\newblock \showarticletitle{PolarFS: An Ultra-Low Latency and Failure Resilient Distributed File System for Shared Storage Cloud Database}.
\newblock \bibinfo{journal}{\emph{Proc. VLDB Endow.}} \bibinfo{volume}{11}, \bibinfo{number}{12} (\bibinfo{date}{aug} \bibinfo{year}{2018}), \bibinfo{pages}{1849–1862}.
\newblock
\showISSN{2150-8097}
\href{https://doi.org/10.14778/3229863.3229872}{doi:\nolinkurl{10.14778/3229863.3229872}}


\bibitem[Cao et~al\mbox{.}(2020)]%
        {rocksdb-study}
\bibfield{author}{\bibinfo{person}{Zhichao Cao}, \bibinfo{person}{Siying Dong}, \bibinfo{person}{Sagar Vemuri}, {and} \bibinfo{person}{David~H.C. Du}.} \bibinfo{year}{2020}\natexlab{}.
\newblock \showarticletitle{Characterizing, Modeling, and Benchmarking {RocksDB} {Key-Value} Workloads at Facebook}. In \bibinfo{booktitle}{\emph{18th USENIX Conference on File and Storage Technologies (FAST 20)}}. \bibinfo{publisher}{USENIX Association}, \bibinfo{address}{Santa Clara, CA}, \bibinfo{pages}{209--223}.
\newblock
\showISBNx{978-1-939133-12-0}
\urldef\tempurl%
\url{https://www.usenix.org/conference/fast20/presentation/cao-zhichao}
\showURL{%
\tempurl}


\bibitem[Castro and Liskov(1999)]%
        {pbft}
\bibfield{author}{\bibinfo{person}{Miguel Castro} {and} \bibinfo{person}{Barbara Liskov}.} \bibinfo{year}{1999}\natexlab{}.
\newblock \showarticletitle{Practical Byzantine Fault Tolerance}. In \bibinfo{booktitle}{\emph{Third Symposium on Operating Systems Design and Implementation (OSDI)}}. \bibinfo{publisher}{USENIX Association, Co-sponsored by IEEE TCOS and ACM SIGOPS}, \bibinfo{address}{New Orleans, Louisiana}.
\newblock


\bibitem[Chandra et~al\mbox{.}(2007)]%
        {paxos-made-live}
\bibfield{author}{\bibinfo{person}{Tushar~D. Chandra}, \bibinfo{person}{Robert Griesemer}, {and} \bibinfo{person}{Joshua Redstone}.} \bibinfo{year}{2007}\natexlab{}.
\newblock \showarticletitle{Paxos Made Live: An Engineering Perspective}. In \bibinfo{booktitle}{\emph{Proceedings of the Twenty-Sixth Annual ACM Symposium on Principles of Distributed Computing}} (Portland, Oregon, USA) \emph{(\bibinfo{series}{PODC '07})}. \bibinfo{publisher}{Association for Computing Machinery}, \bibinfo{address}{New York, NY, USA}, \bibinfo{pages}{398–407}.
\newblock
\showISBNx{9781595936165}
\href{https://doi.org/10.1145/1281100.1281103}{doi:\nolinkurl{10.1145/1281100.1281103}}


\bibitem[Chang et~al\mbox{.}(2008)]%
        {bigtable}
\bibfield{author}{\bibinfo{person}{Fay Chang}, \bibinfo{person}{Jeffrey Dean}, \bibinfo{person}{Sanjay Ghemawat}, \bibinfo{person}{Wilson~C. Hsieh}, \bibinfo{person}{Deborah~A. Wallach}, \bibinfo{person}{Mike Burrows}, \bibinfo{person}{Tushar Chandra}, \bibinfo{person}{Andrew Fikes}, {and} \bibinfo{person}{Robert~E. Gruber}.} \bibinfo{year}{2008}\natexlab{}.
\newblock \showarticletitle{Bigtable: A Distributed Storage System for Structured Data}.
\newblock \bibinfo{journal}{\emph{ACM Trans. Comput. Syst.}} \bibinfo{volume}{26}, \bibinfo{number}{2}, Article \bibinfo{articleno}{4} (\bibinfo{date}{jun} \bibinfo{year}{2008}), \bibinfo{numpages}{26}~pages.
\newblock
\showISSN{0734-2071}
\href{https://doi.org/10.1145/1365815.1365816}{doi:\nolinkurl{10.1145/1365815.1365816}}


\bibitem[Charapko et~al\mbox{.}(2021)]%
        {pigpaxos}
\bibfield{author}{\bibinfo{person}{Aleksey Charapko}, \bibinfo{person}{Ailidani Ailijiang}, {and} \bibinfo{person}{Murat Demirbas}.} \bibinfo{year}{2021}\natexlab{}.
\newblock \showarticletitle{PigPaxos: Devouring the Communication Bottlenecks in Distributed Consensus}. In \bibinfo{booktitle}{\emph{Proceedings of the 2021 International Conference on Management of Data}} (Virtual Event, China) \emph{(\bibinfo{series}{SIGMOD '21})}. \bibinfo{publisher}{Association for Computing Machinery}, \bibinfo{address}{New York, NY, USA}, \bibinfo{pages}{235–247}.
\newblock
\showISBNx{9781450383431}
\href{https://doi.org/10.1145/3448016.3452834}{doi:\nolinkurl{10.1145/3448016.3452834}}


\bibitem[Chintapalli et~al\mbox{.}(2016)]%
        {zookeeper-storm}
\bibfield{author}{\bibinfo{person}{Sanket Chintapalli}, \bibinfo{person}{Derek Dagit}, \bibinfo{person}{Robert Evans}, \bibinfo{person}{Reza Farivar}, \bibinfo{person}{Zhuo Liu}, \bibinfo{person}{Kyle Nusbaum}, \bibinfo{person}{Kishorkumar Patil}, {and} \bibinfo{person}{Boyang Peng}.} \bibinfo{year}{2016}\natexlab{}.
\newblock \showarticletitle{PaceMaker: When ZooKeeper Arteries Get Clogged in Storm Clusters}. In \bibinfo{booktitle}{\emph{2016 IEEE 9th International Conference on Cloud Computing (CLOUD)}}. \bibinfo{pages}{448--455}.
\newblock
\href{https://doi.org/10.1109/CLOUD.2016.0066}{doi:\nolinkurl{10.1109/CLOUD.2016.0066}}


\bibitem[Choi et~al\mbox{.}(2023)]%
        {hydra-network-ordering}
\bibfield{author}{\bibinfo{person}{Inho Choi}, \bibinfo{person}{Ellis Michael}, \bibinfo{person}{Yunfan Li}, \bibinfo{person}{Dan R.~K. Ports}, {and} \bibinfo{person}{Jialin Li}.} \bibinfo{year}{2023}\natexlab{}.
\newblock \showarticletitle{Hydra: {Serialization-Free} Network Ordering for Strongly Consistent Distributed Applications}. In \bibinfo{booktitle}{\emph{20th USENIX Symposium on Networked Systems Design and Implementation (NSDI 23)}}. \bibinfo{publisher}{USENIX Association}, \bibinfo{address}{Boston, MA}, \bibinfo{pages}{293--320}.
\newblock
\showISBNx{978-1-939133-33-5}
\urldef\tempurl%
\url{https://www.usenix.org/conference/nsdi23/presentation/choi}
\showURL{%
\tempurl}


\bibitem[Cooper et~al\mbox{.}(2010)]%
        {ycsb}
\bibfield{author}{\bibinfo{person}{Brian~F. Cooper}, \bibinfo{person}{Adam Silberstein}, \bibinfo{person}{Erwin Tam}, \bibinfo{person}{Raghu Ramakrishnan}, {and} \bibinfo{person}{Russell Sears}.} \bibinfo{year}{2010}\natexlab{}.
\newblock \showarticletitle{Benchmarking Cloud Serving Systems with YCSB}. In \bibinfo{booktitle}{\emph{Proceedings of the 1st ACM Symposium on Cloud Computing}} (Indianapolis, Indiana, USA) \emph{(\bibinfo{series}{SoCC '10})}. \bibinfo{publisher}{Association for Computing Machinery}, \bibinfo{address}{New York, NY, USA}, \bibinfo{pages}{143–154}.
\newblock
\showISBNx{9781450300360}
\href{https://doi.org/10.1145/1807128.1807152}{doi:\nolinkurl{10.1145/1807128.1807152}}


\bibitem[Corbett et~al\mbox{.}(2013)]%
        {spanner}
\bibfield{author}{\bibinfo{person}{James~C. Corbett}, \bibinfo{person}{Jeffrey Dean}, \bibinfo{person}{Michael Epstein}, \bibinfo{person}{Andrew Fikes}, \bibinfo{person}{Christopher Frost}, \bibinfo{person}{J.~J. Furman}, \bibinfo{person}{Sanjay Ghemawat}, \bibinfo{person}{Andrey Gubarev}, \bibinfo{person}{Christopher Heiser}, \bibinfo{person}{Peter Hochschild}, \bibinfo{person}{Wilson Hsieh}, \bibinfo{person}{Sebastian Kanthak}, \bibinfo{person}{Eugene Kogan}, \bibinfo{person}{Hongyi Li}, \bibinfo{person}{Alexander Lloyd}, \bibinfo{person}{Sergey Melnik}, \bibinfo{person}{David Mwaura}, \bibinfo{person}{David Nagle}, \bibinfo{person}{Sean Quinlan}, \bibinfo{person}{Rajesh Rao}, \bibinfo{person}{Lindsay Rolig}, \bibinfo{person}{Yasushi Saito}, \bibinfo{person}{Michal Szymaniak}, \bibinfo{person}{Christopher Taylor}, \bibinfo{person}{Ruth Wang}, {and} \bibinfo{person}{Dale Woodford}.} \bibinfo{year}{2013}\natexlab{}.
\newblock \showarticletitle{Spanner: Google’s Globally Distributed Database}.
\newblock \bibinfo{journal}{\emph{ACM Trans. Comput. Syst.}} \bibinfo{volume}{31}, \bibinfo{number}{3}, Article \bibinfo{articleno}{8} (\bibinfo{date}{aug} \bibinfo{year}{2013}), \bibinfo{numpages}{22}~pages.
\newblock
\showISSN{0734-2071}
\href{https://doi.org/10.1145/2491245}{doi:\nolinkurl{10.1145/2491245}}


\bibitem[Cui et~al\mbox{.}(2015)]%
        {paxos-made-transparent}
\bibfield{author}{\bibinfo{person}{Heming Cui}, \bibinfo{person}{Rui Gu}, \bibinfo{person}{Cheng Liu}, \bibinfo{person}{Tianyu Chen}, {and} \bibinfo{person}{Junfeng Yang}.} \bibinfo{year}{2015}\natexlab{}.
\newblock \showarticletitle{Paxos made transparent}. In \bibinfo{booktitle}{\emph{Proceedings of the 25th Symposium on Operating Systems Principles}} (Monterey, California) \emph{(\bibinfo{series}{SOSP '15})}. \bibinfo{publisher}{Association for Computing Machinery}, \bibinfo{address}{New York, NY, USA}, \bibinfo{pages}{105–120}.
\newblock
\showISBNx{9781450338349}
\href{https://doi.org/10.1145/2815400.2815427}{doi:\nolinkurl{10.1145/2815400.2815427}}


\bibitem[Dang et~al\mbox{.}(2020)]%
        {p4xos}
\bibfield{author}{\bibinfo{person}{Huynh~Tu Dang}, \bibinfo{person}{Pietro Bressana}, \bibinfo{person}{Han Wang}, \bibinfo{person}{Ki~Suh Lee}, \bibinfo{person}{Noa Zilberman}, \bibinfo{person}{Hakim Weatherspoon}, \bibinfo{person}{Marco Canini}, \bibinfo{person}{Fernando Pedone}, {and} \bibinfo{person}{Robert Soul\'{e}}.} \bibinfo{year}{2020}\natexlab{}.
\newblock \showarticletitle{P4xos: Consensus as a Network Service}.
\newblock \bibinfo{journal}{\emph{IEEE/ACM Trans. Netw.}} \bibinfo{volume}{28}, \bibinfo{number}{4} (\bibinfo{date}{Aug.} \bibinfo{year}{2020}), \bibinfo{pages}{1726–1738}.
\newblock
\showISSN{1063-6692}
\href{https://doi.org/10.1109/TNET.2020.2992106}{doi:\nolinkurl{10.1109/TNET.2020.2992106}}


\bibitem[Dang et~al\mbox{.}(2015)]%
        {netpaxos}
\bibfield{author}{\bibinfo{person}{Huynh~Tu Dang}, \bibinfo{person}{Daniele Sciascia}, \bibinfo{person}{Marco Canini}, \bibinfo{person}{Fernando Pedone}, {and} \bibinfo{person}{Robert Soul\'{e}}.} \bibinfo{year}{2015}\natexlab{}.
\newblock \showarticletitle{NetPaxos: consensus at network speed}. In \bibinfo{booktitle}{\emph{Proceedings of the 1st ACM SIGCOMM Symposium on Software Defined Networking Research}} (Santa Clara, California) \emph{(\bibinfo{series}{SOSR '15})}. \bibinfo{publisher}{Association for Computing Machinery}, \bibinfo{address}{New York, NY, USA}, Article \bibinfo{articleno}{5}, \bibinfo{numpages}{7}~pages.
\newblock
\showISBNx{9781450334518}
\href{https://doi.org/10.1145/2774993.2774999}{doi:\nolinkurl{10.1145/2774993.2774999}}


\bibitem[DeCandia et~al\mbox{.}(2007)]%
        {dynamo}
\bibfield{author}{\bibinfo{person}{Giuseppe DeCandia}, \bibinfo{person}{Deniz Hastorun}, \bibinfo{person}{Madan Jampani}, \bibinfo{person}{Gunavardhan Kakulapati}, \bibinfo{person}{Avinash Lakshman}, \bibinfo{person}{Alex Pilchin}, \bibinfo{person}{Swaminathan Sivasubramanian}, \bibinfo{person}{Peter Vosshall}, {and} \bibinfo{person}{Werner Vogels}.} \bibinfo{year}{2007}\natexlab{}.
\newblock \showarticletitle{Dynamo: Amazon's Highly Available Key-Value Store}. In \bibinfo{booktitle}{\emph{Proceedings of Twenty-First ACM SIGOPS Symposium on Operating Systems Principles}} (Stevenson, Washington, USA) \emph{(\bibinfo{series}{SOSP '07})}. \bibinfo{publisher}{Association for Computing Machinery}, \bibinfo{address}{New York, NY, USA}, \bibinfo{pages}{205–220}.
\newblock
\showISBNx{9781595935915}
\href{https://doi.org/10.1145/1294261.1294281}{doi:\nolinkurl{10.1145/1294261.1294281}}


\bibitem[Duplyakin et~al\mbox{.}(2019)]%
        {cloudlab}
\bibfield{author}{\bibinfo{person}{Dmitry Duplyakin}, \bibinfo{person}{Robert Ricci}, \bibinfo{person}{Aleksander Maricq}, \bibinfo{person}{Gary Wong}, \bibinfo{person}{Jonathon Duerig}, \bibinfo{person}{Eric Eide}, \bibinfo{person}{Leigh Stoller}, \bibinfo{person}{Mike Hibler}, \bibinfo{person}{David Johnson}, \bibinfo{person}{Kirk Webb}, \bibinfo{person}{Aditya Akella}, \bibinfo{person}{Kuangching Wang}, \bibinfo{person}{Glenn Ricart}, \bibinfo{person}{Larry Landweber}, \bibinfo{person}{Chip Elliott}, \bibinfo{person}{Michael Zink}, \bibinfo{person}{Emmanuel Cecchet}, \bibinfo{person}{Snigdhaswin Kar}, {and} \bibinfo{person}{Prabodh Mishra}.} \bibinfo{year}{2019}\natexlab{}.
\newblock \showarticletitle{The Design and Operation of {CloudLab}}. In \bibinfo{booktitle}{\emph{Proceedings of the {USENIX} Annual Technical Conference (ATC)}}. \bibinfo{pages}{1--14}.
\newblock
\urldef\tempurl%
\url{https://www.flux.utah.edu/paper/duplyakin-atc19}
\showURL{%
\tempurl}


\bibitem[Enes et~al\mbox{.}(2020)]%
        {atlas}
\bibfield{author}{\bibinfo{person}{Vitor Enes}, \bibinfo{person}{Carlos Baquero}, \bibinfo{person}{Tuanir~Fran\c{c}a Rezende}, \bibinfo{person}{Alexey Gotsman}, \bibinfo{person}{Matthieu Perrin}, {and} \bibinfo{person}{Pierre Sutra}.} \bibinfo{year}{2020}\natexlab{}.
\newblock \showarticletitle{State-Machine Replication for Planet-Scale Systems}. In \bibinfo{booktitle}{\emph{Proceedings of the Fifteenth European Conference on Computer Systems}} (Heraklion, Greece) \emph{(\bibinfo{series}{EuroSys '20})}. \bibinfo{publisher}{Association for Computing Machinery}, \bibinfo{address}{New York, NY, USA}, Article \bibinfo{articleno}{24}, \bibinfo{numpages}{15}~pages.
\newblock
\showISBNx{9781450368827}
\href{https://doi.org/10.1145/3342195.3387543}{doi:\nolinkurl{10.1145/3342195.3387543}}


\bibitem[etcd(2023)]%
        {etcd}
\bibfield{author}{\bibinfo{person}{etcd}.} \bibinfo{year}{2023}\natexlab{}.
\newblock \bibinfo{title}{etcd: A distributed, reliable key-value store for the most critical data of a distributed system}.
\newblock
\newblock
\shownote{\url{https://etcd.io/}, Last accessed on 2023-11-13}.


\bibitem[Fouto et~al\mbox{.}(2022)]%
        {chain-paxos}
\bibfield{author}{\bibinfo{person}{Pedro Fouto}, \bibinfo{person}{Nuno Pregui{\c c}a}, {and} \bibinfo{person}{Joao Leit{\~a}o}.} \bibinfo{year}{2022}\natexlab{}.
\newblock \showarticletitle{High Throughput Replication with Integrated Membership Management}. In \bibinfo{booktitle}{\emph{2022 USENIX Annual Technical Conference (USENIX ATC 22)}}. \bibinfo{publisher}{USENIX Association}, \bibinfo{address}{Carlsbad, CA}, \bibinfo{pages}{575--592}.
\newblock
\showISBNx{978-1-939133-29-67}
\urldef\tempurl%
\url{https://www.usenix.org/conference/atc22/presentation/fouto}
\showURL{%
\tempurl}


\bibitem[Frank et~al\mbox{.}(2025)]%
        {uncertain-consensus}
\bibfield{author}{\bibinfo{person}{Reginald Frank}, \bibinfo{person}{Octavio Lomeli}, \bibinfo{person}{Neil Giridharan}, \bibinfo{person}{Soujanya Ponnapalli}, \bibinfo{person}{Marcos~K. Aguilera}, {and} \bibinfo{person}{Natacha Crooks}.} \bibinfo{year}{2025}\natexlab{}.
\newblock \showarticletitle{Real Life is Uncertain. Consensus Should Be Too!}. In \bibinfo{booktitle}{\emph{Proceedings of the 20th Workshop on Hot Topics in Operating Systems}} (Banff, AB, Canada) \emph{(\bibinfo{series}{HOTOS '25})}. \bibinfo{publisher}{Association for Computing Machinery}, \bibinfo{address}{New York, NY, USA}.
\newblock
\href{https://doi.org/10.1145/3713082.3730374}{doi:\nolinkurl{10.1145/3713082.3730374}}


\bibitem[Gadban and Kunkel(2021)]%
        {s3-hpc-workloads}
\bibfield{author}{\bibinfo{person}{Frank Gadban} {and} \bibinfo{person}{Julian Kunkel}.} \bibinfo{year}{2021}\natexlab{}.
\newblock \showarticletitle{Analyzing the Performance of the S3 Object Storage API for HPC Workloads}.
\newblock \bibinfo{journal}{\emph{Applied Sciences}} \bibinfo{volume}{11}, \bibinfo{number}{18} (\bibinfo{year}{2021}).
\newblock
\showISSN{2076-3417}
\href{https://doi.org/10.3390/app11188540}{doi:\nolinkurl{10.3390/app11188540}}


\bibitem[Ganesan et~al\mbox{.}(2020)]%
        {consistency-aware-durability}
\bibfield{author}{\bibinfo{person}{Aishwarya Ganesan}, \bibinfo{person}{Ramnatthan Alagappan}, \bibinfo{person}{Andrea Arpaci-Dusseau}, {and} \bibinfo{person}{Remzi Arpaci-Dusseau}.} \bibinfo{year}{2020}\natexlab{}.
\newblock \showarticletitle{Strong and Efficient Consistency with {Consistency-Aware} Durability}. In \bibinfo{booktitle}{\emph{18th USENIX Conference on File and Storage Technologies (FAST 20)}}. \bibinfo{publisher}{USENIX Association}, \bibinfo{address}{Santa Clara, CA}, \bibinfo{pages}{323--337}.
\newblock
\showISBNx{978-1-939133-12-0}
\urldef\tempurl%
\url{https://www.usenix.org/conference/fast20/presentation/ganesan}
\showURL{%
\tempurl}


\bibitem[Ganesan et~al\mbox{.}(2021)]%
        {skyros-nil-externality}
\bibfield{author}{\bibinfo{person}{Aishwarya Ganesan}, \bibinfo{person}{Ramnatthan Alagappan}, \bibinfo{person}{Andrea~C. Arpaci-Dusseau}, {and} \bibinfo{person}{Remzi~H. Arpaci-Dusseau}.} \bibinfo{year}{2021}\natexlab{}.
\newblock \showarticletitle{Exploiting Nil-Externality for Fast Replicated Storage}. In \bibinfo{booktitle}{\emph{Proceedings of the ACM SIGOPS 28th Symposium on Operating Systems Principles}} (Virtual Event, Germany) \emph{(\bibinfo{series}{SOSP '21})}. \bibinfo{publisher}{Association for Computing Machinery}, \bibinfo{address}{New York, NY, USA}, \bibinfo{pages}{440–456}.
\newblock
\showISBNx{9781450387095}
\href{https://doi.org/10.1145/3477132.3483543}{doi:\nolinkurl{10.1145/3477132.3483543}}


\bibitem[Ghemawat et~al\mbox{.}(2003)]%
        {gfs}
\bibfield{author}{\bibinfo{person}{Sanjay Ghemawat}, \bibinfo{person}{Howard Gobioff}, {and} \bibinfo{person}{Shun-Tak Leung}.} \bibinfo{year}{2003}\natexlab{}.
\newblock \showarticletitle{The Google File System}. In \bibinfo{booktitle}{\emph{Proceedings of the 19th ACM Symposium on Operating Systems Principles}}. \bibinfo{address}{Bolton Landing, NY}, \bibinfo{pages}{20--43}.
\newblock


\bibitem[Gifford(1979)]%
        {weighted-voting}
\bibfield{author}{\bibinfo{person}{David~K. Gifford}.} \bibinfo{year}{1979}\natexlab{}.
\newblock \showarticletitle{Weighted Voting for Replicated Data}. In \bibinfo{booktitle}{\emph{Proceedings of the Seventh ACM Symposium on Operating Systems Principles}} (Pacific Grove, California, USA) \emph{(\bibinfo{series}{SOSP '79})}. \bibinfo{publisher}{Association for Computing Machinery}, \bibinfo{address}{New York, NY, USA}, \bibinfo{pages}{150–162}.
\newblock
\showISBNx{0897910095}
\href{https://doi.org/10.1145/800215.806583}{doi:\nolinkurl{10.1145/800215.806583}}


\bibitem[Giridharan et~al\mbox{.}(2024)]%
        {autobahn}
\bibfield{author}{\bibinfo{person}{Neil Giridharan}, \bibinfo{person}{Florian Suri-Payer}, \bibinfo{person}{Ittai Abraham}, \bibinfo{person}{Lorenzo Alvisi}, {and} \bibinfo{person}{Natacha Crooks}.} \bibinfo{year}{2024}\natexlab{}.
\newblock \showarticletitle{Autobahn: Seamless high speed BFT}. In \bibinfo{booktitle}{\emph{Proceedings of the ACM SIGOPS 30th Symposium on Operating Systems Principles}} (Austin, TX, USA) \emph{(\bibinfo{series}{SOSP '24})}. \bibinfo{publisher}{Association for Computing Machinery}, \bibinfo{address}{New York, NY, USA}, \bibinfo{pages}{1–23}.
\newblock
\showISBNx{9798400712517}
\href{https://doi.org/10.1145/3694715.3695942}{doi:\nolinkurl{10.1145/3694715.3695942}}


\bibitem[Hadoop(2023)]%
        {hadoop-s3guard}
\bibfield{author}{\bibinfo{person}{Apache Hadoop}.} \bibinfo{year}{2023}\natexlab{}.
\newblock \bibinfo{title}{S3Guard: Consistency and Metadata Caching for S3A}.
\newblock
\newblock
\shownote{\url{https://hadoop.apache.org/docs/r3.0.3/hadoop-aws/tools/hadoop-aws/s3guard.html}, Last accessed on 2023-11-19}.


\bibitem[HashiCorp(2020)]%
        {consul}
\bibfield{author}{\bibinfo{person}{HashiCorp}.} \bibinfo{year}{2020}\natexlab{}.
\newblock \bibinfo{title}{Identity-based networking with Consul}.
\newblock
\newblock
\shownote{\url{https://www.consul.io/}, Last accessed on 2024-10-16}.


\bibitem[Hawblitzel et~al\mbox{.}(2015)]%
        {ironfleet}
\bibfield{author}{\bibinfo{person}{Chris Hawblitzel}, \bibinfo{person}{Jon Howell}, \bibinfo{person}{Manos Kapritsos}, \bibinfo{person}{Jacob~R. Lorch}, \bibinfo{person}{Bryan Parno}, \bibinfo{person}{Michael~L. Roberts}, \bibinfo{person}{Srinath Setty}, {and} \bibinfo{person}{Brian Zill}.} \bibinfo{year}{2015}\natexlab{}.
\newblock \showarticletitle{IronFleet: Proving Practical Distributed Systems Correct}. In \bibinfo{booktitle}{\emph{Proceedings of the 25th Symposium on Operating Systems Principles}} (Monterey, California) \emph{(\bibinfo{series}{SOSP '15})}. \bibinfo{publisher}{Association for Computing Machinery}, \bibinfo{address}{New York, NY, USA}, \bibinfo{pages}{1–17}.
\newblock
\showISBNx{9781450338349}
\href{https://doi.org/10.1145/2815400.2815428}{doi:\nolinkurl{10.1145/2815400.2815428}}


\bibitem[Helt et~al\mbox{.}(2021)]%
        {regular-sequential-consistency}
\bibfield{author}{\bibinfo{person}{Jeffrey Helt}, \bibinfo{person}{Matthew Burke}, \bibinfo{person}{Amit Levy}, {and} \bibinfo{person}{Wyatt Lloyd}.} \bibinfo{year}{2021}\natexlab{}.
\newblock \showarticletitle{Regular Sequential Serializability and Regular Sequential Consistency}. In \bibinfo{booktitle}{\emph{Proceedings of the ACM SIGOPS 28th Symposium on Operating Systems Principles}} (Virtual Event, Germany) \emph{(\bibinfo{series}{SOSP '21})}. \bibinfo{publisher}{Association for Computing Machinery}, \bibinfo{address}{New York, NY, USA}, \bibinfo{pages}{163–179}.
\newblock
\showISBNx{9781450387095}
\href{https://doi.org/10.1145/3477132.3483566}{doi:\nolinkurl{10.1145/3477132.3483566}}


\bibitem[Herlihy and Wing(1990)]%
        {linearizability}
\bibfield{author}{\bibinfo{person}{Maurice~P. Herlihy} {and} \bibinfo{person}{Jeannette~M. Wing}.} \bibinfo{year}{1990}\natexlab{}.
\newblock \showarticletitle{Linearizability: A Correctness Condition for Concurrent Objects}.
\newblock \bibinfo{journal}{\emph{ACM Trans. Program. Lang. Syst.}} \bibinfo{volume}{12}, \bibinfo{number}{3} (\bibinfo{date}{jul} \bibinfo{year}{1990}), \bibinfo{pages}{463–492}.
\newblock
\showISSN{0164-0925}
\href{https://doi.org/10.1145/78969.78972}{doi:\nolinkurl{10.1145/78969.78972}}


\bibitem[Hildebrand and Serenyi(2023)]%
        {colossus}
\bibfield{author}{\bibinfo{person}{Dean Hildebrand} {and} \bibinfo{person}{Denis Serenyi}.} \bibinfo{year}{2023}\natexlab{}.
\newblock \bibinfo{title}{Colossus under the hood: a peek into Google’s scalable storage system}.
\newblock
\newblock
\shownote{\url{https://cloud.google.com/blog/products/storage-data-transfer/a-peek-behind-colossus-googles-file-system}, Last accessed on 2023-11-19}.


\bibitem[Hoang~Le et~al\mbox{.}(2019)]%
        {dynastar}
\bibfield{author}{\bibinfo{person}{Long Hoang~Le}, \bibinfo{person}{Enrique Fynn}, \bibinfo{person}{Mojtaba Eslahi-Kelorazi}, \bibinfo{person}{Robert Soulé}, {and} \bibinfo{person}{Fernando Pedone}.} \bibinfo{year}{2019}\natexlab{}.
\newblock \showarticletitle{DynaStar: Optimized Dynamic Partitioning for Scalable State Machine Replication}. In \bibinfo{booktitle}{\emph{2019 IEEE 39th International Conference on Distributed Computing Systems (ICDCS)}}. \bibinfo{pages}{1453--1465}.
\newblock
\href{https://doi.org/10.1109/ICDCS.2019.00145}{doi:\nolinkurl{10.1109/ICDCS.2019.00145}}


\bibitem[Howard et~al\mbox{.}(2016)]%
        {fpaxos}
\bibfield{author}{\bibinfo{person}{Heidi Howard}, \bibinfo{person}{Dahlia Malkhi}, {and} \bibinfo{person}{Alexander Spiegelman}.} \bibinfo{year}{2016}\natexlab{}.
\newblock \bibinfo{title}{Flexible Paxos: Quorum intersection revisited}.
\newblock
\showeprint[arxiv]{1608.06696}~[cs.DC]


\bibitem[Hu et~al\mbox{.}(2024a)]%
        {practical-consistency-summary}
\bibfield{author}{\bibinfo{person}{Guanzhou Hu}, \bibinfo{person}{Andrea Arpaci-Dusseau}, {and} \bibinfo{person}{Remzi Arpaci-Dusseau}.} \bibinfo{year}{2024}\natexlab{a}.
\newblock \bibinfo{title}{A Unified, Practical, and Understandable Summary of Non-transactional Consistency Levels in Distributed Replication}.
\newblock
\showeprint[arxiv]{2409.01576}~[cs.DC]
\urldef\tempurl%
\url{https://arxiv.org/abs/2409.01576}
\showURL{%
\tempurl}


\bibitem[Hu et~al\mbox{.}(2021)]%
        {network-coding}
\bibfield{author}{\bibinfo{person}{Yuchong Hu}, \bibinfo{person}{Xiaoyang Zhang}, \bibinfo{person}{Patrick P.~C. Lee}, {and} \bibinfo{person}{Pan Zhou}.} \bibinfo{year}{2021}\natexlab{}.
\newblock \showarticletitle{NCScale: Toward Optimal Storage Scaling via Network Coding}.
\newblock \bibinfo{journal}{\emph{IEEE/ACM Trans. Netw.}} \bibinfo{volume}{30}, \bibinfo{number}{1} (\bibinfo{date}{Aug.} \bibinfo{year}{2021}), \bibinfo{pages}{271–284}.
\newblock
\showISSN{1063-6692}
\href{https://doi.org/10.1109/TNET.2021.3106394}{doi:\nolinkurl{10.1109/TNET.2021.3106394}}


\bibitem[Hu et~al\mbox{.}(2024b)]%
        {aceso-disaggregated-memory}
\bibfield{author}{\bibinfo{person}{Zhisheng Hu}, \bibinfo{person}{Pengfei Zuo}, \bibinfo{person}{Yizou Chen}, \bibinfo{person}{Chao Wang}, \bibinfo{person}{Junliang Hu}, {and} \bibinfo{person}{Ming-Chang Yang}.} \bibinfo{year}{2024}\natexlab{b}.
\newblock \showarticletitle{Aceso: Achieving Efficient Fault Tolerance in Memory-Disaggregated Key-Value Stores}. In \bibinfo{booktitle}{\emph{Proceedings of the ACM SIGOPS 30th Symposium on Operating Systems Principles}} (Austin, TX, USA) \emph{(\bibinfo{series}{SOSP '24})}. \bibinfo{publisher}{Association for Computing Machinery}, \bibinfo{address}{New York, NY, USA}, \bibinfo{pages}{127–143}.
\newblock
\showISBNx{9798400712517}
\href{https://doi.org/10.1145/3694715.3695951}{doi:\nolinkurl{10.1145/3694715.3695951}}


\bibitem[Huang et~al\mbox{.}(2012)]%
        {erasure-coding-azure}
\bibfield{author}{\bibinfo{person}{Cheng Huang}, \bibinfo{person}{Huseyin Simitci}, \bibinfo{person}{Yikang Xu}, \bibinfo{person}{Aaron Ogus}, \bibinfo{person}{Brad Calder}, \bibinfo{person}{Parikshit Gopalan}, \bibinfo{person}{Jin Li}, {and} \bibinfo{person}{Sergey Yekhanin}.} \bibinfo{year}{2012}\natexlab{}.
\newblock \showarticletitle{Erasure Coding in Windows Azure Storage}. In \bibinfo{booktitle}{\emph{2012 USENIX Annual Technical Conference (USENIX ATC 12)}}. \bibinfo{publisher}{USENIX Association}, \bibinfo{address}{Boston, MA}, \bibinfo{pages}{15--26}.
\newblock
\showISBNx{978-931971-93-5}
\urldef\tempurl%
\url{https://www.usenix.org/conference/atc12/technical-sessions/presentation/huang}
\showURL{%
\tempurl}


\bibitem[Huang et~al\mbox{.}(2020)]%
        {tidb}
\bibfield{author}{\bibinfo{person}{Dongxu Huang}, \bibinfo{person}{Qi Liu}, \bibinfo{person}{Qiu Cui}, \bibinfo{person}{Zhuhe Fang}, \bibinfo{person}{Xiaoyu Ma}, \bibinfo{person}{Fei Xu}, \bibinfo{person}{Li Shen}, \bibinfo{person}{Liu Tang}, \bibinfo{person}{Yuxing Zhou}, \bibinfo{person}{Menglong Huang}, \bibinfo{person}{Wan Wei}, \bibinfo{person}{Cong Liu}, \bibinfo{person}{Jian Zhang}, \bibinfo{person}{Jianjun Li}, \bibinfo{person}{Xuelian Wu}, \bibinfo{person}{Lingyu Song}, \bibinfo{person}{Ruoxi Sun}, \bibinfo{person}{Shuaipeng Yu}, \bibinfo{person}{Lei Zhao}, \bibinfo{person}{Nicholas Cameron}, \bibinfo{person}{Liquan Pei}, {and} \bibinfo{person}{Xin Tang}.} \bibinfo{year}{2020}\natexlab{}.
\newblock \showarticletitle{TiDB: A Raft-Based HTAP Database}.
\newblock \bibinfo{journal}{\emph{Proc. VLDB Endow.}} \bibinfo{volume}{13}, \bibinfo{number}{12} (\bibinfo{date}{aug} \bibinfo{year}{2020}), \bibinfo{pages}{3072–3084}.
\newblock
\showISSN{2150-8097}
\href{https://doi.org/10.14778/3415478.3415535}{doi:\nolinkurl{10.14778/3415478.3415535}}


\bibitem[Huang et~al\mbox{.}(2017)]%
        {gray-failures}
\bibfield{author}{\bibinfo{person}{Peng Huang}, \bibinfo{person}{Chuanxiong Guo}, \bibinfo{person}{Lidong Zhou}, \bibinfo{person}{Jacob~R. Lorch}, \bibinfo{person}{Yingnong Dang}, \bibinfo{person}{Murali Chintalapati}, {and} \bibinfo{person}{Randolph Yao}.} \bibinfo{year}{2017}\natexlab{}.
\newblock \showarticletitle{Gray Failure: The Achilles' Heel of Cloud-Scale Systems}. In \bibinfo{booktitle}{\emph{Proceedings of the 16th Workshop on Hot Topics in Operating Systems}} (Whistler, BC, Canada) \emph{(\bibinfo{series}{HotOS '17})}. \bibinfo{publisher}{Association for Computing Machinery}, \bibinfo{address}{New York, NY, USA}, \bibinfo{pages}{150–155}.
\newblock
\showISBNx{9781450350686}
\href{https://doi.org/10.1145/3102980.3103005}{doi:\nolinkurl{10.1145/3102980.3103005}}


\bibitem[Hunt et~al\mbox{.}(2010)]%
        {zookeeper}
\bibfield{author}{\bibinfo{person}{Patrick Hunt}, \bibinfo{person}{Mahadev Konar}, \bibinfo{person}{Flavio~P. Junqueira}, {and} \bibinfo{person}{Benjamin Reed}.} \bibinfo{year}{2010}\natexlab{}.
\newblock \showarticletitle{ZooKeeper: Wait-Free Coordination for Internet-Scale Systems}. In \bibinfo{booktitle}{\emph{Proceedings of the 2010 USENIX Conference on USENIX Annual Technical Conference}} (Boston, MA) \emph{(\bibinfo{series}{USENIXATC'10})}. \bibinfo{publisher}{USENIX Association}, \bibinfo{address}{USA}, \bibinfo{pages}{11}.
\newblock


\bibitem[Inc.(2023)]%
        {firescroll}
\bibfield{author}{\bibinfo{person}{Redpanda~Data Inc.}} \bibinfo{year}{2023}\natexlab{}.
\newblock \bibinfo{title}{FireScroll: The config database to deploy everywhere}.
\newblock
\newblock
\shownote{\url{https://github.com/FireScroll/FireScroll}, Last accessed on 2024-09-05}.


\bibitem[Jha et~al\mbox{.}(2019)]%
        {derecho}
\bibfield{author}{\bibinfo{person}{Sagar Jha}, \bibinfo{person}{Jonathan Behrens}, \bibinfo{person}{Theo Gkountouvas}, \bibinfo{person}{Mae Milano}, \bibinfo{person}{Weijia Song}, \bibinfo{person}{Edward Tremel}, \bibinfo{person}{Robbert~Van Renesse}, \bibinfo{person}{Sydney Zink}, {and} \bibinfo{person}{Kenneth~P. Birman}.} \bibinfo{year}{2019}\natexlab{}.
\newblock \showarticletitle{Derecho: Fast State Machine Replication for Cloud Services}.
\newblock \bibinfo{journal}{\emph{ACM Trans. Comput. Syst.}} \bibinfo{volume}{36}, \bibinfo{number}{2}, Article \bibinfo{articleno}{4} (\bibinfo{date}{apr} \bibinfo{year}{2019}), \bibinfo{numpages}{49}~pages.
\newblock
\showISSN{0734-2071}
\href{https://doi.org/10.1145/3302258}{doi:\nolinkurl{10.1145/3302258}}


\bibitem[Jia et~al\mbox{.}(2022)]%
        {hraft}
\bibfield{author}{\bibinfo{person}{Yulei Jia}, \bibinfo{person}{Guangping Xu}, \bibinfo{person}{Chi~Wan Sung}, \bibinfo{person}{Salwa Mostafa}, {and} \bibinfo{person}{Yulei Wu}.} \bibinfo{year}{2022}\natexlab{}.
\newblock \showarticletitle{HRaft: Adaptive Erasure Coded Data Maintenance for Consensus in Distributed Networks}. In \bibinfo{booktitle}{\emph{2022 IEEE International Parallel and Distributed Processing Symposium (IPDPS)}}. \bibinfo{pages}{1316--1326}.
\newblock
\href{https://doi.org/10.1109/IPDPS53621.2022.00130}{doi:\nolinkurl{10.1109/IPDPS53621.2022.00130}}


\bibitem[Jiang et~al\mbox{.}(2023)]%
        {non-blocking-raft}
\bibfield{author}{\bibinfo{person}{Tian Jiang}, \bibinfo{person}{Xiangdong Huang}, \bibinfo{person}{Shaoxu Song}, \bibinfo{person}{Chen Wang}, \bibinfo{person}{Jianmin Wang}, \bibinfo{person}{Ruibo Li}, {and} \bibinfo{person}{Jincheng Sun}.} \bibinfo{year}{2023}\natexlab{}.
\newblock \showarticletitle{Non-Blocking Raft for High Throughput IoT Data}. In \bibinfo{booktitle}{\emph{2023 IEEE 39th International Conference on Data Engineering (ICDE)}}. \bibinfo{pages}{1140--1152}.
\newblock
\href{https://doi.org/10.1109/ICDE55515.2023.00092}{doi:\nolinkurl{10.1109/ICDE55515.2023.00092}}


\bibitem[Kadekodi et~al\mbox{.}(2023)]%
        {wide-lrcs}
\bibfield{author}{\bibinfo{person}{Saurabh Kadekodi}, \bibinfo{person}{Shashwat Silas}, \bibinfo{person}{David Clausen}, {and} \bibinfo{person}{Arif Merchant}.} \bibinfo{year}{2023}\natexlab{}.
\newblock \showarticletitle{Practical Design Considerations for Wide Locally Recoverable Codes ({{{{{LRCs}}}}})}. In \bibinfo{booktitle}{\emph{21st USENIX Conference on File and Storage Technologies (FAST 23)}}. \bibinfo{publisher}{USENIX Association}, \bibinfo{address}{Santa Clara, CA}, \bibinfo{pages}{1--16}.
\newblock
\showISBNx{978-1-939133-32-8}
\urldef\tempurl%
\url{https://www.usenix.org/conference/fast23/presentation/kadekodi}
\showURL{%
\tempurl}


\bibitem[Kreps et~al\mbox{.}(2011)]%
        {kafka}
\bibfield{author}{\bibinfo{person}{Jay Kreps}, \bibinfo{person}{Neha Narkhede}, \bibinfo{person}{Jun Rao}, {et~al\mbox{.}}} \bibinfo{year}{2011}\natexlab{}.
\newblock \showarticletitle{Kafka: A distributed messaging system for log processing}. In \bibinfo{booktitle}{\emph{Proceedings of the NetDB}}, Vol.~\bibinfo{volume}{11}. Athens, Greece, \bibinfo{pages}{1--7}.
\newblock


\bibitem[Kubiatowicz et~al\mbox{.}(2000)]%
        {oceanstore}
\bibfield{author}{\bibinfo{person}{John Kubiatowicz}, \bibinfo{person}{David Bindel}, \bibinfo{person}{Yan Chen}, \bibinfo{person}{Steven Czerwinski}, \bibinfo{person}{Patrick Eaton}, \bibinfo{person}{Dennis Geels}, \bibinfo{person}{Ramakrishan Gummadi}, \bibinfo{person}{Sean Rhea}, \bibinfo{person}{Hakim Weatherspoon}, \bibinfo{person}{Westley Weimer}, \bibinfo{person}{Chris Wells}, {and} \bibinfo{person}{Ben Zhao}.} \bibinfo{year}{2000}\natexlab{}.
\newblock \showarticletitle{OceanStore: an architecture for global-scale persistent storage}.
\newblock \bibinfo{journal}{\emph{SIGPLAN Not.}} \bibinfo{volume}{35}, \bibinfo{number}{11} (\bibinfo{date}{Nov.} \bibinfo{year}{2000}), \bibinfo{pages}{190–201}.
\newblock
\showISSN{0362-1340}
\href{https://doi.org/10.1145/356989.357007}{doi:\nolinkurl{10.1145/356989.357007}}


\bibitem[Lamport(1979)]%
        {sequential-consistency}
\bibfield{author}{\bibinfo{person}{Leslie Lamport}.} \bibinfo{year}{1979}\natexlab{}.
\newblock \showarticletitle{How to Make a Multiprocessor Computer That Correctly Executes Multiprocess Programs}.
\newblock \bibinfo{journal}{\emph{IEEE Transactions on Computers C-28}}  \bibinfo{volume}{9} (\bibinfo{date}{September} \bibinfo{year}{1979}), \bibinfo{pages}{690--691}.
\newblock
\urldef\tempurl%
\url{https://www.microsoft.com/en-us/research/publication/make-multiprocessor-computer-correctly-executes-multiprocess-programs/}
\showURL{%
\tempurl}


\bibitem[Lamport(1998)]%
        {paxos-parliament}
\bibfield{author}{\bibinfo{person}{Leslie Lamport}.} \bibinfo{year}{1998}\natexlab{}.
\newblock \showarticletitle{The Part-Time Parliament}.
\newblock \bibinfo{journal}{\emph{ACM Trans. Comput. Syst.}} \bibinfo{volume}{16}, \bibinfo{number}{2} (\bibinfo{date}{may} \bibinfo{year}{1998}), \bibinfo{pages}{133–169}.
\newblock
\showISSN{0734-2071}
\href{https://doi.org/10.1145/279227.279229}{doi:\nolinkurl{10.1145/279227.279229}}


\bibitem[Lamport(2001)]%
        {paxos-made-simple}
\bibfield{author}{\bibinfo{person}{Leslie Lamport}.} \bibinfo{year}{2001}\natexlab{}.
\newblock \showarticletitle{Paxos Made Simple}.
\newblock \bibinfo{journal}{\emph{ACM SIGACT News (Distributed Computing Column) 32, 4 (Whole Number 121, December 2001)}} (\bibinfo{date}{December} \bibinfo{year}{2001}), \bibinfo{pages}{51--58}.
\newblock
\urldef\tempurl%
\url{https://www.microsoft.com/en-us/research/publication/paxos-made-simple/}
\showURL{%
\tempurl}


\bibitem[Lamport(2005)]%
        {generalized-paxos}
\bibfield{author}{\bibinfo{person}{Leslie Lamport}.} \bibinfo{year}{2005}\natexlab{}.
\newblock \showarticletitle{Generalized consensus and Paxos}.
\newblock \bibinfo{journal}{\emph{Microsoft Research Technical Report}} (\bibinfo{year}{2005}).
\newblock


\bibitem[Lamport(2006)]%
        {fast-paxos}
\bibfield{author}{\bibinfo{person}{Leslie Lamport}.} \bibinfo{year}{2006}\natexlab{}.
\newblock \showarticletitle{Fast Paxos}.
\newblock \bibinfo{journal}{\emph{Distrib. Comput.}} \bibinfo{volume}{19}, \bibinfo{number}{2} (\bibinfo{date}{oct} \bibinfo{year}{2006}), \bibinfo{pages}{79–103}.
\newblock
\showISSN{0178-2770}
\href{https://doi.org/10.1007/s00446-006-0005-x}{doi:\nolinkurl{10.1007/s00446-006-0005-x}}


\bibitem[Lattuada et~al\mbox{.}(2024)]%
        {verus}
\bibfield{author}{\bibinfo{person}{Andrea Lattuada}, \bibinfo{person}{Travis Hance}, \bibinfo{person}{Jay Bosamiya}, \bibinfo{person}{Matthias Brun}, \bibinfo{person}{Chanhee Cho}, \bibinfo{person}{Hayley LeBlanc}, \bibinfo{person}{Pranav Srinivasan}, \bibinfo{person}{Reto Achermann}, \bibinfo{person}{Tej Chajed}, \bibinfo{person}{Chris Hawblitzel}, \bibinfo{person}{Jon Howell}, \bibinfo{person}{Jacob~R. Lorch}, \bibinfo{person}{Oded Padon}, {and} \bibinfo{person}{Bryan Parno}.} \bibinfo{year}{2024}\natexlab{}.
\newblock \showarticletitle{Verus: A Practical Foundation for Systems Verification}. In \bibinfo{booktitle}{\emph{Proceedings of the ACM SIGOPS 30th Symposium on Operating Systems Principles}} (Austin, TX, USA) \emph{(\bibinfo{series}{SOSP '24})}. \bibinfo{publisher}{Association for Computing Machinery}, \bibinfo{address}{New York, NY, USA}, \bibinfo{pages}{438–454}.
\newblock
\showISBNx{9798400712517}
\href{https://doi.org/10.1145/3694715.3695952}{doi:\nolinkurl{10.1145/3694715.3695952}}


\bibitem[Leutenegger and Dias(1993)]%
        {tpc-c}
\bibfield{author}{\bibinfo{person}{Scott~T. Leutenegger} {and} \bibinfo{person}{Daniel Dias}.} \bibinfo{year}{1993}\natexlab{}.
\newblock \showarticletitle{A Modeling Study of the TPC-C Benchmark}. In \bibinfo{booktitle}{\emph{Proceedings of the 1993 ACM SIGMOD International Conference on Management of Data}} (Washington, D.C., USA) \emph{(\bibinfo{series}{SIGMOD '93})}. \bibinfo{publisher}{Association for Computing Machinery}, \bibinfo{address}{New York, NY, USA}, \bibinfo{pages}{22–31}.
\newblock
\showISBNx{0897915925}
\href{https://doi.org/10.1145/170035.170042}{doi:\nolinkurl{10.1145/170035.170042}}


\bibitem[Li et~al\mbox{.}(2016)]%
        {nopaxos}
\bibfield{author}{\bibinfo{person}{Jialin Li}, \bibinfo{person}{Ellis Michael}, \bibinfo{person}{Naveen~Kr. Sharma}, \bibinfo{person}{Adriana Szekeres}, {and} \bibinfo{person}{Dan R.~K. Ports}.} \bibinfo{year}{2016}\natexlab{}.
\newblock \showarticletitle{Just Say {NO} to Paxos Overhead: Replacing Consensus with Network Ordering}. In \bibinfo{booktitle}{\emph{12th USENIX Symposium on Operating Systems Design and Implementation (OSDI 16)}}. \bibinfo{publisher}{USENIX Association}, \bibinfo{address}{Savannah, GA}, \bibinfo{pages}{467--483}.
\newblock
\showISBNx{978-1-931971-33-1}
\urldef\tempurl%
\url{https://www.usenix.org/conference/osdi16/technical-sessions/presentation/li}
\showURL{%
\tempurl}


\bibitem[Li and Lee(2014)]%
        {stair-codes}
\bibfield{author}{\bibinfo{person}{Mingqiang Li} {and} \bibinfo{person}{Patrick P.~C. Lee}.} \bibinfo{year}{2014}\natexlab{}.
\newblock \showarticletitle{{STAIR} Codes: A General Family of Erasure Codes for Tolerating Device and Sector Failures in Practical Storage Systems}. In \bibinfo{booktitle}{\emph{12th USENIX Conference on File and Storage Technologies (FAST 14)}}. \bibinfo{publisher}{USENIX Association}, \bibinfo{address}{Santa Clara, CA}, \bibinfo{pages}{147--162}.
\newblock
\showISBNx{ISBN 978-1-931971-08-9}
\urldef\tempurl%
\url{https://www.usenix.org/conference/fast14/technical-sessions/presentation/li-mingqiang}
\showURL{%
\tempurl}


\bibitem[Li et~al\mbox{.}(2010)]%
        {memory-errors-eval}
\bibfield{author}{\bibinfo{person}{Xin Li}, \bibinfo{person}{Michael~C. Huang}, \bibinfo{person}{Kai Shen}, {and} \bibinfo{person}{Lingkun Chu}.} \bibinfo{year}{2010}\natexlab{}.
\newblock \showarticletitle{A Realistic Evaluation of Memory Hardware Errors and Software System Susceptibility}. In \bibinfo{booktitle}{\emph{Proceedings of the 2010 USENIX Conference on USENIX Annual Technical Conference}} (Boston, MA) \emph{(\bibinfo{series}{USENIXATC'10})}. \bibinfo{publisher}{USENIX Association}, \bibinfo{address}{USA}, \bibinfo{pages}{6}.
\newblock


\bibitem[Li et~al\mbox{.}(2019)]%
        {openec}
\bibfield{author}{\bibinfo{person}{Xiaolu Li}, \bibinfo{person}{Runhui Li}, \bibinfo{person}{Patrick P.~C. Lee}, {and} \bibinfo{person}{Yuchong Hu}.} \bibinfo{year}{2019}\natexlab{}.
\newblock \showarticletitle{{OpenEC}: Toward Unified and Configurable Erasure Coding Management in Distributed Storage Systems}. In \bibinfo{booktitle}{\emph{17th USENIX Conference on File and Storage Technologies (FAST 19)}}. \bibinfo{publisher}{USENIX Association}, \bibinfo{address}{Boston, MA}, \bibinfo{pages}{331--344}.
\newblock
\showISBNx{978-1-939133-09-0}
\urldef\tempurl%
\url{https://www.usenix.org/conference/fast19/presentation/li}
\showURL{%
\tempurl}


\bibitem[{Linux man pages}(2011)]%
        {tc-netem}
\bibfield{author}{\bibinfo{person}{{Linux man pages}}.} \bibinfo{year}{2011}\natexlab{}.
\newblock \bibinfo{title}{tc-netem(8) — Linux manual page}.
\newblock \bibinfo{howpublished}{\url{https://man7.org/linux/man-pages/man8/tc-netem.8.html}}.
\newblock
\newblock
\shownote{[Online; accessed 29-November-2023]}.


\bibitem[Liu et~al\mbox{.}(2020)]%
        {fine-grained-rsms}
\bibfield{author}{\bibinfo{person}{Ming Liu}, \bibinfo{person}{Arvind Krishnamurthy}, \bibinfo{person}{Harsha~V. Madhyastha}, \bibinfo{person}{Rishi Bhardwaj}, \bibinfo{person}{Karan Gupta}, \bibinfo{person}{Chinmay Kamat}, \bibinfo{person}{Huapeng Yuan}, \bibinfo{person}{Aditya Jaltade}, \bibinfo{person}{Roger Liao}, \bibinfo{person}{Pavan Konka}, {and} \bibinfo{person}{Anoop Jawahar}.} \bibinfo{year}{2020}\natexlab{}.
\newblock \showarticletitle{{Fine-Grained} Replicated State Machines for a Cluster Storage System}. In \bibinfo{booktitle}{\emph{17th USENIX Symposium on Networked Systems Design and Implementation (NSDI 20)}}. \bibinfo{publisher}{USENIX Association}, \bibinfo{address}{Santa Clara, CA}, \bibinfo{pages}{305--323}.
\newblock
\showISBNx{978-1-939133-13-7}
\urldef\tempurl%
\url{https://www.usenix.org/conference/nsdi20/presentation/liu-ming}
\showURL{%
\tempurl}


\bibitem[Lloyd et~al\mbox{.}(2011)]%
        {cops}
\bibfield{author}{\bibinfo{person}{Wyatt Lloyd}, \bibinfo{person}{Michael~J. Freedman}, \bibinfo{person}{Michael Kaminsky}, {and} \bibinfo{person}{David~G. Andersen}.} \bibinfo{year}{2011}\natexlab{}.
\newblock \showarticletitle{Don't settle for eventual: scalable causal consistency for wide-area storage with COPS}. In \bibinfo{booktitle}{\emph{Proceedings of the Twenty-Third ACM Symposium on Operating Systems Principles}} (Cascais, Portugal) \emph{(\bibinfo{series}{SOSP '11})}. \bibinfo{publisher}{Association for Computing Machinery}, \bibinfo{address}{New York, NY, USA}, \bibinfo{pages}{401–416}.
\newblock
\showISBNx{9781450309776}
\href{https://doi.org/10.1145/2043556.2043593}{doi:\nolinkurl{10.1145/2043556.2043593}}


\bibitem[Lockerman et~al\mbox{.}(2018)]%
        {fuzzylog}
\bibfield{author}{\bibinfo{person}{Joshua Lockerman}, \bibinfo{person}{Jose~M. Faleiro}, \bibinfo{person}{Juno Kim}, \bibinfo{person}{Soham Sankaran}, \bibinfo{person}{Daniel~J. Abadi}, \bibinfo{person}{James Aspnes}, \bibinfo{person}{Siddhartha Sen}, {and} \bibinfo{person}{Mahesh Balakrishnan}.} \bibinfo{year}{2018}\natexlab{}.
\newblock \showarticletitle{The {FuzzyLog}: A Partially Ordered Shared Log}. In \bibinfo{booktitle}{\emph{13th USENIX Symposium on Operating Systems Design and Implementation (OSDI 18)}}. \bibinfo{publisher}{USENIX Association}, \bibinfo{address}{Carlsbad, CA}, \bibinfo{pages}{357--372}.
\newblock
\showISBNx{978-1-939133-08-3}
\urldef\tempurl%
\url{https://www.usenix.org/conference/osdi18/presentation/lockerman}
\showURL{%
\tempurl}


\bibitem[Lucas(1971)]%
        {perf-eval-monitor}
\bibfield{author}{\bibinfo{person}{Henry Lucas}.} \bibinfo{year}{1971}\natexlab{}.
\newblock \showarticletitle{Performance Evaluation and Monitoring}.
\newblock \bibinfo{journal}{\emph{ACM Comput. Surv.}} \bibinfo{volume}{3}, \bibinfo{number}{3} (\bibinfo{date}{sep} \bibinfo{year}{1971}), \bibinfo{pages}{79–91}.
\newblock
\showISSN{0360-0300}
\href{https://doi.org/10.1145/356589.356590}{doi:\nolinkurl{10.1145/356589.356590}}


\bibitem[Luo et~al\mbox{.}(2024)]%
        {lazylog}
\bibfield{author}{\bibinfo{person}{Xuhao Luo}, \bibinfo{person}{Shreesha~G. Bhat}, \bibinfo{person}{Jiyu Hu}, \bibinfo{person}{Ramnatthan Alagappan}, {and} \bibinfo{person}{Aishwarya Ganesan}.} \bibinfo{year}{2024}\natexlab{}.
\newblock \showarticletitle{LazyLog: A New Shared Log Abstraction for Low-Latency Applications}. In \bibinfo{booktitle}{\emph{Proceedings of the ACM SIGOPS 30th Symposium on Operating Systems Principles}} (Austin, TX, USA) \emph{(\bibinfo{series}{SOSP '24})}. \bibinfo{publisher}{Association for Computing Machinery}, \bibinfo{address}{New York, NY, USA}, \bibinfo{pages}{296–312}.
\newblock
\showISBNx{9798400712517}
\href{https://doi.org/10.1145/3694715.3695983}{doi:\nolinkurl{10.1145/3694715.3695983}}


\bibitem[Ma et~al\mbox{.}(2022)]%
        {sift-veri}
\bibfield{author}{\bibinfo{person}{Haojun Ma}, \bibinfo{person}{Hammad Ahmad}, \bibinfo{person}{Aman Goel}, \bibinfo{person}{Eli Goldweber}, \bibinfo{person}{Jean-Baptiste Jeannin}, \bibinfo{person}{Manos Kapritsos}, {and} \bibinfo{person}{Baris Kasikci}.} \bibinfo{year}{2022}\natexlab{}.
\newblock \showarticletitle{Sift: Using Refinement-guided Automation to Verify Complex Distributed Systems}. In \bibinfo{booktitle}{\emph{2022 USENIX Annual Technical Conference (USENIX ATC 22)}}. \bibinfo{publisher}{USENIX Association}, \bibinfo{address}{Carlsbad, CA}, \bibinfo{pages}{151--166}.
\newblock
\showISBNx{978-1-939133-29-64}
\urldef\tempurl%
\url{https://www.usenix.org/conference/atc22/presentation/ma}
\showURL{%
\tempurl}


\bibitem[Ma et~al\mbox{.}(2019)]%
        {I4}
\bibfield{author}{\bibinfo{person}{Haojun Ma}, \bibinfo{person}{Aman Goel}, \bibinfo{person}{Jean-Baptiste Jeannin}, \bibinfo{person}{Manos Kapritsos}, \bibinfo{person}{Baris Kasikci}, {and} \bibinfo{person}{Karem~A. Sakallah}.} \bibinfo{year}{2019}\natexlab{}.
\newblock \showarticletitle{I4: Incremental Inference of Inductive Invariants for Verification of Distributed Protocols}. In \bibinfo{booktitle}{\emph{Proceedings of the 27th ACM Symposium on Operating Systems Principles}} (Huntsville, Ontario, Canada) \emph{(\bibinfo{series}{SOSP '19})}. \bibinfo{publisher}{Association for Computing Machinery}, \bibinfo{address}{New York, NY, USA}, \bibinfo{pages}{370–384}.
\newblock
\showISBNx{9781450368735}
\href{https://doi.org/10.1145/3341301.3359651}{doi:\nolinkurl{10.1145/3341301.3359651}}


\bibitem[Ma et~al\mbox{.}(2024)]%
        {noctua}
\bibfield{author}{\bibinfo{person}{Kai Ma}, \bibinfo{person}{Cheng Li}, \bibinfo{person}{Enzuo Zhu}, \bibinfo{person}{Ruichuan Chen}, \bibinfo{person}{Feng Yan}, {and} \bibinfo{person}{Kang Chen}.} \bibinfo{year}{2024}\natexlab{}.
\newblock \showarticletitle{Noctua: Towards Automated and Practical Fine-grained Consistency Analysis}. In \bibinfo{booktitle}{\emph{Proceedings of the Nineteenth European Conference on Computer Systems}} (Athens, Greece) \emph{(\bibinfo{series}{EuroSys '24})}. \bibinfo{publisher}{Association for Computing Machinery}, \bibinfo{address}{New York, NY, USA}, \bibinfo{pages}{704–719}.
\newblock
\showISBNx{9798400704376}
\href{https://doi.org/10.1145/3627703.3629570}{doi:\nolinkurl{10.1145/3627703.3629570}}


\bibitem[Maccormick et~al\mbox{.}(2008)]%
        {niobe}
\bibfield{author}{\bibinfo{person}{John Maccormick}, \bibinfo{person}{Chandramohan~A. Thekkath}, \bibinfo{person}{Marcus Jager}, \bibinfo{person}{Kristof Roomp}, \bibinfo{person}{Lidong Zhou}, {and} \bibinfo{person}{Ryan Peterson}.} \bibinfo{year}{2008}\natexlab{}.
\newblock \showarticletitle{Niobe: A Practical Replication Protocol}.
\newblock \bibinfo{journal}{\emph{ACM Trans. Storage}} \bibinfo{volume}{3}, \bibinfo{number}{4}, Article \bibinfo{articleno}{1} (\bibinfo{date}{feb} \bibinfo{year}{2008}), \bibinfo{numpages}{43}~pages.
\newblock
\showISSN{1553-3077}
\href{https://doi.org/10.1145/1326542.1326543}{doi:\nolinkurl{10.1145/1326542.1326543}}


\bibitem[Mao et~al\mbox{.}(2008)]%
        {mencius}
\bibfield{author}{\bibinfo{person}{Yanhua Mao}, \bibinfo{person}{Flavio~P. Junqueira}, {and} \bibinfo{person}{Keith Marzullo}.} \bibinfo{year}{2008}\natexlab{}.
\newblock \showarticletitle{Mencius: Building Efficient Replicated State Machines for WANs}. In \bibinfo{booktitle}{\emph{Proceedings of the 8th USENIX Conference on Operating Systems Design and Implementation}} (San Diego, California) \emph{(\bibinfo{series}{OSDI'08})}. \bibinfo{publisher}{USENIX Association}, \bibinfo{address}{USA}, \bibinfo{pages}{369–384}.
\newblock


\bibitem[Marandi et~al\mbox{.}(2010)]%
        {ring-paxos}
\bibfield{author}{\bibinfo{person}{Parisa~Jalili Marandi}, \bibinfo{person}{Marco Primi}, \bibinfo{person}{Nicolas Schiper}, {and} \bibinfo{person}{Fernando Pedone}.} \bibinfo{year}{2010}\natexlab{}.
\newblock \showarticletitle{Ring Paxos: A high-throughput atomic broadcast protocol}. In \bibinfo{booktitle}{\emph{2010 IEEE/IFIP International Conference on Dependable Systems and Networks (DSN)}}. \bibinfo{pages}{527--536}.
\newblock
\href{https://doi.org/10.1109/DSN.2010.5544272}{doi:\nolinkurl{10.1109/DSN.2010.5544272}}


\bibitem[Mehdi et~al\mbox{.}(2017)]%
        {occult}
\bibfield{author}{\bibinfo{person}{Syed~Akbar Mehdi}, \bibinfo{person}{Cody Littley}, \bibinfo{person}{Natacha Crooks}, \bibinfo{person}{Lorenzo Alvisi}, \bibinfo{person}{Nathan Bronson}, {and} \bibinfo{person}{Wyatt Lloyd}.} \bibinfo{year}{2017}\natexlab{}.
\newblock \showarticletitle{I {Can{\textquoteright}t} Believe {It{\textquoteright}s} Not Causal! Scalable Causal Consistency with No Slowdown Cascades}. In \bibinfo{booktitle}{\emph{14th USENIX Symposium on Networked Systems Design and Implementation (NSDI 17)}}. \bibinfo{publisher}{USENIX Association}, \bibinfo{address}{Boston, MA}, \bibinfo{pages}{453--468}.
\newblock
\showISBNx{978-1-931971-37-9}
\urldef\tempurl%
\url{https://www.usenix.org/conference/nsdi17/technical-sessions/presentation/mehdi}
\showURL{%
\tempurl}


\bibitem[Moraru et~al\mbox{.}(2013)]%
        {epaxos}
\bibfield{author}{\bibinfo{person}{Iulian Moraru}, \bibinfo{person}{David~G. Andersen}, {and} \bibinfo{person}{Michael Kaminsky}.} \bibinfo{year}{2013}\natexlab{}.
\newblock \showarticletitle{There is More Consensus in Egalitarian Parliaments}. In \bibinfo{booktitle}{\emph{Proceedings of the Twenty-Fourth ACM Symposium on Operating Systems Principles}} (Farminton, Pennsylvania) \emph{(\bibinfo{series}{SOSP '13})}. \bibinfo{publisher}{Association for Computing Machinery}, \bibinfo{address}{New York, NY, USA}, \bibinfo{pages}{358–372}.
\newblock
\showISBNx{9781450323888}
\href{https://doi.org/10.1145/2517349.2517350}{doi:\nolinkurl{10.1145/2517349.2517350}}


\bibitem[Moraru et~al\mbox{.}(2014)]%
        {quorum-leases}
\bibfield{author}{\bibinfo{person}{Iulian Moraru}, \bibinfo{person}{David~G. Andersen}, {and} \bibinfo{person}{Michael Kaminsky}.} \bibinfo{year}{2014}\natexlab{}.
\newblock \showarticletitle{Paxos Quorum Leases: Fast Reads Without Sacrificing Writes}. In \bibinfo{booktitle}{\emph{Proceedings of the ACM Symposium on Cloud Computing}} (Seattle, WA, USA) \emph{(\bibinfo{series}{SOCC '14})}. \bibinfo{publisher}{Association for Computing Machinery}, \bibinfo{address}{New York, NY, USA}, \bibinfo{pages}{1–13}.
\newblock
\showISBNx{9781450332521}
\href{https://doi.org/10.1145/2670979.2671001}{doi:\nolinkurl{10.1145/2670979.2671001}}


\bibitem[Mu et~al\mbox{.}(2014)]%
        {rs-paxos}
\bibfield{author}{\bibinfo{person}{Shuai Mu}, \bibinfo{person}{Kang Chen}, \bibinfo{person}{Yongwei Wu}, {and} \bibinfo{person}{Weimin Zheng}.} \bibinfo{year}{2014}\natexlab{}.
\newblock \showarticletitle{When Paxos Meets Erasure Code: Reduce Network and Storage Cost in State Machine Replication}. In \bibinfo{booktitle}{\emph{Proceedings of the 23rd International Symposium on High-Performance Parallel and Distributed Computing}} (Vancouver, BC, Canada) \emph{(\bibinfo{series}{HPDC '14})}. \bibinfo{publisher}{Association for Computing Machinery}, \bibinfo{address}{New York, NY, USA}, \bibinfo{pages}{61–72}.
\newblock
\showISBNx{9781450327497}
\href{https://doi.org/10.1145/2600212.2600218}{doi:\nolinkurl{10.1145/2600212.2600218}}


\bibitem[Murat et~al\mbox{.}(2024)]%
        {swarm-disaggregated-memory}
\bibfield{author}{\bibinfo{person}{Antoine Murat}, \bibinfo{person}{Clément Burgelin}, \bibinfo{person}{Athanasios Xygkis}, \bibinfo{person}{Igor Zablotchi}, \bibinfo{person}{Marcos~Kawazoe Aguilera}, {and} \bibinfo{person}{Rachid Guerraoui}.} \bibinfo{year}{2024}\natexlab{}.
\newblock \showarticletitle{SWARM: Replicating Shared Disaggregated-Memory Data in No Time}. In \bibinfo{booktitle}{\emph{Proceedings of the ACM SIGOPS 30th Symposium on Operating Systems Principles}} \emph{(\bibinfo{series}{SOSP ’24})}. \bibinfo{publisher}{ACM}, \bibinfo{pages}{24–45}.
\newblock
\href{https://doi.org/10.1145/3694715.3695945}{doi:\nolinkurl{10.1145/3694715.3695945}}


\bibitem[Ngo et~al\mbox{.}(2020)]%
        {copilots}
\bibfield{author}{\bibinfo{person}{Khiem Ngo}, \bibinfo{person}{Siddhartha Sen}, {and} \bibinfo{person}{Wyatt Lloyd}.} \bibinfo{year}{2020}\natexlab{}.
\newblock \showarticletitle{Tolerating Slowdowns in Replicated State Machines using Copilots}. In \bibinfo{booktitle}{\emph{14th USENIX Symposium on Operating Systems Design and Implementation (OSDI 20)}}. \bibinfo{publisher}{USENIX Association}, \bibinfo{pages}{583--598}.
\newblock
\showISBNx{978-1-939133-19-9}
\urldef\tempurl%
\url{https://www.usenix.org/conference/osdi20/presentation/ngo}
\showURL{%
\tempurl}


\bibitem[Oki and Liskov(1988)]%
        {viewstamped-replication}
\bibfield{author}{\bibinfo{person}{Brian~M. Oki} {and} \bibinfo{person}{Barbara~H. Liskov}.} \bibinfo{year}{1988}\natexlab{}.
\newblock \showarticletitle{Viewstamped Replication: A New Primary Copy Method to Support Highly-Available Distributed Systems}. In \bibinfo{booktitle}{\emph{Proceedings of the Seventh Annual ACM Symposium on Principles of Distributed Computing}} (Toronto, Ontario, Canada) \emph{(\bibinfo{series}{PODC '88})}. \bibinfo{publisher}{Association for Computing Machinery}, \bibinfo{address}{New York, NY, USA}, \bibinfo{pages}{8–17}.
\newblock
\showISBNx{0897912772}
\href{https://doi.org/10.1145/62546.62549}{doi:\nolinkurl{10.1145/62546.62549}}


\bibitem[Ongaro and Ousterhout(2014)]%
        {raft}
\bibfield{author}{\bibinfo{person}{Diego Ongaro} {and} \bibinfo{person}{John Ousterhout}.} \bibinfo{year}{2014}\natexlab{}.
\newblock \showarticletitle{In Search of an Understandable Consensus Algorithm}. In \bibinfo{booktitle}{\emph{Proceedings of the 2014 USENIX Conference on USENIX Annual Technical Conference}} (Philadelphia, PA) \emph{(\bibinfo{series}{USENIX ATC'14})}. \bibinfo{publisher}{USENIX Association}, \bibinfo{address}{USA}, \bibinfo{pages}{305–320}.
\newblock
\showISBNx{9781931971102}


\bibitem[Pan et~al\mbox{.}(2021)]%
        {rabia}
\bibfield{author}{\bibinfo{person}{Haochen Pan}, \bibinfo{person}{Jesse Tuglu}, \bibinfo{person}{Neo Zhou}, \bibinfo{person}{Tianshu Wang}, \bibinfo{person}{Yicheng Shen}, \bibinfo{person}{Xiong Zheng}, \bibinfo{person}{Joseph Tassarotti}, \bibinfo{person}{Lewis Tseng}, {and} \bibinfo{person}{Roberto Palmieri}.} \bibinfo{year}{2021}\natexlab{}.
\newblock \showarticletitle{Rabia: Simplifying State-Machine Replication Through Randomization}. In \bibinfo{booktitle}{\emph{Proceedings of the ACM SIGOPS 28th Symposium on Operating Systems Principles}} (Virtual Event, Germany) \emph{(\bibinfo{series}{SOSP '21})}. \bibinfo{publisher}{Association for Computing Machinery}, \bibinfo{address}{New York, NY, USA}, \bibinfo{pages}{472–487}.
\newblock
\showISBNx{9781450387095}
\href{https://doi.org/10.1145/3477132.3483582}{doi:\nolinkurl{10.1145/3477132.3483582}}


\bibitem[Papailiopoulos and Dimakis(2014)]%
        {lrc-codes}
\bibfield{author}{\bibinfo{person}{Dimitris~S. Papailiopoulos} {and} \bibinfo{person}{Alexandros~G. Dimakis}.} \bibinfo{year}{2014}\natexlab{}.
\newblock \bibinfo{title}{Locally Repairable Codes}.
\newblock
\showeprint[arxiv]{1206.3804}~[cs.IT]
\urldef\tempurl%
\url{https://arxiv.org/abs/1206.3804}
\showURL{%
\tempurl}


\bibitem[Patterson et~al\mbox{.}(1988)]%
        {raid}
\bibfield{author}{\bibinfo{person}{David~A. Patterson}, \bibinfo{person}{Garth Gibson}, {and} \bibinfo{person}{Randy~H. Katz}.} \bibinfo{year}{1988}\natexlab{}.
\newblock \showarticletitle{A Case for Redundant Arrays of Inexpensive Disks (RAID)}. In \bibinfo{booktitle}{\emph{Proceedings of the 1988 ACM SIGMOD International Conference on Management of Data}} (Chicago, Illinois, USA) \emph{(\bibinfo{series}{SIGMOD '88})}. \bibinfo{publisher}{Association for Computing Machinery}, \bibinfo{address}{New York, NY, USA}, \bibinfo{pages}{109–116}.
\newblock
\showISBNx{0897912683}
\href{https://doi.org/10.1145/50202.50214}{doi:\nolinkurl{10.1145/50202.50214}}


\bibitem[Ports et~al\mbox{.}(2015)]%
        {speculative-paxos}
\bibfield{author}{\bibinfo{person}{Dan R.~K. Ports}, \bibinfo{person}{Jialin Li}, \bibinfo{person}{Vincent Liu}, \bibinfo{person}{Naveen~Kr. Sharma}, {and} \bibinfo{person}{Arvind Krishnamurthy}.} \bibinfo{year}{2015}\natexlab{}.
\newblock \showarticletitle{Designing Distributed Systems Using Approximate Synchrony in Data Center Networks}. In \bibinfo{booktitle}{\emph{12th USENIX Symposium on Networked Systems Design and Implementation (NSDI 15)}}. \bibinfo{publisher}{USENIX Association}, \bibinfo{address}{Oakland, CA}, \bibinfo{pages}{43--57}.
\newblock
\showISBNx{978-1-931971-218}
\urldef\tempurl%
\url{https://www.usenix.org/conference/nsdi15/technical-sessions/presentation/ports}
\showURL{%
\tempurl}


\bibitem[Qi et~al\mbox{.}(2021)]%
        {bidl-blockchain}
\bibfield{author}{\bibinfo{person}{Ji Qi}, \bibinfo{person}{Xusheng Chen}, \bibinfo{person}{Yunpeng Jiang}, \bibinfo{person}{Jianyu Jiang}, \bibinfo{person}{Tianxiang Shen}, \bibinfo{person}{Shixiong Zhao}, \bibinfo{person}{Sen Wang}, \bibinfo{person}{Gong Zhang}, \bibinfo{person}{Li Chen}, \bibinfo{person}{Man~Ho Au}, {and} \bibinfo{person}{Heming Cui}.} \bibinfo{year}{2021}\natexlab{}.
\newblock \showarticletitle{Bidl: A High-throughput, Low-latency Permissioned Blockchain Framework for Datacenter Networks}. In \bibinfo{booktitle}{\emph{Proceedings of the ACM SIGOPS 28th Symposium on Operating Systems Principles}} (Virtual Event, Germany) \emph{(\bibinfo{series}{SOSP '21})}. \bibinfo{publisher}{Association for Computing Machinery}, \bibinfo{address}{New York, NY, USA}, \bibinfo{pages}{18–34}.
\newblock
\showISBNx{9781450387095}
\href{https://doi.org/10.1145/3477132.3483574}{doi:\nolinkurl{10.1145/3477132.3483574}}


\bibitem[Redpanda(2024)]%
        {redpanda}
\bibfield{author}{\bibinfo{person}{Redpanda}.} \bibinfo{year}{2024}\natexlab{}.
\newblock \bibinfo{title}{Redpanda: The Unified Streaming Data Platform}.
\newblock
\newblock
\shownote{\url{https://www.redpanda.com/}, Last accessed on 2024-09-05}.


\bibitem[Reed and Solomon(1960)]%
        {rs-coding}
\bibfield{author}{\bibinfo{person}{I.~S. Reed} {and} \bibinfo{person}{G. Solomon}.} \bibinfo{year}{1960}\natexlab{}.
\newblock \showarticletitle{Polynomial Codes Over Certain Finite Fields}.
\newblock \bibinfo{journal}{\emph{J. Soc. Indust. Appl. Math.}} \bibinfo{volume}{8}, \bibinfo{number}{2} (\bibinfo{year}{1960}), \bibinfo{pages}{300--304}.
\newblock
\href{https://doi.org/10.1137/0108018}{doi:\nolinkurl{10.1137/0108018}}
\showeprint{https://doi.org/10.1137/0108018}


\bibitem[Ren et~al\mbox{.}(2022a)]%
        {crs-raft}
\bibfield{author}{\bibinfo{person}{Donglin Ren}, \bibinfo{person}{Jun Tu}, {and} \bibinfo{person}{Wei Xie}.} \bibinfo{year}{2022}\natexlab{a}.
\newblock \showarticletitle{An Improved Raft Protocol Combined with Cauchy Reed-Solomon Codes}. In \bibinfo{booktitle}{\emph{2022 5th International Conference on Artificial Intelligence and Big Data (ICAIBD)}}. \bibinfo{pages}{563--568}.
\newblock
\href{https://doi.org/10.1109/ICAIBD55127.2022.9820425}{doi:\nolinkurl{10.1109/ICAIBD55127.2022.9820425}}


\bibitem[Ren et~al\mbox{.}(2022b)]%
        {adraft}
\bibfield{author}{\bibinfo{person}{Donglin Ren}, \bibinfo{person}{Jun Tu}, \bibinfo{person}{Wei Xie}, {and} \bibinfo{person}{Changyin Wu}.} \bibinfo{year}{2022}\natexlab{b}.
\newblock \showarticletitle{An Optimized Raft Protocol Combined with Redundant Residue Number System}. In \bibinfo{booktitle}{\emph{2022 5th International Conference on Data Science and Information Technology (DSIT)}}. \bibinfo{pages}{1--6}.
\newblock
\href{https://doi.org/10.1109/DSIT55514.2022.9943823}{doi:\nolinkurl{10.1109/DSIT55514.2022.9943823}}


\bibitem[Renesse and Schneider(2004)]%
        {chain-replication}
\bibfield{author}{\bibinfo{person}{Robbert~Van Renesse} {and} \bibinfo{person}{Fred~B. Schneider}.} \bibinfo{year}{2004}\natexlab{}.
\newblock \showarticletitle{Chain Replication for Supporting High Throughput and Availability}. In \bibinfo{booktitle}{\emph{6th Symposium on Operating Systems Design \& Implementation (OSDI 04)}}. \bibinfo{publisher}{USENIX Association}, \bibinfo{address}{San Francisco, CA}.
\newblock
\urldef\tempurl%
\url{https://www.usenix.org/conference/osdi-04/chain-replication-supporting-high-throughput-and-availability}
\showURL{%
\tempurl}


\bibitem[Rensin(2015)]%
        {kubernetes}
\bibfield{author}{\bibinfo{person}{David~K. Rensin}.} \bibinfo{year}{2015}\natexlab{}.
\newblock \bibinfo{booktitle}{\emph{Kubernetes - Scheduling the Future at Cloud Scale}}.
\newblock \bibinfo{publisher}{O'Reilly and Associates}, \bibinfo{address}{1005 Gravenstein Highway North Sebastopol, CA 95472}. All pages.
\newblock
\urldef\tempurl%
\url{http://www.oreilly.com/webops-perf/free/kubernetes.csp}
\showURL{%
\tempurl}


\bibitem[rqlite(2024)]%
        {rqlite}
\bibfield{author}{\bibinfo{person}{rqlite}.} \bibinfo{year}{2024}\natexlab{}.
\newblock \bibinfo{title}{rqlite is a distributed relational database that combines the simplicity of SQLite with the robustness of a fault-tolerant, highly available system}.
\newblock
\newblock
\shownote{\url{https://rqlite.io/}, Last accessed on 2024-11-13}.


\bibitem[Ryabinin et~al\mbox{.}(2024)]%
        {swiftpaxos}
\bibfield{author}{\bibinfo{person}{Fedor Ryabinin}, \bibinfo{person}{Alexey Gotsman}, {and} \bibinfo{person}{Pierre Sutra}.} \bibinfo{year}{2024}\natexlab{}.
\newblock \showarticletitle{{SwiftPaxos}: Fast {Geo-Replicated} State Machines}. In \bibinfo{booktitle}{\emph{21st USENIX Symposium on Networked Systems Design and Implementation (NSDI 24)}}. \bibinfo{publisher}{USENIX Association}, \bibinfo{address}{Santa Clara, CA}, \bibinfo{pages}{345--369}.
\newblock
\showISBNx{978-1-939133-39-7}
\urldef\tempurl%
\url{https://www.usenix.org/conference/nsdi24/presentation/ryabinin}
\showURL{%
\tempurl}


\bibitem[Schneider(1990)]%
        {smr-approach}
\bibfield{author}{\bibinfo{person}{Fred~B. Schneider}.} \bibinfo{year}{1990}\natexlab{}.
\newblock \showarticletitle{Implementing Fault-Tolerant Services Using the State Machine Approach: A Tutorial}.
\newblock \bibinfo{journal}{\emph{ACM Comput. Surv.}} \bibinfo{volume}{22}, \bibinfo{number}{4} (\bibinfo{date}{dec} \bibinfo{year}{1990}), \bibinfo{pages}{299–319}.
\newblock
\showISSN{0360-0300}
\href{https://doi.org/10.1145/98163.98167}{doi:\nolinkurl{10.1145/98163.98167}}


\bibitem[ScyllaDB(2023)]%
        {scylladb}
\bibfield{author}{\bibinfo{person}{ScyllaDB}.} \bibinfo{year}{2023}\natexlab{}.
\newblock \bibinfo{title}{Beyond Legacy NoSQL: 7 Design Principles Behind ScyllaDB}.
\newblock
\newblock
\shownote{\url{https://lp.scylladb.com/real-time-big-data-database-principles-thanks.html}, Last accessed on 2023-11-13}.


\bibitem[Shan et~al\mbox{.}(2021)]%
        {geometric-partitioning}
\bibfield{author}{\bibinfo{person}{Yingdi Shan}, \bibinfo{person}{Kang Chen}, \bibinfo{person}{Tuoyu Gong}, \bibinfo{person}{Lidong Zhou}, \bibinfo{person}{Tai Zhou}, {and} \bibinfo{person}{Yongwei Wu}.} \bibinfo{year}{2021}\natexlab{}.
\newblock \showarticletitle{Geometric Partitioning: Explore the Boundary of Optimal Erasure Code Repair}. In \bibinfo{booktitle}{\emph{Proceedings of the ACM SIGOPS 28th Symposium on Operating Systems Principles}} (Virtual Event, Germany) \emph{(\bibinfo{series}{SOSP '21})}. \bibinfo{publisher}{Association for Computing Machinery}, \bibinfo{address}{New York, NY, USA}, \bibinfo{pages}{457–471}.
\newblock
\showISBNx{9781450387095}
\href{https://doi.org/10.1145/3477132.3483558}{doi:\nolinkurl{10.1145/3477132.3483558}}


\bibitem[Shin et~al\mbox{.}(2016)]%
        {stalestores}
\bibfield{author}{\bibinfo{person}{Ji-Yong Shin}, \bibinfo{person}{Mahesh Balakrishnan}, \bibinfo{person}{Tudor Marian}, \bibinfo{person}{Jakub Szefer}, {and} \bibinfo{person}{Hakim Weatherspoon}.} \bibinfo{year}{2016}\natexlab{}.
\newblock \showarticletitle{Towards Weakly Consistent Local Storage Systems}. In \bibinfo{booktitle}{\emph{Proceedings of the Seventh ACM Symposium on Cloud Computing}} (Santa Clara, CA, USA) \emph{(\bibinfo{series}{SoCC '16})}. \bibinfo{publisher}{Association for Computing Machinery}, \bibinfo{address}{New York, NY, USA}, \bibinfo{pages}{294–306}.
\newblock
\showISBNx{9781450345255}
\href{https://doi.org/10.1145/2987550.2987579}{doi:\nolinkurl{10.1145/2987550.2987579}}


\bibitem[Shute et~al\mbox{.}(2013)]%
        {f1}
\bibfield{author}{\bibinfo{person}{Jeff Shute}, \bibinfo{person}{Radek Vingralek}, \bibinfo{person}{Bart Samwel}, \bibinfo{person}{Ben Handy}, \bibinfo{person}{Chad Whipkey}, \bibinfo{person}{Eric Rollins}, \bibinfo{person}{Mircea Oancea}, \bibinfo{person}{Kyle Littleﬁeld}, \bibinfo{person}{David Menestrina}, \bibinfo{person}{Stephan Ellner}, \bibinfo{person}{John Cieslewicz}, \bibinfo{person}{Ian Rae}, \bibinfo{person}{Traian Stancescu}, {and} \bibinfo{person}{Himani Apte}.} \bibinfo{year}{2013}\natexlab{}.
\newblock \showarticletitle{F1: A Distributed SQL Database That Scales}. In \bibinfo{booktitle}{\emph{VLDB}}.
\newblock


\bibitem[Stathakopoulou et~al\mbox{.}(2022)]%
        {insanely-scalable-smr}
\bibfield{author}{\bibinfo{person}{Chrysoula Stathakopoulou}, \bibinfo{person}{Matej Pavlovic}, {and} \bibinfo{person}{Marko Vukoli\'{c}}.} \bibinfo{year}{2022}\natexlab{}.
\newblock \showarticletitle{State Machine Replication Scalability Made Simple}. In \bibinfo{booktitle}{\emph{Proceedings of the Seventeenth European Conference on Computer Systems}} (Rennes, France) \emph{(\bibinfo{series}{EuroSys '22})}. \bibinfo{publisher}{Association for Computing Machinery}, \bibinfo{address}{New York, NY, USA}, \bibinfo{pages}{17–33}.
\newblock
\showISBNx{9781450391627}
\href{https://doi.org/10.1145/3492321.3519579}{doi:\nolinkurl{10.1145/3492321.3519579}}


\bibitem[Sun et~al\mbox{.}(2024)]%
        {anvil}
\bibfield{author}{\bibinfo{person}{Xudong Sun}, \bibinfo{person}{Wenjie Ma}, \bibinfo{person}{Jiawei~Tyler Gu}, \bibinfo{person}{Zicheng Ma}, \bibinfo{person}{Tej Chajed}, \bibinfo{person}{Jon Howell}, \bibinfo{person}{Andrea Lattuada}, \bibinfo{person}{Oded Padon}, \bibinfo{person}{Lalith Suresh}, \bibinfo{person}{Adriana Szekeres}, {and} \bibinfo{person}{Tianyin Xu}.} \bibinfo{year}{2024}\natexlab{}.
\newblock \showarticletitle{Anvil: Verifying Liveness of Cluster Management Controllers}. In \bibinfo{booktitle}{\emph{18th USENIX Symposium on Operating Systems Design and Implementation (OSDI 24)}}. \bibinfo{publisher}{USENIX Association}, \bibinfo{address}{Santa Clara, CA}, \bibinfo{pages}{649--666}.
\newblock
\showISBNx{978-1-939133-40-3}
\urldef\tempurl%
\url{https://www.usenix.org/conference/osdi24/presentation/sun-xudong}
\showURL{%
\tempurl}


\bibitem[Suri-Payer et~al\mbox{.}(2021)]%
        {basil}
\bibfield{author}{\bibinfo{person}{Florian Suri-Payer}, \bibinfo{person}{Matthew Burke}, \bibinfo{person}{Zheng Wang}, \bibinfo{person}{Yunhao Zhang}, \bibinfo{person}{Lorenzo Alvisi}, {and} \bibinfo{person}{Natacha Crooks}.} \bibinfo{year}{2021}\natexlab{}.
\newblock \showarticletitle{Basil: Breaking up BFT with ACID (Transactions)}. In \bibinfo{booktitle}{\emph{Proceedings of the ACM SIGOPS 28th Symposium on Operating Systems Principles}} (Virtual Event, Germany) \emph{(\bibinfo{series}{SOSP '21})}. \bibinfo{publisher}{Association for Computing Machinery}, \bibinfo{address}{New York, NY, USA}, \bibinfo{pages}{1–17}.
\newblock
\showISBNx{9781450387095}
\href{https://doi.org/10.1145/3477132.3483552}{doi:\nolinkurl{10.1145/3477132.3483552}}


\bibitem[Terrace and Freedman(2009)]%
        {craq}
\bibfield{author}{\bibinfo{person}{Jeff Terrace} {and} \bibinfo{person}{Michael~J. Freedman}.} \bibinfo{year}{2009}\natexlab{}.
\newblock \showarticletitle{Object Storage on {CRAQ}: {High-Throughput} Chain Replication for {Read-Mostly} Workloads}. In \bibinfo{booktitle}{\emph{2009 USENIX Annual Technical Conference (USENIX ATC 09)}}. \bibinfo{publisher}{USENIX Association}, \bibinfo{address}{San Diego, CA}.
\newblock
\urldef\tempurl%
\url{https://www.usenix.org/conference/usenix-09/object-storage-craq-high-throughput-chain-replication-read-mostly-workloads}
\showURL{%
\tempurl}


\bibitem[Terry et~al\mbox{.}(1994)]%
        {session-guarantees}
\bibfield{author}{\bibinfo{person}{Douglas~B. Terry}, \bibinfo{person}{Alan~J. Demers}, \bibinfo{person}{Karin Petersen}, \bibinfo{person}{Mike~J. Spreitzer}, \bibinfo{person}{Marvin~M. Theimer}, {and} \bibinfo{person}{Brent~B. Welch}.} \bibinfo{year}{1994}\natexlab{}.
\newblock \showarticletitle{Session Guarantees for Weakly Consistent Replicated Data}. In \bibinfo{booktitle}{\emph{Proceedings of the Third International Conference on on Parallel and Distributed Information Systems}} (Autin, Texas, USA) \emph{(\bibinfo{series}{PDIS '94})}. \bibinfo{publisher}{IEEE Computer Society Press}, \bibinfo{address}{Washington, DC, USA}, \bibinfo{pages}{140–150}.
\newblock
\showISBNx{0818664010}


\bibitem[Terry et~al\mbox{.}(2013)]%
        {consistency-based-sla}
\bibfield{author}{\bibinfo{person}{Douglas~B. Terry}, \bibinfo{person}{Vijayan Prabhakaran}, \bibinfo{person}{Ramakrishna Kotla}, \bibinfo{person}{Mahesh Balakrishnan}, \bibinfo{person}{Marcos~K. Aguilera}, {and} \bibinfo{person}{Hussam Abu-Libdeh}.} \bibinfo{year}{2013}\natexlab{}.
\newblock \showarticletitle{Consistency-based service level agreements for cloud storage}. In \bibinfo{booktitle}{\emph{Proceedings of the Twenty-Fourth ACM Symposium on Operating Systems Principles}} (Farminton, Pennsylvania) \emph{(\bibinfo{series}{SOSP '13})}. \bibinfo{publisher}{Association for Computing Machinery}, \bibinfo{address}{New York, NY, USA}, \bibinfo{pages}{309–324}.
\newblock
\showISBNx{9781450323888}
\href{https://doi.org/10.1145/2517349.2522731}{doi:\nolinkurl{10.1145/2517349.2522731}}


\bibitem[Tollman et~al\mbox{.}(2021)]%
        {epaxos-revisited}
\bibfield{author}{\bibinfo{person}{Sarah Tollman}, \bibinfo{person}{Seo~Jin Park}, {and} \bibinfo{person}{John Ousterhout}.} \bibinfo{year}{2021}\natexlab{}.
\newblock \showarticletitle{{EPaxos} Revisited}. In \bibinfo{booktitle}{\emph{18th USENIX Symposium on Networked Systems Design and Implementation (NSDI 21)}}. \bibinfo{publisher}{USENIX Association}, \bibinfo{pages}{613--632}.
\newblock
\showISBNx{978-1-939133-21-2}
\urldef\tempurl%
\url{https://www.usenix.org/conference/nsdi21/presentation/tollman}
\showURL{%
\tempurl}


\bibitem[Uluyol et~al\mbox{.}(2020)]%
        {tradeoffs-geo-distributed}
\bibfield{author}{\bibinfo{person}{Muhammed Uluyol}, \bibinfo{person}{Anthony Huang}, \bibinfo{person}{Ayush Goel}, \bibinfo{person}{Mosharaf Chowdhury}, {and} \bibinfo{person}{Harsha~V. Madhyastha}.} \bibinfo{year}{2020}\natexlab{}.
\newblock \showarticletitle{{Near-Optimal} Latency Versus Cost Tradeoffs in {Geo-Distributed} Storage}. In \bibinfo{booktitle}{\emph{17th USENIX Symposium on Networked Systems Design and Implementation (NSDI 20)}}. \bibinfo{publisher}{USENIX Association}, \bibinfo{address}{Santa Clara, CA}, \bibinfo{pages}{157--180}.
\newblock
\showISBNx{978-1-939133-13-7}
\urldef\tempurl%
\url{https://www.usenix.org/conference/nsdi20/presentation/uluyol}
\showURL{%
\tempurl}


\bibitem[VanBenschoten et~al\mbox{.}(2022)]%
        {cockroachdb}
\bibfield{author}{\bibinfo{person}{Nathan VanBenschoten}, \bibinfo{person}{Arul Ajmani}, \bibinfo{person}{Marcus Gartner}, \bibinfo{person}{Andrei Matei}, \bibinfo{person}{Aayush Shah}, \bibinfo{person}{Irfan Sharif}, \bibinfo{person}{Alexander Shraer}, \bibinfo{person}{Adam Storm}, \bibinfo{person}{Rebecca Taft}, \bibinfo{person}{Oliver Tan}, \bibinfo{person}{Andy Woods}, {and} \bibinfo{person}{Peyton Walters}.} \bibinfo{year}{2022}\natexlab{}.
\newblock \showarticletitle{Enabling the Next Generation of Multi-Region Applications with CockroachDB}. In \bibinfo{booktitle}{\emph{Proceedings of the 2022 International Conference on Management of Data}} (Philadelphia, PA, USA) \emph{(\bibinfo{series}{SIGMOD '22})}. \bibinfo{publisher}{Association for Computing Machinery}, \bibinfo{address}{New York, NY, USA}, \bibinfo{pages}{2312–2325}.
\newblock
\showISBNx{9781450392495}
\href{https://doi.org/10.1145/3514221.3526053}{doi:\nolinkurl{10.1145/3514221.3526053}}


\bibitem[Vogels(2008)]%
        {eventual-consistency}
\bibfield{author}{\bibinfo{person}{Werner Vogels}.} \bibinfo{year}{2008}\natexlab{}.
\newblock \showarticletitle{Eventually Consistent: Building Reliable Distributed Systems at a Worldwide Scale Demands Trade-Offs?Between Consistency and Availability.}
\newblock \bibinfo{journal}{\emph{Queue}} \bibinfo{volume}{6}, \bibinfo{number}{6} (\bibinfo{date}{oct} \bibinfo{year}{2008}), \bibinfo{pages}{14–19}.
\newblock
\showISSN{1542-7730}
\href{https://doi.org/10.1145/1466443.1466448}{doi:\nolinkurl{10.1145/1466443.1466448}}


\bibitem[Wang et~al\mbox{.}(2017)]%
        {apus-rdma-paxos}
\bibfield{author}{\bibinfo{person}{Cheng Wang}, \bibinfo{person}{Jianyu Jiang}, \bibinfo{person}{Xusheng Chen}, \bibinfo{person}{Ning Yi}, {and} \bibinfo{person}{Heming Cui}.} \bibinfo{year}{2017}\natexlab{}.
\newblock \showarticletitle{APUS: fast and scalable paxos on RDMA}. In \bibinfo{booktitle}{\emph{Proceedings of the 2017 Symposium on Cloud Computing}} (Santa Clara, California) \emph{(\bibinfo{series}{SoCC '17})}. \bibinfo{publisher}{Association for Computing Machinery}, \bibinfo{address}{New York, NY, USA}, \bibinfo{pages}{94–107}.
\newblock
\showISBNx{9781450350280}
\href{https://doi.org/10.1145/3127479.3128609}{doi:\nolinkurl{10.1145/3127479.3128609}}


\bibitem[Wang et~al\mbox{.}(2012)]%
        {gnothi}
\bibfield{author}{\bibinfo{person}{Yang Wang}, \bibinfo{person}{Lorenzo Alvisi}, {and} \bibinfo{person}{Mike Dahlin}.} \bibinfo{year}{2012}\natexlab{}.
\newblock \showarticletitle{Gnothi: Separating Data and Metadata for Efficient and Available Storage Replication}. In \bibinfo{booktitle}{\emph{2012 USENIX Annual Technical Conference (USENIX ATC 12)}}. \bibinfo{publisher}{USENIX Association}, \bibinfo{address}{Boston, MA}, \bibinfo{pages}{413--424}.
\newblock
\showISBNx{978-931971-93-5}
\urldef\tempurl%
\url{https://www.usenix.org/conference/atc12/technical-sessions/presentation/wang}
\showURL{%
\tempurl}


\bibitem[Wang et~al\mbox{.}(2020)]%
        {craft}
\bibfield{author}{\bibinfo{person}{Zizhong Wang}, \bibinfo{person}{Tongliang Li}, \bibinfo{person}{Haixia Wang}, \bibinfo{person}{Airan Shao}, \bibinfo{person}{Yunren Bai}, \bibinfo{person}{Shangming Cai}, \bibinfo{person}{Zihan Xu}, {and} \bibinfo{person}{Dongsheng Wang}.} \bibinfo{year}{2020}\natexlab{}.
\newblock \showarticletitle{{CRaft}: An Erasure-coding-supported Version of Raft for Reducing Storage Cost and Network Cost}. In \bibinfo{booktitle}{\emph{18th USENIX Conference on File and Storage Technologies (FAST 20)}}. \bibinfo{publisher}{USENIX Association}, \bibinfo{address}{Santa Clara, CA}, \bibinfo{pages}{297--308}.
\newblock
\showISBNx{978-1-939133-12-0}
\urldef\tempurl%
\url{https://www.usenix.org/conference/fast20/presentation/wang-zizhong}
\showURL{%
\tempurl}


\bibitem[Wang et~al\mbox{.}(2019)]%
        {parallel-paxos-raft}
\bibfield{author}{\bibinfo{person}{Zhaoguo Wang}, \bibinfo{person}{Changgeng Zhao}, \bibinfo{person}{Shuai Mu}, \bibinfo{person}{Haibo Chen}, {and} \bibinfo{person}{Jinyang Li}.} \bibinfo{year}{2019}\natexlab{}.
\newblock \showarticletitle{On the Parallels between Paxos and Raft, and how to Port Optimizations}. In \bibinfo{booktitle}{\emph{Proceedings of the 2019 ACM Symposium on Principles of Distributed Computing}} (Toronto ON, Canada) \emph{(\bibinfo{series}{PODC '19})}. \bibinfo{publisher}{Association for Computing Machinery}, \bibinfo{address}{New York, NY, USA}, \bibinfo{pages}{445–454}.
\newblock
\showISBNx{9781450362177}
\href{https://doi.org/10.1145/3293611.3331595}{doi:\nolinkurl{10.1145/3293611.3331595}}


\bibitem[Wei et~al\mbox{.}(2023)]%
        {off-path-smartnic}
\bibfield{author}{\bibinfo{person}{Xingda Wei}, \bibinfo{person}{Rongxin Cheng}, \bibinfo{person}{Yuhan Yang}, \bibinfo{person}{Rong Chen}, {and} \bibinfo{person}{Haibo Chen}.} \bibinfo{year}{2023}\natexlab{}.
\newblock \showarticletitle{Characterizing Off-path {SmartNIC} for Accelerating Distributed Systems}. In \bibinfo{booktitle}{\emph{17th USENIX Symposium on Operating Systems Design and Implementation (OSDI 23)}}. \bibinfo{publisher}{USENIX Association}, \bibinfo{address}{Boston, MA}, \bibinfo{pages}{987--1004}.
\newblock
\showISBNx{978-1-939133-34-2}
\urldef\tempurl%
\url{https://www.usenix.org/conference/osdi23/presentation/wei-smartnic}
\showURL{%
\tempurl}


\bibitem[Whittaker et~al\mbox{.}(2021)]%
        {compartmentalization}
\bibfield{author}{\bibinfo{person}{Michael Whittaker}, \bibinfo{person}{Ailidani Ailijiang}, \bibinfo{person}{Aleksey Charapko}, \bibinfo{person}{Murat Demirbas}, \bibinfo{person}{Neil Giridharan}, \bibinfo{person}{Joseph~M. Hellerstein}, \bibinfo{person}{Heidi Howard}, \bibinfo{person}{Ion Stoica}, {and} \bibinfo{person}{Adriana Szekeres}.} \bibinfo{year}{2021}\natexlab{}.
\newblock \showarticletitle{Scaling Replicated State Machines with Compartmentalization}.
\newblock \bibinfo{journal}{\emph{Proc. VLDB Endow.}} \bibinfo{volume}{14}, \bibinfo{number}{11} (\bibinfo{date}{jul} \bibinfo{year}{2021}), \bibinfo{pages}{2203–2215}.
\newblock
\showISSN{2150-8097}
\href{https://doi.org/10.14778/3476249.3476273}{doi:\nolinkurl{10.14778/3476249.3476273}}


\bibitem[{Wikipedia contributors}(2023)]%
        {ordinary-least-squares}
\bibfield{author}{\bibinfo{person}{{Wikipedia contributors}}.} \bibinfo{year}{2023}\natexlab{}.
\newblock \bibinfo{title}{Ordinary least squares --- {Wikipedia}{,} The Free Encyclopedia}.
\newblock \bibinfo{howpublished}{\url{https://en.wikipedia.org/w/index.php?title=Ordinary_least_squares&oldid=1184283716}}.
\newblock
\newblock
\shownote{[Online; accessed 28-November-2023]}.


\bibitem[WintelGuy(2020)]%
        {wan-latency-estimator}
\bibfield{author}{\bibinfo{person}{WintelGuy}.} \bibinfo{year}{2020}\natexlab{}.
\newblock \bibinfo{title}{WAN Latency Estimator}.
\newblock
\newblock
\shownote{\url{https://wintelguy.com/wanlat.html}, Last accessed on 2024-04-29}.


\bibitem[Xu et~al\mbox{.}(2014)]%
        {springfs}
\bibfield{author}{\bibinfo{person}{Lianghong Xu}, \bibinfo{person}{James Cipar}, \bibinfo{person}{Elie Krevat}, \bibinfo{person}{Alexey Tumanov}, \bibinfo{person}{Nitin Gupta}, \bibinfo{person}{Michael~A. Kozuch}, {and} \bibinfo{person}{Gregory~R. Ganger}.} \bibinfo{year}{2014}\natexlab{}.
\newblock \showarticletitle{{SpringFS}: Bridging Agility and Performance in Elastic Distributed Storage}. In \bibinfo{booktitle}{\emph{12th USENIX Conference on File and Storage Technologies (FAST 14)}}. \bibinfo{publisher}{USENIX Association}, \bibinfo{address}{Santa Clara, CA}, \bibinfo{pages}{243--255}.
\newblock
\showISBNx{ISBN 978-1-931971-08-9}
\urldef\tempurl%
\url{https://www.usenix.org/conference/fast14/technical-sessions/presentation/xu}
\showURL{%
\tempurl}


\bibitem[Xu et~al\mbox{.}(2021)]%
        {ecraft}
\bibfield{author}{\bibinfo{person}{MingWei Xu}, \bibinfo{person}{Yu Zhou}, \bibinfo{person}{Yuan~Yuan Qiao}, \bibinfo{person}{Kai Xu}, \bibinfo{person}{Yu Wang}, {and} \bibinfo{person}{Jie Yang}.} \bibinfo{year}{2021}\natexlab{}.
\newblock \showarticletitle{ECRaft: A Raft Based Consensus Protocol for Highly Available and Reliable Erasure-Coded Storage Systems}. In \bibinfo{booktitle}{\emph{2021 IEEE 27th International Conference on Parallel and Distributed Systems (ICPADS)}}. \bibinfo{pages}{707--714}.
\newblock
\href{https://doi.org/10.1109/ICPADS53394.2021.00094}{doi:\nolinkurl{10.1109/ICPADS53394.2021.00094}}


\bibitem[Xu et~al\mbox{.}(2022)]%
        {blackwater-raft}
\bibfield{author}{\bibinfo{person}{Zichen Xu}, \bibinfo{person}{Yunxiao Du}, \bibinfo{person}{Kanqi Zhang}, \bibinfo{person}{Jiacheng Huang}, \bibinfo{person}{Jie Liu}, \bibinfo{person}{Jingxiong Gao}, {and} \bibinfo{person}{Christopher Stewart}.} \bibinfo{year}{2022}\natexlab{}.
\newblock \bibinfo{title}{Cost-effective BlackWater Raft on Highly Unreliable Nodes at Scale Out}.
\newblock
\showeprint[arxiv]{2203.07920}~[cs.DC]


\bibitem[Yin et~al\mbox{.}(2019)]%
        {hotstuff}
\bibfield{author}{\bibinfo{person}{Maofan Yin}, \bibinfo{person}{Dahlia Malkhi}, \bibinfo{person}{Michael~K. Reiter}, \bibinfo{person}{Guy~Golan Gueta}, {and} \bibinfo{person}{Ittai Abraham}.} \bibinfo{year}{2019}\natexlab{}.
\newblock \showarticletitle{HotStuff: BFT Consensus with Linearity and Responsiveness}. In \bibinfo{booktitle}{\emph{Proceedings of the 2019 ACM Symposium on Principles of Distributed Computing}} (Toronto ON, Canada) \emph{(\bibinfo{series}{PODC '19})}. \bibinfo{publisher}{Association for Computing Machinery}, \bibinfo{address}{New York, NY, USA}, \bibinfo{pages}{347–356}.
\newblock
\showISBNx{9781450362177}
\href{https://doi.org/10.1145/3293611.3331591}{doi:\nolinkurl{10.1145/3293611.3331591}}


\bibitem[Yu(2000)]%
        {tact-continuous}
\bibfield{author}{\bibinfo{person}{Haifeng Yu}.} \bibinfo{year}{2000}\natexlab{}.
\newblock \showarticletitle{Design and Evaluation of a Continuous Consistency Model for Replicated Services}. In \bibinfo{booktitle}{\emph{Fourth Symposium on Operating Systems Design and Implementation (OSDI 2000)}}. \bibinfo{publisher}{USENIX Association}, \bibinfo{address}{San Diego, CA}.
\newblock
\urldef\tempurl%
\url{https://www.usenix.org/conference/osdi-2000/design-and-evaluation-continuous-consistency-model-replicated-services}
\showURL{%
\tempurl}


\bibitem[Zarnstorff et~al\mbox{.}(2024)]%
        {racos}
\bibfield{author}{\bibinfo{person}{Jonathan Zarnstorff}, \bibinfo{person}{Lucas Lebow}, \bibinfo{person}{Christopher Siems}, \bibinfo{person}{Dillon Remuck}, \bibinfo{person}{Colin Ruiz}, {and} \bibinfo{person}{Lewis Tseng}.} \bibinfo{year}{2024}\natexlab{}.
\newblock \showarticletitle{Racos: Improving Erasure Coding State Machine Replication using Leaderless Consensus}. In \bibinfo{booktitle}{\emph{Proceedings of the 2024 ACM Symposium on Cloud Computing}} (Redmond, WA, USA) \emph{(\bibinfo{series}{SoCC '24})}. \bibinfo{publisher}{Association for Computing Machinery}, \bibinfo{address}{New York, NY, USA}, \bibinfo{pages}{600–617}.
\newblock
\showISBNx{9798400712869}
\href{https://doi.org/10.1145/3698038.3698511}{doi:\nolinkurl{10.1145/3698038.3698511}}


\bibitem[Zhang et~al\mbox{.}(2024)]%
        {fisslock}
\bibfield{author}{\bibinfo{person}{Hanze Zhang}, \bibinfo{person}{Ke Cheng}, \bibinfo{person}{Rong Chen}, {and} \bibinfo{person}{Haibo Chen}.} \bibinfo{year}{2024}\natexlab{}.
\newblock \showarticletitle{Fast and Scalable In-network Lock Management Using Lock Fission}. In \bibinfo{booktitle}{\emph{18th USENIX Symposium on Operating Systems Design and Implementation (OSDI 24)}}. \bibinfo{publisher}{USENIX Association}, \bibinfo{address}{Santa Clara, CA}, \bibinfo{pages}{251--268}.
\newblock
\showISBNx{978-1-939133-40-3}
\urldef\tempurl%
\url{https://www.usenix.org/conference/osdi24/presentation/zhang-hanze}
\showURL{%
\tempurl}


\bibitem[Zhang et~al\mbox{.}(2016)]%
        {cocytus}
\bibfield{author}{\bibinfo{person}{Heng Zhang}, \bibinfo{person}{Mingkai Dong}, {and} \bibinfo{person}{Haibo Chen}.} \bibinfo{year}{2016}\natexlab{}.
\newblock \showarticletitle{Efficient and Available In-memory {KV-Store} with Hybrid Erasure Coding and Replication}. In \bibinfo{booktitle}{\emph{14th USENIX Conference on File and Storage Technologies (FAST 16)}}. \bibinfo{publisher}{USENIX Association}, \bibinfo{address}{Santa Clara, CA}, \bibinfo{pages}{167--180}.
\newblock
\showISBNx{978-1-931971-28-7}
\urldef\tempurl%
\url{https://www.usenix.org/conference/fast16/technical-sessions/presentation/zhang-heng}
\showURL{%
\tempurl}


\bibitem[Zhang et~al\mbox{.}(2023)]%
        {flex-raft}
\bibfield{author}{\bibinfo{person}{Mi Zhang}, \bibinfo{person}{Qihan Kang}, {and} \bibinfo{person}{Patrick P.~C. Lee}.} \bibinfo{year}{2023}\natexlab{}.
\newblock \showarticletitle{Minimizing Network and Storage Costs for Consensus with Flexible Erasure Coding}. In \bibinfo{booktitle}{\emph{Proceedings of the 52nd International Conference on Parallel Processing}} (Salt Lake City, UT, USA) \emph{(\bibinfo{series}{ICPP '23})}. \bibinfo{publisher}{Association for Computing Machinery}, \bibinfo{address}{New York, NY, USA}, \bibinfo{pages}{41–50}.
\newblock
\showISBNx{9798400708435}
\href{https://doi.org/10.1145/3605573.3605619}{doi:\nolinkurl{10.1145/3605573.3605619}}


\bibitem[Zhao et~al\mbox{.}(2018)]%
        {sdpaxos}
\bibfield{author}{\bibinfo{person}{Hanyu Zhao}, \bibinfo{person}{Quanlu Zhang}, \bibinfo{person}{Zhi Yang}, \bibinfo{person}{Ming Wu}, {and} \bibinfo{person}{Yafei Dai}.} \bibinfo{year}{2018}\natexlab{}.
\newblock \showarticletitle{SDPaxos: Building Efficient Semi-Decentralized Geo-replicated State Machines}. In \bibinfo{booktitle}{\emph{Proceedings of the ACM Symposium on Cloud Computing}} (Carlsbad, CA, USA) \emph{(\bibinfo{series}{SoCC '18})}. \bibinfo{publisher}{Association for Computing Machinery}, \bibinfo{address}{New York, NY, USA}, \bibinfo{pages}{68–81}.
\newblock
\showISBNx{9781450360111}
\href{https://doi.org/10.1145/3267809.3267837}{doi:\nolinkurl{10.1145/3267809.3267837}}


\bibitem[Zhou and Mu(2021)]%
        {pull-based-consensus-mongodb}
\bibfield{author}{\bibinfo{person}{Siyuan Zhou} {and} \bibinfo{person}{Shuai Mu}.} \bibinfo{year}{2021}\natexlab{}.
\newblock \showarticletitle{{Fault-Tolerant} Replication with {Pull-Based} Consensus in {MongoDB}}. In \bibinfo{booktitle}{\emph{18th USENIX Symposium on Networked Systems Design and Implementation (NSDI 21)}}. \bibinfo{publisher}{USENIX Association}, \bibinfo{pages}{687--703}.
\newblock
\showISBNx{978-1-939133-21-2}
\urldef\tempurl%
\url{https://www.usenix.org/conference/nsdi21/presentation/zhou}
\showURL{%
\tempurl}


\bibitem[Zhou et~al\mbox{.}(2023)]%
        {electrode}
\bibfield{author}{\bibinfo{person}{Yang Zhou}, \bibinfo{person}{Zezhou Wang}, \bibinfo{person}{Sowmya Dharanipragada}, {and} \bibinfo{person}{Minlan Yu}.} \bibinfo{year}{2023}\natexlab{}.
\newblock \showarticletitle{Electrode: Accelerating Distributed Protocols with {eBPF}}. In \bibinfo{booktitle}{\emph{20th USENIX Symposium on Networked Systems Design and Implementation (NSDI 23)}}. \bibinfo{publisher}{USENIX Association}, \bibinfo{address}{Boston, MA}, \bibinfo{pages}{1391--1407}.
\newblock
\showISBNx{978-1-939133-33-5}
\urldef\tempurl%
\url{https://www.usenix.org/conference/nsdi23/presentation/zhou}
\showURL{%
\tempurl}


\end{thebibliography}

\appendix
\section{Formal \tlaplus Specification}
\label{sec:appendix-tla-spec}

We append the \tlaplus specification of \crossword written as a PlusCal algorithm that can be auto-translated into \tlaplus. It has been model-checked for both \textit{linearizability} and \textit{fault-tolerance} properties on symbolic inputs of 3 machines (thus 1 allowed failure), 2 ballot numbers, 3 distinct client requests, and all Balanced Round-Robin assignment policies across 3 shards (with 2 being data shards). These inputs should be large enough to explore all the distinct, interesting execution paths of the protocol. Please refer to the inlined comments for details of this specification.

\clearpage
\pagenumbering{gobble}



\section*{Supplementary material}

--------- MODULE *Crossword* ---------

EXTENDS *FiniteSets*, *Sequences*, *Integers*, *TLC*

Model inputs & assumptions.

CONSTANT *Replicas*,           symmetric set of server nodes
         *Writes*,              symmetric set of write commands (each w/ unique value)
         *Reads*,               symmetric set of read commands
         *MaxBallot*,           maximum ballot pickable for leader preemption
         *CommitNoticeOn*,      if true, turn on *CommitNotice* messages
         *NodeFailuresOn*       if true, turn on node failures injection

$ReplicasAssumption \triangleq \land IsFiniteSet(Replicas)$
$\phantom{ReplicasAssumption \triangleq\ } \land Cardinality(Replicas) \geq 1$

$WritesAssumption \triangleq \land IsFiniteSet(Writes)$
$\phantom{WritesAssumption \triangleq\ } \land Cardinality(Writes) \geq 1$
$\phantom{WritesAssumption \triangleq\ } \land \text{``nil''} \notin Writes$

a write command model value serves as both the
*ID* of the command and the value to be written

$ReadsAssumption \triangleq \land IsFiniteSet(Reads)$
$\phantom{ReadsAssumption \triangleq\ } \land Cardinality(Reads) \geq 0$
$\phantom{ReadsAssumption \triangleq\ } \land \text{``nil''} \notin Writes$

$MaxBallotAssumption \triangleq \land MaxBallot \in Nat$
$\phantom{MaxBallotAssumption \triangleq\ } \land MaxBallot \geq 2$

$CommitNoticeOnAssumption \triangleq CommitNoticeOn \in \text{BOOLEAN}$

$NodeFailuresOnAssumption \triangleq NodeFailuresOn \in \text{BOOLEAN}$

ASSUME $\land ReplicasAssumption$
       $\land WritesAssumption$
       $\land ReadsAssumption$
       $\land MaxBallotAssumption$
       $\land CommitNoticeOnAssumption$
       $\land NodeFailuresOnAssumption$

Useful constants & typedefs.

$Commands \triangleq Writes \cup Reads$

$NumCommands \triangleq Cardinality(Commands)$

$Population \triangleq Cardinality(Replicas)$

$MajorityNum \triangleq (Population \div 2) + 1$

$Shards \triangleq Replicas$

$NumDataShards \triangleq MajorityNum$

$Range(func) \triangleq \{func[i] : i \in \text{DOMAIN } func\}$

Client observable events.

$ClientEvents \triangleq \quad [type : \{\text{``Req''}\}, cmd : Commands]$
$\phantom{ClientEvents \triangleq\ } \cup \quad [type : \{\text{``Ack''}\}, cmd : Commands,$
$\phantom{ClientEvents \triangleq\ \cup \quad [} val \ : \{\text{``nil''}\} \cup Writes]$

$ReqEvent(c) \triangleq [type \mapsto \text{"Req"}, cmd \mapsto c]$

$AckEvent(c, v) \triangleq [type \mapsto \text{"Ack"}, cmd \mapsto c, val \mapsto v]$
    *val* is the old value for a write command

$InitPending \triangleq$     $(\text{CHOOSE } ws \in [1 .. Cardinality(Writes) \rightarrow Writes]$
              : $Range(ws) = Writes)$
    $\circ$  $(\text{CHOOSE } rs \in [1 .. Cardinality(Reads) \rightarrow Reads]$
             : $Range(rs) = Reads)$
    *W.L.O.G.*, choose any sequence concatenating writes
    commands and read commands as the sequence of reqs;
    all other cases are either symmetric or less useful
    than this one

Server-side constants & states.

$Ballots \triangleq 1 .. MaxBallot$

$Slots \triangleq 1 .. NumCommands$

$Statuses \triangleq \{\text{"Preparing"}, \text{"Accepting"}, \text{"Committed"}\}$

$InstStates \triangleq [status : \{\text{"Empty"}\} \cup Statuses,$
                 $cmd : \{\text{"nil"}\} \cup Commands,$
                 $shards : \text{SUBSET } Shards,$
                 $voted : [bal \ \ : \{0\} \cup Ballots,$
                           $cmd : \{\text{"nil"}\} \cup Commands,$
                           $shards : \text{SUBSET } Shards]]$

$NullInst \triangleq [status \mapsto \text{"Empty"},$
             $cmd \mapsto \text{"nil"},$
             $shards \mapsto \{\},$
             $voted \mapsto [bal \mapsto 0, cmd \mapsto \text{"nil"}, shards \mapsto \{\}]]$

$NodeStates \triangleq [leader : \{\text{"none"}\} \cup Replicas,$
                 $kvalue : \{\text{"nil"}\} \quad \cup Writes,$
                 $commitUpTo : \{0\} \cup Slots,$
                 $balPrepared \quad : \{0\} \cup Ballots,$
                 $balMaxKnown : \{0\} \cup Ballots,$
                 $insts : [Slots \rightarrow InstStates]]$

$NullNode \triangleq [leader \mapsto \text{"none"},$
              $kvalue \mapsto \text{"nil"},$
              $commitUpTo \mapsto 0,$
              $balPrepared \quad \mapsto 0,$
              $balMaxKnown \mapsto 0,$
              $insts \mapsto [s \in Slots \mapsto NullInst]]$

$FirstEmptySlot(insts) \triangleq$
    CHOOSE $s \in Slots$ :
        $\wedge$   $insts[s].status = \text{"Empty"}$
        $\wedge$   $\forall t \in 1 .. (s - 1) : insts[t].status \neq \text{"Empty"}$

Erasure-coding related expressions.

$BigEnoughUnderFaults(g, u) \triangleq$
    Is $g$ a large enough subset of $u$ under promised fault-tolerance?
    $Cardinality(g) \geq (Cardinality(u) + MajorityNum - Population)$

$SubsetsUnderFaults(u) \triangleq$
   Set of subsets of $u$ we consider under promised fault-tolerance.
   $\{g \in \text{SUBSET } u : BigEnoughUnderFaults(g, u)\}$

$IsGoodCoverageSet(cs) \triangleq$
   Is $cs$ a coverage set (*i.e.*, a set of sets of *shards*) from which
   we can reconstruct the original data?
   $Cardinality(\text{UNION } cs) \geq NumDataShards$

$ShardToIdx \triangleq \text{CHOOSE } map \in [Shards \to 1\,..\,Cardinality(Shards)] :$
$\qquad\qquad\qquad Cardinality(Range(map)) = Cardinality(Shards)$

$IdxToShard \triangleq [i \in 1\,..\,Cardinality(Shards) \mapsto$
$\qquad\qquad\qquad \text{CHOOSE } r \in Shards : ShardToIdx[r] = i]$

$ValidAssignments \triangleq$
   Set of all valid shard assignments.
   $\{[r \;\in\; Replicas \mapsto \{IdxToShard[((i-1)\%\,Cardinality(Shards))+1] \quad :$
   $\qquad\qquad\qquad i \in (ShardToIdx[r])\,..\,(ShardToIdx[r]+na-1)\}] :$
   $\quad na \in 1\,..\,MajorityNum\}$

Service-internal messages.

$PrepareMsgs \triangleq [type : \{\text{"Prepare"}\}, src : Replicas,$
$\qquad\qquad\qquad\qquad\qquad\qquad bal : Ballots]$

$PrepareMsg(r, b) \triangleq [type \mapsto \text{"Prepare"}, src \mapsto r,$
$\qquad\qquad\qquad\qquad\qquad\qquad bal \mapsto b]$

$InstsVotes \triangleq [Slots \to [bal \;\; : \{0\} \cup Ballots,$
$\qquad\qquad\qquad\qquad cmd : \{\text{"nil"}\} \cup Commands,$
$\qquad\qquad\qquad\qquad shards : \text{SUBSET } Shards]]$

$VotesByNode(n) \triangleq [s \in Slots \mapsto n.insts[s].voted]$

$PrepareReplyMsgs \triangleq [type : \{\text{"PrepareReply"}\}, src : Replicas,$
$\qquad\qquad\qquad\qquad\qquad\qquad\qquad bal : Ballots,$
$\qquad\qquad\qquad\qquad\qquad\qquad\qquad votes : InstsVotes]$

$PrepareReplyMsg(r, b, iv) \triangleq$
$\quad [type \mapsto \text{"PrepareReply"}, src \mapsto r,$
$\qquad\qquad\qquad\qquad bal \mapsto b,$
$\qquad\qquad\qquad\qquad votes \mapsto iv]$

$PreparedConditionAndCommand(prs, s) \triangleq$
   examines a set of *PrepareReplies* and returns a tuple:
   (if the given slot can be decided as prepared,
    the prepared command if forced,
    known *shards* of the command if forced)
   $\text{LET } ppr \triangleq \text{CHOOSE } ppr \in prs :$
   $\qquad\qquad\qquad \forall\, pr \quad\; \in prs : pr.votes[s].bal \leq ppr.votes[s].bal$
   $\text{IN } \text{IF } \quad \wedge BigEnoughUnderFaults(prs, Replicas)$
   $\qquad\qquad \wedge \forall\, pr \in prs : pr.votes[s].bal = 0$
   $\qquad \text{THEN } [prepared \mapsto \text{TRUE}, cmd \mapsto \text{"nil"}, shards \mapsto \{\}]$
   $\qquad\qquad\qquad$ prepared, can choose any
   $\quad \text{ELSE IF } \wedge BigEnoughUnderFaults(prs, Replicas)$
   $\qquad\qquad\qquad \wedge IsGoodCoverageSet($
   $\qquad\qquad\qquad\qquad \{pr.votes[s].shards :$

$$pr \in \{pr \in prs \quad : \\ pr.votes[s].cmd = ppr.votes[s].cmd\}\})$$
$$\text{THEN } [prepared \mapsto \text{TRUE}, \\ cmd \mapsto ppr.votes[s].cmd, \\ shards \mapsto \text{UNION} \\ \{pr.votes[s].shards : \\ pr \in \{pr \in prs \quad : \\ pr.votes[s].cmd = ppr.votes[s].cmd\}\}]$$

prepared, command forced

$$\text{ELSE IF } \land BigEnoughUnderFaults(prs, Replicas) \\ \land \neg IsGoodCoverageSet( \\ \{pr.votes[s].shards : \\ pr \in \{pr \in prs \quad : \\ pr.votes[s].cmd = ppr.votes[s].cmd\}\})$$
$$\text{THEN } [prepared \mapsto \text{TRUE}, cmd \mapsto \text{"nil"}, shards \mapsto \{\}]$$

prepared, can choose any

$$\text{ELSE } [prepared \mapsto \text{FALSE}, cmd \mapsto \text{"nil"}, shard \mapsto \{\}]$$

not prepared

$$AcceptMsgs \triangleq [type : \{\text{"Accept"}\}, src : Replicas, \\ dst : Replicas, \\ bal : Ballots, \\ slot : Slots, \\ cmd : Commands, \\ shards : \text{SUBSET } Shards]$$

$$AcceptMsg(r, d, b, s, c, sds) \triangleq [type \mapsto \text{"Accept"}, src \mapsto r, \\ dst \mapsto d, \\ bal \mapsto b, \\ slot \mapsto s, \\ cmd \mapsto c, \\ shards \mapsto sds]$$

$$AcceptReplyMsgs \triangleq [type : \{\text{"AcceptReply"}\}, src : Replicas, \\ bal : Ballots, \\ slot : Slots, \\ shards : \text{SUBSET } Shards]$$

$$AcceptReplyMsg(r, b, s, sds) \triangleq \\ [type \mapsto \text{"AcceptReply"}, src \mapsto r, \\ bal \mapsto b, \\ slot \mapsto s, \\ shards \mapsto sds]$$

$$CommittedCondition(ars, s) \triangleq$$

the condition which decides if a set of $AcceptReplies$ makes an instance committed

$$\land BigEnoughUnderFaults(ars, Replicas) \\ \land \forall group \in SubsetsUnderFaults(ars) : \\ IsGoodCoverageSet(\{ar.shards : ar \in group\})$$

$$CommitNoticeMsgs \triangleq [type : \{\text{"CommitNotice"}\}, upto : Slots]$$

$$CommitNoticeMsg(u) \triangleq [type \mapsto \text{"CommitNotice"}, upto \mapsto u]$$

$$Messages \triangleq \quad PrepareMsgs \\ \cup \quad PrepareReplyMsgs$$

$\cup$    $AcceptMsgs$
$\cup$    $AcceptReplyMsgs$
$\cup$    $CommitNoticeMsgs$

Main algorithm in *PlusCal*.

**algorithm** *Crossword*

**variable** $msgs = \{\}$,                           messages in the network
            $node = [r \in Replicas \mapsto NullNode]$,   replica node state
            $pending = InitPending$,                 sequence of pending reqs
            $observed = \langle\rangle$,             client observed events
            $crashed = [r \in Replicas \mapsto \text{FALSE}]$;  replica crashed flag

**define**
   $UnseenPending(insts) \triangleq$
      LET $filter(c) \triangleq c \notin \{insts[s].cmd : s \in Slots\}$
      IN  $SelectSeq(pending, filter)$

   $RemovePending(cmd) \triangleq$
      LET $filter(c) \triangleq c \neq cmd$
      IN  $SelectSeq(pending, filter)$

   $reqsMade \triangleq \{e.cmd : e \in \{e \in Range(observed) : e.type = \text{"Req"}\}\}$

   $acksRecv \triangleq \{e.cmd : e \in \{e \in Range(observed) : e.type = \text{"Ack"}\}\}$

   $terminated \triangleq \land Len(pending) = 0$
                      $\land Cardinality(reqsMade) = NumCommands$
                      $\land Cardinality(acksRecv) = NumCommands$

   $numCrashed \triangleq Cardinality(\{r \in Replicas : crashed[r]\})$
**end define**;

Send a set of messages helper.

**macro** $Send(set)$ **begin**
   $msgs := msgs \cup set$;
**end macro**;

Observe a client event helper.

**macro** $Observe(e)$ **begin**
   **if** $e \notin Range(observed)$ **then**
      $observed := Append(observed, e)$;
   **end if**;
**end macro**;

Resolve a pending command helper.

**macro** $Resolve(c)$ **begin**
   $pending := RemovePending(c)$;
**end macro**;

Someone steps up as leader and sends *Prepare* message to followers.

**macro** $BecomeLeader(r)$ **begin**
   if I'm not a leader
   **await** $node[r].leader \neq r$;
   pick a greater ballot number

```
    with b ∈ Ballots do
        await ∧ b > node[r].balMaxKnown
              ∧ ¬∃ m ∈ msgs : (m.type = "Prepare") ∧ (m.bal = b);
                    W.L.O.G., using this clause to model that ballot
                    numbers from different proposers be unique
        update states and restart Prepare phase for in-progress instances
        node[r].leader := r ∥
        node[r].balPrepared   := 0 ∥
        node[r].balMaxKnown := b ∥
        node[r].insts :=
            [s ∈ Slots ↦
                [node[r].insts[s]
                    EXCEPT !.status = IF @ = "Accepting"
                                      THEN "Preparing"
                                      ELSE  @]];
        broadcast Prepare and reply to myself instantly
        Send({PrepareMsg(r, b),
              PrepareReplyMsg(r, b, VotesByNode(node[r]))});
    end with ;
end macro ;
```

Replica replies to a *Prepare* message.

```
macro HandlePrepare(r) begin
    if receiving a Prepare message with larger ballot than ever seen
    with m ∈ msgs do
        await ∧ m.type = "Prepare"
              ∧ m.bal > node[r].balMaxKnown ;
        update states and reset statuses
        node[r].leader := m.src ∥
        node[r].balMaxKnown := m.bal ∥
        node[r].insts :=
            [s ∈ Slots ↦
                [node[r].insts[s]
                    EXCEPT !.status = IF @ = "Accepting"
                                      THEN "Preparing"
                                      ELSE  @]];
        send back PrepareReply with my voted list
        Send({PrepareReplyMsg(r, m.bal, VotesByNode(node[r]))});
    end with ;
end macro ;
```

Leader gathers *PrepareReply* messages until condition met, then marks the corresponding ballot as prepared and saves highest voted commands.

```
macro HandlePrepareReplies(r) begin
    if I'm waiting for PrepareReplies
    await ∧ node[r].leader = r
          ∧ node[r].balPrepared = 0 ;
    when there are a set of PrepareReplies of desired ballot that satisfy
    the prepared condition
    with prs = {m ∈ msgs : ∧ m.type = "PrepareReply"
                           ∧ m.bal = node[r].balMaxKnown},
         exam = [s ∈ Slots ↦ PreparedConditionAndCommand(prs, s)]
    do
        await ∀ s ∈ Slots : exam[s].prepared ;
        marks this ballot as prepared and saves highest voted command
```

in each slot if any
$node[r].balPrepared := node[r].balMaxKnown \parallel$
$node[r].insts :=$
    $[s \in Slots \mapsto$
       $[node[r].insts[s]$
         EXCEPT $!.status =$ IF $\land \lor @ =$ "Empty"
                                             $\lor @ =$ "Preparing"
                                             $\lor @ =$ "Accepting"
                                  $\land exam[s].cmd \neq$ "nil"
                               THEN "Accepting"
                             ELSE  IF $@ =$ "Committed"
                                       THEN "Committed"
                                       ELSE "Empty",
                  $!.cmd\ \ \ = exam[s].cmd,$
                  $!.shards = exam[s].shards]];$

pick a reasonable shard assignment and send *Accept* messages for
in-progress instances according to it

**with** $assign \in ValidAssignments$ **do**
    $Send(\{AcceptMsg(r, d, node[r].balPrepared, s,$
                            $node[r].insts[s].cmd, assign[d]) :$
        $s \in \{s \in Slots :$
                $node[r].insts[s].status =$ "Accepting"$\},$
        $d \in Replicas\}$
    $\cup \{AcceptReplyMsg(r, node[r].balPrepared, s, assign[r]) :$
        $s \in \{s \in Slots :$
                $node[r].insts[s].status =$ "Accepting"$\}\});$
**end with** ;
**end with** ;
**end macro** ;

A prepared leader takes a new request to fill the next empty slot.

**macro** $TakeNewRequest(r)$ **begin**
    if I'm a prepared leader and there's pending request
    **await** $\land node[r].leader = r$
              $\land node[r].balPrepared = node[r].balMaxKnown$
              $\land \exists s \in Slots : node[r].insts[s].status =$ "Empty"
              $\land Len(UnseenPending(node[r].insts)) > 0;$
    find the next empty slot and pick a pending request
    **with** $s = FirstEmptySlot(node[r].insts),$
         $c = Head(UnseenPending(node[r].insts))$
                    $W.L.O.G.$, only pick a command not seen in current
                    prepared log to have smaller state space; in practice,
                    duplicated client requests should be treated by some
                    idempotency mechanism such as using request $IDs$
    **do**
        update slot status and voted
        $node[r].insts[s].status :=$ "Accepting" $\parallel$
        $node[r].insts[s].cmd := c \parallel$
        $node[r].insts[s].voted.bal\ \ := node[r].balPrepared \parallel$
        $node[r].insts[s].voted.cmd := c \parallel$
        $node[r].insts[s].voted.shards := Shards;$
        pick a reasonable shard assignment, send *Accept* messages, and
        reply to myself instantly
        **with** $assign \in ValidAssignments$ **do**

```
                    Send({AcceptMsg(r, d, node[r].balPrepared, s, c, assign[d]) :
                            d ∈ Replicas}
                        ∪ {AcceptReplyMsg(r, node[r].balPrepared, s, assign[r])}) ;
            end with ;
            append to observed events sequence if haven't yet
            Observe(ReqEvent(c)) ;
        end with ;
end macro ;
```

Replica replies to an *Accept* message.

**macro** *HandleAccept*(r) **begin**
    if receiving an unreplied *Accept* message with valid ballot
    **with** $m \in msgs$ **do**
        **await** $\land\ m.type =$ "Accept"
                $\land\ m.dst = r$
                $\land\ m.bal \geq node[r].balMaxKnown$
                $\land\ m.bal > node[r].insts[m.slot].voted.bal$ ;
        update node states and corresponding instance's states
        $node[r].leader := m.src$ ||
        $node[r].balMaxKnown := m.bal$ ||
        $node[r].insts[m.slot].status :=$ "Accepting" ||
        $node[r].insts[m.slot].cmd := m.cmd$ ||
        $node[r].insts[m.slot].shards := m.shards$ ||
        $node[r].insts[m.slot].voted.bal := m.bal$ ||
        $node[r].insts[m.slot].voted.cmd := m.cmd$ ||
        $node[r].insts[m.slot].voted.shards := m.shards$ ;
        send back *AcceptReply*
        $Send(\{AcceptReplyMsg(r, m.bal, m.slot, m.shards)\})$ ;
    **end with** ;
**end macro** ;

Leader gathers *AcceptReply* messages for a slot until condition met, then marks the slot as committed and acknowledges the client.

**macro** *HandleAcceptReplies*(r) **begin**
    if I think I'm a current leader
    **await** $\land\ node[r].leader = r$
            $\land\ node[r].balPrepared = node[r].balMaxKnown$
            $\land\ node[r].commitUpTo < NumCommands$
            $\land\ node[r].insts[node[r].commitUpTo + 1].status =$ "Accepting" ;
                *W.L.O.G.*, only enabling the next slot after *commitUpTo*
                here to make the body of this macro simpler
    for this slot, when there is a set of *AcceptReplies* that satisfy the
    committed condition
    **with** $s = node[r].commitUpTo + 1,$
          $c = node[r].insts[s].cmd,$
          $v = node[r].kvalue,$
          $ars = \{m \in msgs :\ \land\ m.type =$ "AcceptReply"
                                $\land\ m.slot = s$
                                $\land\ m.bal = node[r].balPrepared\}$
    **do**
        **await** *CommittedCondition*(ars, s) ;
        marks this slot as committed and apply command
        $node[r].insts[s].status :=$ "Committed" ||
        $node[r].commitUpTo := s$ ||
        $node[r].kvalue :=$ IF $c \in Writes$ THEN $c$ ELSE @ ;

append to observed events sequence if haven't yet, and remove
                the command from pending
                $Observe(AckEvent(c, v))$;
                $Resolve(c)$;
                broadcast *CommitNotice* to followers
                $Send(\{CommitNoticeMsg(s)\})$;
        **end with**;
**end macro**;

Replica receives new commit notification.

**macro** $HandleCommitNotice(r)$ **begin**
        if I'm a follower waiting on *CommitNotice*
        **await** $\land\, node[r].leader \neq r$
                $\land\, node[r].commitUpTo < NumCommands$
                $\land\, node[r].insts[node[r].commitUpTo + 1].status = \text{“Accepting”}$;
                        *W.L.O.G.*, only enabling the next slot after *commitUpTo*
                        here to make the body of this macro simpler
        for this slot, when there's a *CommitNotice* message
        **with** $s\ = node[r].commitUpTo + 1,$
                $c\ = node[r].insts[s].cmd,$
                $m \in msgs$
        **do**
                **await** $\land\, m.type = \text{“CommitNotice”}$
                        $\land\, m.upto = s$;
                marks this slot as committed and apply command
                $node[r].insts[s].status := \text{“Committed”} \,\|$
                $node[r].commitUpTo := s \,\|$
                $node[r].kvalue := \text{IF } c \in Writes \text{ THEN } c \text{ ELSE } @$;
        **end with**;
**end macro**;

Replica node crashes itself under promised conditions.

**macro** $ReplicaCrashes(r)$ **begin**
        if less than $(N - \text{majority})$ number of replicas have failed
        **await** $\land\, MajorityNum + numCrashed < Population$
                $\land\, \neg crashed[r]$
                $\land\, node[r].balMaxKnown < MaxBallot$;
                        this clause is needed only because we have an upper
                        bound ballot number for modeling checking; in practice
                        someone else could always come up with a higher ballot
        mark myself as crashed
        $crashed[r] := \text{TRUE}$;
**end macro**;

Replica server node main loop.

**process** $Replica \in Replicas$
**begin**
        $rloop$: **while** $(\neg terminated) \land (\neg crashed[self])$ **do**
                **either**
                        $BecomeLeader(self)$;
                **or**
                        $HandlePrepare(self)$;
                **or**
                        $HandlePrepareReplies(self)$;

```
            or
                TakeNewRequest(self);
            or
                HandleAccept(self);
            or
                HandleAcceptReplies(self);
            or
                if CommitNoticeOn then
                    HandleCommitNotice(self);
                end if;
            or
                if NodeFailuresOn then
                    ReplicaCrashes(self);
                end if;
            end either;
        end while;
    end process;

end algorithm;
```

―――――――――――――――――――― MODULE $Crossword\_MC$ ――――――――――――――――――――

EXTENDS $Crossword$

*TLC* config-related defs.

$ConditionalPerm(set) \triangleq$ IF $Cardinality(set) > 1$
$\qquad\qquad\qquad\qquad\qquad$ THEN $Permutations(set)$
$\qquad\qquad\qquad\qquad\qquad$ ELSE $\{\}$

$SymmetricPerms \triangleq \quad ConditionalPerm(Replicas)$
$\qquad\qquad\qquad\quad \cup \quad ConditionalPerm(Writes)$
$\qquad\qquad\qquad\quad \cup \quad ConditionalPerm(Reads)$

$ConfigEmptySet \triangleq \{\}$

$ConstMaxBallot \triangleq 2$

―――――――――――――――――――――――――――――――――――――――――――――

Type check invariant.

$TypeOK \triangleq \land \forall m \in msgs : m \in Messages$
$\qquad\qquad \land \forall r \in Replicas : node[r] \in NodeStates$
$\qquad\qquad \land Len(pending) \leq NumCommands$
$\qquad\qquad \land Cardinality(Range(pending)) = Len(pending)$
$\qquad\qquad \land \forall c \in Range(pending) : c \in Commands$
$\qquad\qquad \land Len(observed) \leq 2 * NumCommands$
$\qquad\qquad \land Cardinality(Range(observed)) = Len(observed)$
$\qquad\qquad \land Cardinality(reqsMade) \geq Cardinality(acksRecv)$
$\qquad\qquad \land \forall e \in Range(observed) : e \in ClientEvents$
$\qquad\qquad \land \forall r \in Replicas : crashed[r] \in$ BOOLEAN

THEOREM $Spec \Rightarrow \Box TypeOK$

―――――――――――――――――――――――――――――――――――――――――――――

*Linearizability* constraint.

$ReqPosOfCmd(c) \triangleq \text{CHOOSE } i \in 1 \mathbin{..} Len(observed):$
$\qquad\qquad\qquad\quad \wedge observed[i].type = \text{``Req''}$
$\qquad\qquad\qquad\quad \wedge observed[i].cmd = c$

$AckPosOfCmd(c) \triangleq \text{CHOOSE } i \in 1 \mathbin{..} Len(observed):$
$\qquad\qquad\qquad\quad \wedge observed[i].type = \text{``Ack''}$
$\qquad\qquad\qquad\quad \wedge observed[i].cmd = c$

$ResultOfCmd(c) \triangleq observed[AckPosOfCmd(c)].val$

$OrderIdxOfCmd(order, c) \triangleq \text{CHOOSE } j \in 1 \mathbin{..} Len(order) : order[j] = c$

$LastWriteBefore(order, j) \triangleq$
$\quad \text{LET } k \triangleq \text{CHOOSE } k \in 0 \mathbin{..} (j-1):$
$\qquad\qquad\qquad\quad \wedge (k = 0 \vee order[k] \in Writes)$
$\qquad\qquad\qquad\quad \wedge \forall l \in (k+1) \mathbin{..} (j-1) : order[l] \in Reads$
$\quad \text{IN } \text{IF } k = 0 \text{ THEN ``nil'' ELSE } order[k]$

$IsLinearOrder(order) \triangleq$
$\quad \wedge \{order[j] : j \in 1 \mathbin{..} Len(order)\} = Commands$
$\quad \wedge \forall j \in 1 \mathbin{..} Len(order):$
$\qquad ResultOfCmd(order[j]) = LastWriteBefore(order, j)$

$ObeysRealTime(order) \triangleq$
$\quad \forall c1, c2 \in Commands:$
$\qquad (AckPosOfCmd(c1) < ReqPosOfCmd(c2))$
$\qquad\quad \Rightarrow (OrderIdxOfCmd(order, c1) < OrderIdxOfCmd(order, c2))$

$Linearizability \triangleq$
$\quad terminated \Rightarrow$
$\qquad \exists\, order \in [1 \mathbin{..} NumCommands \rightarrow Commands]:$
$\qquad\quad \wedge IsLinearOrder(order)$
$\qquad\quad \wedge ObeysRealTime(order)$

THEOREM $Spec \Rightarrow Linearizability$

---

— Crossword_MC.cfg —

```
SPECIFICATION Spec

CONSTANTS
    Replicas = {s1, s2, s3}
    Writes = {w1, w2, w3}
    Reads <- ConfigEmptySet
    MaxBallot <- ConstMaxBallot
    CommitNoticeOn <- FALSE
    NodeFailuresOn <- TRUE

SYMMETRY SymmetricPerms

INVARIANTS
    TypeOK
    Linearizability

CHECK_DEADLOCK TRUE
```



\end{document}